\documentclass[onecolumn,apj,numberedappendix]{emulateapj}
\usepackage{srcltx}
\usepackage{graphicx}

\newcommand{\be}{\begin{equation}}
\newcommand{\ee}{\end{equation}}
\newcommand{\ba}{\begin{eqnarray}}
\newcommand{\ea}{\end{eqnarray}}

\begin{document}
\title{Statistical Description of Synchrotron Intensity Fluctuations: Studies of
Astrophysical Magnetic Turbulence}

\author{A. Lazarian}
\affil{Department of Astronomy, University of Wisconsin,
Madison, US}
\author{D. Pogosyan}
\affil{Physics Department, University of
Alberta, Edmonton, Canada}

\begin{abstract}
We provide a theoretical description of synchrotron fluctuations arising from
magnetic turbulence. We derive an expression that relates the correlation of
synchrotron fluctuations for an arbitrary index of relativistic electrons to
the correlations arising from a particular $\gamma=2$ index that provides 
synchrotron emissivity proportional to the squared intensity of perpendicular
to the line of sight component of magnetic field. We provide a detailed study of the
statistics in the latter case assuming that the underlying magnetic turbulence
is axisymmetric. We obtain general relations valid for an arbitrary model of
magnetic axisymmetric turbulence and analyze the relations for the
particular example of magnetic turbulence that is supported by numerical simulations. 
We predict that the synchrotron intensity fluctuations
are anisotropic with larger correlation present along the direction of magnetic field. This
anisotropy is dominated by the quadrupole component with the ratio between quadrupole and
monopole parts being sensitive to the compressibility of underlying turbulence.
Our work opens avenues for quantitative studies of magnetic turbulence in our galaxy and beyond
using synchrotron emission. It also outlines the directions of how synchrotron
foreground emission can be separated from cosmological signal, i.e. from CMB or
highly redshifted HI emission. For the sake of completeness we also provide the expressions
for the synchrotron polarization (Stocks parameters and their combinations) for the model
of axisymmetric magnetic turbulence. 

\end{abstract}


\section{Introduction}
Synchrotron fluctuations both carry information about magnetic fields and
interfere with the measurements of cosmic microwave emission (CMB) as well as
with attempts to measure enigmatic emission from atomic hydrogen (HI) in the
early Universe. Therefore it is essential to have proper description of the
statistics of synchrotron fluctuations. This paper concentrates on the
statistical properties of the synchrotron intensities arising from turbulent
magnetic fields.  

Galactic and extragalactic synchrotron emission arises from the interaction of
astrophysical magnetic fields and cosmic rays \citep[see][]{Ginzburg1981}. In terms of
CMB and high redshift HI studies, the most important are galactic synchrotron
emission. This emission provides the largest range of scales for studying
magnetic fields. 
Magnetic fields are turbulent as observations testify that turbulence is ubiquitous in
astrophysics \citep{1995ApJ...443..209A,2009SSRv..143..357L,2010ApJ...710..853C}. As 
relativistic electrons are present in most cases, the turbulence  results in
synchrotron fluctuations, which may provide
detailed information about magnetic fields at different scales, but, at the same
time, impede measures of CMB and high redshift HI. The latter
has become a direction of intensive discussion recently 
\citep[see][]{LoebZaldarriaga2004, Pen.etal2008, LoebWyithe2008, Liu2009}.

Present and future big telescopes, e.g. Square Kilometer Array (SKA), LOw
Frequency ARray (LOFAR), are designed to map synchrotron emission with
unprecedented resolution and sensitivity. Therefore it is absolutely crucial
that the information available with these instruments is being used to get
better insight fundamental astrophysical processes.

Properties of turbulent magnetic fields that influences most of the processes in
diffuse astrophysical environments are very much desired through the
corresponding studies. The obstacle for such studies is that the relation
between the underlying turbulence and the synchrotron fluctuations is not
trivial because the synchrotron intensity depends both on magnetic field
$H$ and on the power index $\alpha$ of relativistic electrons, which emit
radiation spiraling in magnetic field. The physics of emission is rather
non-trivial which results in the expression of synchrotron intensity which are
proportional to magnetic field in a fractional power, $H^{\gamma}$,
$\gamma=1/2(\alpha+1)$ that has been an impediment for a correct description of
synchrotron fluctuations.

For instance, studies of synchrotron intensity fluctuations in the Milky Way
(see Cho \& Lazarian 2010 and references therein) reveal power law spectra that
can be related to magnetic turbulence. How are these spectra related to the
underlying spectra of magnetic fluctuations?  What is the effect of the electron
spectral index on the result? Are the fluctuations of magnetic field related to
compressions or have Alfvenic nature? Are there additional information apart
from the magnetic spectrum one can obtain from synchrotron fluctuations? These
are the questions that we are going to address in the paper.  

A number of attempts to describe synchrotron fluctuations has been attempted
assuming that $\gamma=2$, which provides an approximation to the actual index of
galactic synchrotron emission 
\citep[see][]{Getmantsev,Chibisov,LazarianShutenkov,LazarianChibisov,1998A&AT...17..281C}.
Some of these papers \citep[e.g.][]{Getmantsev,1998A&AT...17..281C} did not
consider mean magnetic field while others dealt with the spectrum of
fluctuations that arise from unrealistic idealized isotropic magnetic turbulence.
As we discuss further, such an isotropic model contradicts the current
understanding of magnetic turbulence. 
 
An attempt to take anisotropy into account was made in Lazarian \&
Shutenkov (1990), where the anisotropy of the synchrotron fluctuations was
analyzed and the quadruple component of the spatial decomposition of the
synchrotron correlations was used to study the spectrum of synchrotron
fluctuations. The work used the explicit integral equations
solving the inverse problem for the magnetic fluctuations, but the anisotropy
was introduced in the problem by assuming that the magnetic field can be
presented as a superposition of a regular and isotropic random magnetic fields.
The latter assumption was not supported by the later numerical and theoretical
research.

In short, the limitations of all earlier studies were related both to the
incorrect models of astrophysical turbulence and to the constrains arising from
doing calculation just for a single cosmic ray electron index $\gamma=2$. The
magnetic turbulence is known to be anisotropic and the actual index may
substantially differs from that single chosen value and it is not clear how
accurate the description of synchrotron fluctuations stays for $\gamma$ not
equal to 2. 

The goal of this paper is to address both of the deficiencies of the earlier
studies. First of all, recent years have been marked by important advances in
the theory of compressible MHD turbulence and its statistical description 
\citep[e.g][]{1995ApJ...438..763G,1999ApJ...517..700L,2000ApJ...539..273C,2001ApJ...554.1175M,
2001ApJ...562..279L,2002ApJ...564..291C,
2002PhRvL..88x5001C,2003MNRAS.345..325C,2010ApJ...720..742K}.
This understanding of
turbulence should be included in the description of synchrotron fluctuations.
In addition, it is absolutely vital to understand what is happening with the
synchrotron statistics for $\gamma$ different from the single value of
$\gamma=2$ that was the focus of the earlier studies.

Our paper relays synchrotron fluctuations with realistic magnetic turbulence for
arbitrary electron power law index extending the reliability of magnetic field
studies. We also propose new ways of studying magnetic fields, in particular,
separating contributions from compressible and incompressible magnetic
fluctuations.   

In this paper we do not directly address the synchrotron polarization studies.
We, however, should mention that the existing studies attempting to answer
important questions about magnetic fields, e.g. question of magnetic helicity,
using synchrotron (see Waelkens, Schekochihin \& En{\ss}lin 2009, Junklewitz \&
En{\ss}lin 2011) suffer the from same limitations as the earlier 
statistical studies of synchrotron intensity, i.e. they also assume $\gamma=2$
and isotropy of magnetic turbulence. We believe that our approach should be
useful for these studies. To illustrate our point we address the problem of
Faraday rotation measure fluctuations arising from anisotropic turbulence. 

Our present paper is complementary to our work on probing turbulence using
spectral measurements. In terms of those studies focused on recovering the
spectrum of velocity turbulence from Doppler broadened absorption and emission
lines \citep{2000ApJ...537..720L,2004ApJ...616..943L,2006ApJ...652.1348L,2008ApJ...686..350L}
the present paper is
intended to provide additional information about magnetic field fluctuations.
This study is intended to provide the basis for the future development of
techniques that recover turbulent statistics from observations using not only
intensity, but other Stocks parameters of synchrotron emission. 

Our study is intended to provide foundations for studying magnetic turbulence in
the halo of Milky Way, in supernova remnants, in external galaxies as well as in
clusters of galaxies (see En{\ss}lin et al. 2010, 2009, En{\ss}lin \& Vogt
2006). As the resolution of telescopes gets higher detailed studies of
turbulence in astrophysical jets, synchrotron emitting lobes get possible.

In \S 2 we justify the model of magnetic turbulence that we adopt, in \S 3 we
describe the fluctuations of synchrotron emissivity and obtain general
expressions for synchrotron fluctuations
for an arbitrary spectral index of relativistic electrons. In \S 4 we analyse
the anisotropy of
synchrotron fluctuations arising from an arbitrary model of axisymmetric
turbulence, while in \S 5 we show
how this general treatment gets modified when a model of MHD turbulence that
follows from numerical simulations
is used. We discuss our results in \S 6 and provide our summary in \S 7. Appendixes
are important part of these work. Apart from containing derivations related
to the synchrotron intensity correlations they also contain the statistics of correlations
of other Stocks parameters and useful combinations of the Stocks parameters (see Appendix E).

\section{Statistical description of MHD turbulence: perspective of an observer}

MHD Turbulence plays a crucial role for the processes of cosmic ray propagation
\citep[see][]{2002cra..book.....S,Longair}, star formation 
\citep[see][]{2004ARA&A..42..211E,2007ARA&A..45..565M},
heat transfer in magnetized plasmas 
\citep[see][]{2001ApJ...562L.129N,2006ApJ...645L..25L},
magnetic reconnection \citep[see][]{1999ApJ...517..700L,2009ApJ...700...63K}.
Statistical description allows to reveal
regular features within chaotic picture
of turbulent fluctuations. The famous Kolmogorov description of incompressible
hydrodynamic turbulence provides a vivid example of how turbulence complexity
can be reduced to a simple formula
$E(k)\sim k^{5/3}$, where $E(k)$ is an energy spectrum of turbulent motions.

MHD turbulence is more complex than the hydrodynamical one. Magnetic field
defines the chosen direction of anisotropy 
\citep{1981PhFl...24..825M,1983JPlPh..29..525S,1984ApJ...285..109H}.
For small scale motions this is true even in the absence of
the mean magnetic field in the system. In this situation the magnetic field of
large eddies defines the direction of anisotropy for smaller eddies. This
observation brings us to the notion of {\it local system of reference}, which is
one of the major pillars of the modern theory of MHD turbulence. Therefore a
correct formulation of the theory requires wavelet description 
\citep[see][]{2010ApJ...720..742K}.
Indeed, a customary description of anisotropic turbulence using
parallel and perpendicular wavenumbers assumes that the direction is fixed in
space. In this situation, however, the turbulence loses its universality in the
sense that, for instance, the critical balance condition of the widely accepted
model incompressible MHD turbulence \citep[][henceforth GS95]{1995ApJ...438..763G}
which expresses the equality of the time of wave
transfer along the magnetic field lines and the eddy turnover time is not
satisfied\footnote{The original GS95 model was formulated in terms of global
system of reference and used closure relations which are only valid in this
system of reference. This point was corrected in further publications 
\citep{1999ApJ...517..700L,2000ApJ...539..273C,2001ApJ...554.1175M,2002ApJ...564..291C}
where the necessity of using local reference system defined by magnetic field on the scale
of turbulence study was suggested, justified and tested. When in the literature
the critical balance and the relation between parallel and perpendicular scales
are still expressed in terms of parallel and perpendicular wavenumbers, the
corresponding wavenumbers should be understood as shorthand notations of the
inverse scales calculated in the local reference frame, rather than wavenumbers
in the system of reference related to the mean magnetic field.}.

MHD turbulence is a developing field with its ongoing debates\footnote{Recent
debates, for instance, were centered on the role of dynamical alignment or
polarization intermittency that could change the slope of MHD turbulence from
the Kolmogorov slope predicted in the GS95 to a more shallow slope observed in
numerical simulations \citep[see][]{2005ApJ...626L..37B,2006PhRvL..96k5002B,2006ApJ...640L.175B}.
More recent studies \citep[see][]{2009ApJ...702.1190B,2010ApJ...722L.110B} indicate that
numerical simulations may not have enough dynamical range to test the actual
spectrum of turbulence and the flattening of the spectrum measured in the
numerical simulations is expected due to MHD turbulence being less local than
its hydro counterpart.}. However, we feel that among all the existing models the
GS95 provides the best correspondence to the existing numerical and
observational data \citep[see][]{2010ApJ...722L.110B,2010ApJ...710..853C}.
A limitation of the original GS95 model is that it assumes that the injection of
energy happens with $V_L$ equal to the Alfv\'en velocity $V_A$. In astrophysical
situations turbulence can be injected with both $V_L<V_A$ and $V_L>V_A$. For the
first case the turbulence starts initially in the regime of weak turbulence when
Alfv\'enic perturbations weakly interact with each other. As scales of motions
decrease, the strength of interactions, paradoxically, increases and the
turbulence gets stronger. This regime was described analytically in Lazarian \&
Vishniac (1999, henceforth LV99). The regime $V_L>V_A$ is the regime of super-Alfv\'enic turbulence,
when turbulence initially follows the Kolmogorov cascade and magnetic fields are
not dynamically important. At smaller scales the turbulence gets into the strong
MHD turbulence regime \citep[see][]{2006ApJ...645L..25L}.  We should add that the exact form of
the magnetic correlation tensor does not follow from the GS95 model and should
be obtained from numerical experiments. This was done in \citet{2002ApJ...564..291C}
and \citet{2003MNRAS.345..325C}.

The local system of reference is not accessible to an observer who deals with
projection of magnetic fields from the volume to the pictorial plane. The
projection effects inevitably mask the actual direction of magnetic field
within individual eddies along the line of sight. As a result, the scale
dependent anisotropy predicted in the GS95, which taking into our considerations
above (see footnote 1)
 should be written as $\lambda_{\|}\sim L^{1/3}
\lambda_{\bot}^{2/3}$, where $\lambda_{\|}$ and $\lambda_{\bot}$ are the
parallel and perpendicular scale of the eddy, respectively, is not valid for the
observer measuring parallel and perpendicular scales of projected and averaged
along the line of sight eddies. As the observer maps the projected magnetic
field in the global reference, e.g. system of reference of the mean field, the
anisotropy of eddies becomes {\it scale-independent} and the degree of
anisotropy
gets determined by the anisotropy of the largest eddies which projections are
mapped. This property of eddies being scale independent in the global system of
reference is discussed in the earlier works 
\citep[e.g.][]{2002ApJ...564..291C,2005ApJ...631..320E}.

The testing of how ideas of incompressible MHD turbulence are applicable in the
presence of compressibility was performed in 
\citet{2002PhRvL..88x5001C,2003MNRAS.345..325C} \citep[also see][for a review]{2005ThCFD..19..127C}.
There, a decomposition of motions into basic MHD modes (Alfv\'enic, slow and fast)
was performed to show that Alfv\'enic and slow modes keep the scaling of
the incompressible MHD, while fast modes exhibit isotropy
\citep[see GS95,][]{2001ApJ...562..279L}.

In a number of situations  magnetic turbulence becomes isotropic for the
observer.
One of them is when the turbulence is super-Alfv\'enic, i.e. the velocity at the
injection scale $V_L$ is larger than the Alfv\'en speed $V_A$. Even in the
opposite case, $V_A>V_L$, the
observer may still detect isotropic turbulence if the line of sight is directed
along the mean field in the volume. Super-Alfv\'enic
turbulence may take place for molecular clouds \citep[see][]{2004ApJ...604L..49P} and 
clusters of galaxies \citep{2007MNRAS.378..245B}.

In the presence of the mean magnetic field in the volume under study, an
observer will see anisotropic turbulence, where statistical properties of magnetic field
differ in the directions orthogonal and parallel to the mean magnetic field.
The description of axisymmetric turbulence was
given by \citet{1946RSPSA.186..480B}, \citet{1950RSPTA.242..557C} and later 
\citet{1981PhRvA..24.2135M} and \citet{1997PhRvE..56.2875O}.
In our case it is
natural to identify the axis of symmetry with the mean magnetic field and then
the index-symmetric part of the correlation tensor can be presented in the
following form:
\begin{equation}
\langle H_i({\bf x_1}) H_j({\bf x_2}) \rangle = A_\xi(r, \mu)
\hat r_i \hat r_j +  B_\xi(r, \mu) \delta_{ij} + C_\xi(r, \mu) \hat \lambda_i
\hat \lambda_j + D_\xi(r, \mu) \left( \hat r_i \hat \lambda_j + \hat r_j
\hat \lambda_i \right)
\end{equation}
where  the separation vector ${\bf r} = {\bf x}_1 - {\bf x}_2$ has the
magnitude $r$ and the direction specified by the unit vector $\hat {\bf r}$.
The direction of the symmetry axis set by the  mean magnetic field is given by
the unit
vector
$\hat {\bf \lambda}$ and $\mu = \hat {\bf r} \cdot \hat {\bf \lambda}$.  The
magnetic field correlation tensor may
also have antisymmetric, helical part, however it does not contribute to
synchrotron correlations. Hence, further on
we shall consider only the index-symmetric part of  $\langle H_i({\bf x_1})
H_j({\bf x_2}) \rangle$.

The structure function of the field has the same representation
\begin{eqnarray}
\frac{1}{2} \langle \left( H_i({\bf x_1}) - H_i({\bf x_2}) \right) \left(
H_j({\bf x_1}) - H_j({\bf x_2}) \right) \rangle =
A(r, \mu) \hat r_i \hat r_j +  B(r, \mu) \delta_{ij} + C(r, \mu) \hat \lambda_i
\hat \lambda_j + D(r, \mu) \left( \hat r_i \hat \lambda_j + \hat r_j
\hat \lambda_i
\right)
\label{eq:Dij}
\end{eqnarray}
with coefficients $A(r,\mu) = A_\xi(0,\mu) - A_\xi(r,\mu) ~, \ldots$ etc.  In
case of power-law spectra, one can use either the correlation function or the
structure function, depending on the spectral slope. In this case the structure
function coefficients can be thought of as renormalized correlation
coefficients.

In Appendix~\ref{sec:3Dpower_to_corr} we discuss the relation of the correlation
tensor and spectrum
tensor representation of the axisymmetric turbulent fields. An observer studying
synchrotron does not map projected magnetic fluctuations directly, as the
synchrotron emission is a non-linear function of the magnetic field component
perpendicular to the line of sight. This is the subject of the next section.

\section{3D spatial correlations of synchrotron emissivity}

The goal of this section is to show that one can introduce measures of synchrotron correlation which
very weakly depend on the electron spectral distribution, i.e. on $\gamma$. Whenever possible we provide the
prove analytically for selected cases. We also show results of numerical calculations.

\subsection{Synchrotron correlations}

Synchrotron emission arises from relativistic electron spiraling about magnetic
fields. The emission has been discussed in many monographs 
\citep[see Pacholczyk 1970,][and references therein]{Fleishman}. Careful study of the formation of the
synchrotron signal \citep[see][]{1959ApJ...130..241W} revealed that the
signal is {\it essentially} non-linear in the magnetic field $H$ ($B$).
The origin of nonlinearity is in relativistic effects.
Nonlinearity comes from the fact that the signal is formed only over the narrow
fraction of the electron cycle, and the two leading orders in the
deviation of the trajectory from the straight line give the same contribution
in terms of $1/\gamma$. Summation of the result over ``flashes'' is also
non-straightforward,
and produces spread frequency spectrum \citep{1959ApJ...130..241W}.
Situation is remarkably different from cyclotron (non relativistic)
emission where emission is monochromatic and has intensity just quadratic in the
magnetic field.

If the distribution of relativistic electrons is
\begin{equation}
N_e({\cal E})d{\cal E}\sim {\cal E}^{\alpha} d{\cal E}
\end{equation}
then the observer sees intensity of the synchrotron emission is, thus,
\begin{equation}
I_{sync}({\bf X}) \propto \int dz H_{\perp}^\gamma({\bf x})
\end{equation}
where ${\bf X} = (x,y)$ is the 2D position vector on the sky and
$H_{\perp} = \sqrt{H_x^2 + H_y^2} $ is the magnitude of the magnetic
field perpendicular to the line-of-sight $z$. Note that
$\gamma=\onehalf(\alpha+1)$ is, generally,
a fractional power.

The relativistic electron power law index $\alpha$ changes from object to object
and also varies with energy of electrons.
For galactic radio halo tested at meter wavelengths, observations indicate that
$\alpha\approx 2.7$ \citep[see][]{1996A&A...307L..57P}. The index $\alpha$
and  therefore $\gamma$ may vary due to processes of acceleration and loses. In
the simplest models of shock acceleration $\alpha=2$ \citep[see][]{Longair} and
in the acceleration in turbulent reconnection $\alpha=2.5$ 
\citep[see][]{2005A&A...441..845D}. Those should produce $\gamma=1.5$ and $\gamma=1.75$,
respectively. In reality,
the acceleration of particles is a more sophisticated process which in case of a
shock includes the formation of the precursor and its interaction with the media
\citep[see][]{2009ApJ...707.1541B,2009ApJ...692.1571M}
as well as various feedback processes. As a result
of these and propagation effects $\gamma$ will vary. In this paper we consider
variations of $\gamma$ in the range from 1 to  4, which covers the
astrophysically important cases we are aware of.

In observations, one measures the  correlation function of the synchrotron
intensity
\begin{equation}
\xi_{sync}({\bf R}) \equiv \left\langle I_{sync}({\bf X_1}) I_{sync}({\bf X_2})
\right\rangle
\propto \int dz_1 dz_2 \left\langle
H_{\perp}^\gamma \left({\bf x}_1 \right)
H_{\perp}^\gamma \left({\bf x}_2 \right)
\right\rangle
\label{eq:synchro_gen}
\end{equation}
by averaging over an ensemble of the pairs of the
sky measurements at fixed two dimensional separation ${\bf R}=\bf {X_1} - {\bf
X_2} $.
This function is the projection of the three-dimensional correlation of
emissivity
\begin{equation}
\xi_{H_\perp^\gamma}({\bf R},z) = \left\langle
H_{\perp}^\gamma \left({\bf x}_1 \right)
H_{\perp}^\gamma \left({\bf x}_2 \right)
\right\rangle
\label{eq:3Dcorr_emissivity}
\end{equation}
which for homogeneous turbulence depends only on ${\bf x_1} - {\bf x_2} = ({\bf
R}, z)$.

The structure function is formally related to the correlation one
\begin{equation}
D_{sync}({\bf R}) \equiv  \left\langle \left(I_{sync}({\bf X_1}) - 
I_{sync}({\bf X_2})\right)^2 \right\rangle
\propto 2 \int dz_1  \int dz
\left( \xi_{H_\perp^\gamma}(0,z) - \xi_{H_\perp^\gamma}({\bf R},z) \right)
\label{eq:synchro_strucut}
\end{equation}
but it can often be defined even when the correlation function itself has a
divergent behaviour.

Thus, we are presented with the problem of describing the correlation  between
the fractional powers of the magnitudes of the orthogonal projections of
a vector field,  $\xi_{H_\perp^\gamma}$.
Our main aim is to demonstrate that one
can develop statistical measures that are insensitive to the parameter $\gamma$.
Using these statistics simplifies the interpretation of the observational signal
in terms of
the properties of the magnetized turbulence.

Our approach is the following.  Let us note that however complex the statistics
of the magnetic
field is, the observed correlations are functions of just three functional
quantities,
$\xi_{xx}({\bf r}) = \left\langle H_x({\bf x_1}) H_x({\bf x_2})\right\rangle $,
$\xi_{yy}({\bf r}) = \left\langle H_y({\bf x_1}) H_y({\bf x_2}) \right\rangle $
and
$\xi_{xy}({\bf r}) = \left\langle H_x({\bf x_1}) H_y({\bf x_2}) \right\rangle $.
To study the $\gamma$ dependence of the correlation of emissivity
$\xi_{H_\perp^\gamma}$, we shall consider it
in the space spanned by $\xi_{xx}, \xi_{yy}, \xi_{xy}$ as independent variables,
$\xi_{H_\perp^\gamma}(\xi_{xx},\xi_{yy},\xi_{xy})$.

\subsection{Statistically isotropic magnetic field and synchrotron correlations}

First we consider the case when statistics of the magnetic field is isotropic,
which may, for instance,
correspond to the super-Alfv\'enic turbulence, i.e. for the turbulence with the
injection velocity much
in excess of the Alfv\'enic one.
The structure tensor of a Gaussian isotropic vector field, a special case of
Eq.~(\ref{eq:Dij}),
is usually written in the form
\begin{equation}
\left\langle \left( H_i ( {\bf x}_1 )- H_i ({\bf x}_2) \right) \left( H_j ( {\bf
x}_1 )- H_j ({\bf x}_2) \right) \right\rangle
= \left( D_{LL} - D_{NN}  \right) \hat r_i \hat r_j + D_{NN} \delta_{ij} \quad,
\end{equation}
where $D_{LL}(r)$ and $D_{NN}(r)$ are structure functions that describe,
respectively, the correlation of the vector components
parallel and orthogonal to point separation ${\bf r}$.
In case of solenoidal vector field, in particular the magnetic field,
two structure functions are related by
\begin{equation}
\frac{d}{d r} D_{LL} = - \frac{2}{r} \left( D_{LL} - D_{NN} \right)
\label{eq:solenoid_condition}
\end{equation}
which in the regime of the power-law behaviour $D_{NN} \propto r^m$ leads to
both functions being proportional to each other
$D_{LL} = \frac{2}{2+m} D_{NN}$.

To describe our system we chose the x-axis
to coincide with the direction of the projected separation ${\bf R}$.
This can be done for analysis of isotropic statistics since there is no
other preferred direction. In this case, we find the following two correlations
\begin{eqnarray}
d_1 \equiv \left\langle \left[ H_x({\bf X}_1,z_1) - H_x({\bf X}_2,z_2)\right]^2
\right\rangle
&=& \left( D_{LL} - D_{NN}  \right) \sin^2\theta + D_{NN}  \\
d_2 \equiv \left\langle \left[ H_y({\bf X}_1,z_1) - H_y({\bf X}_2,z_2) \right]^2
\right\rangle  &=&  D_{NN}  \\
\left\langle \left[ H_x({\bf X}_1,z_1) - H_y({\bf X}_2,z_2)\right]^2
\right\rangle &=& 0
\end{eqnarray}
to determine completely the correlations of the orthogonal components of the
magnetic field.
Here $\theta$ is the angle between the 3D separation vector ${\bf r}$ and the
line of sight.

For convenience we shall also introduce the normalized correlation coefficients
\begin{eqnarray}
c_1 &\equiv&  \left\langle H_x({\bf X}_1,z_1) H_x({\bf X}_2,z_2) \right\rangle /
\left\langle H_x({\bf X}_1,z_1)^2 \right\rangle
= 1 - \frac{1}{2} d_1 \\
c_2 &\equiv&  \left\langle H_y({\bf X}_1,z_1) H_y({\bf X}_2,z_2) \right\rangle /
\left\langle H_y({\bf X}_1,z_1)^2 \right\rangle
= 1 - \frac{1}{2} d_2
\end{eqnarray}
which vary from unity for coincident points to zero at large separation.
Normalized structure functions,
conversely, are zero at small separations and reach asymptotically the value of
two where the
correlation vanishes.

Let us now investigate the {\it normalized} correlations
\begin{eqnarray}
\tilde \xi_{H_\perp^\gamma}&=&\left(\left\langle
H_{\perp}^\gamma \left({\bf x}_1 \right)
H_{\perp}^\gamma \left({\bf x}_2 \right)
\right\rangle -
\left\langle H_{\perp}^\gamma \left({\bf x} \right)
\right\rangle^2 \right) /
\left(
\left\langle H_{\perp}^\gamma \left({\bf x} \right)^2
\right\rangle
-
\left\langle H_{\perp}^\gamma \left({\bf x} \right)
\right\rangle^2
\right) \\
\tilde D_{H_\perp^\gamma} &=& 2 \left( 1 - \tilde \xi_{H_\perp^\gamma} \right)
\end{eqnarray}
where the mean and the variance can be trivially expressed through Gamma
functions
\begin{eqnarray}
\left\langle H_{\perp}^\gamma \left({\bf x} \right) \right\rangle &=&
2^{\gamma/2} \Gamma\left[1 + \gamma/2\right] H^\gamma\\
\left\langle H_{\perp}^\gamma \left({\bf x} \right)^2 \right\rangle &=&
2^{\gamma} \Gamma\left[1 + \gamma \right] H^{2\gamma}
\label{disper}
\end{eqnarray}
where $H$ is the amplitude of the magnetic field.
Both the mean $\left\langle H_{\perp}^\gamma \left({\bf x} \right)
\right\rangle$ and the variance
$\left\langle H_{\perp}^\gamma \left({\bf x} \right)^2 \right\rangle
-\left\langle H_{\perp}^\gamma \left({\bf x} \right) \right\rangle^2$
that determine the average intensity and the rms fluctuations of the synchrotron
emission are strong
functions of $\gamma$ index. However, as we shall demonstrate, the
normalized correlation is not, thus providing
a direct measure to the correlation properties of the magnetic field.

Below we identify the normalized correlations of synchrotron emission as the measures which 
depend on the index $\gamma$ only weakly. The goal is to reduce the problem of relating magnetic field
and synchrotron statistics for an arbitrary $\gamma$ to the mathematically trackable case of $\gamma=2$.

\subsubsection{Special cases of spectral index: $\gamma=2$ and $4$}

In case of even integer $\gamma$, among which of the most interest are cases
$\gamma=2$ and $\gamma=4$,
the correlation of the emissivity can be found analytically to yield
\begin{eqnarray}
\tilde \xi_{H_\perp^2} &=& \frac{1}{2} \left( {c_1}^2 + {c_2}^2
\right)
\quad\quad\quad\quad\quad\quad\quad\quad\quad\quad\quad\quad\quad\quad\gamma=2
\label{eq:alpha=0}\\
 \tilde \xi_{H_\perp^4} &=& \frac{2}{5} \left( {c_1}^2 + {c_2}^2
\right)  \label{eq:alpha=2}
+ \frac{3}{40} \left( {c_1}^4 + {c_2}^4 \right) + \frac{1}{20} {c_1}^2 {c_2}^2
\quad\quad \gamma=4
\end{eqnarray}
Starting at unity when $c_1=c_2=1$, the correlations show remarkably little
difference throughout the whole
range of $c_1$ and $c_2$, although $\xi_{H_\perp^4} \sim \frac{4}{5}
\xi_{H_\perp^2} $
as  $c_1 \to 0, c_2 \to 0$.

In terms of the structure functions
\begin{eqnarray}
\tilde D_{H_\perp^2} &=& \left( d_1 + d_2\right) - \frac{1}{4} \left(d_1^2
+ d_2^2 \right)
\quad\quad\quad\quad\quad\quad\quad\quad\quad\quad\gamma=2
\label{eq:Dalpha=0}\\
\tilde D_{H_\perp^4}  &=& \frac{6}{5} \left(d_1 + d_2\right) -
\frac{1}{20}\left(d_1+d_2\right)^2 \label{eq:Dalpha=2} \ldots
\quad\quad\quad\quad\quad\quad\quad \gamma=4
\end{eqnarray}
Although a different index leads to a change of the amplitude at small
separation
where linear term dominates by a factor $6/5$ it does not affect the slope, and
as we shall see numerically the difference in the transition to
non-linear terms will be difficult to observe.

\subsubsection{Case of equal correlation coefficients $c_1 = c_2 $}
\begin{figure}[ht]
\centerline{\includegraphics[width=0.4\textwidth]{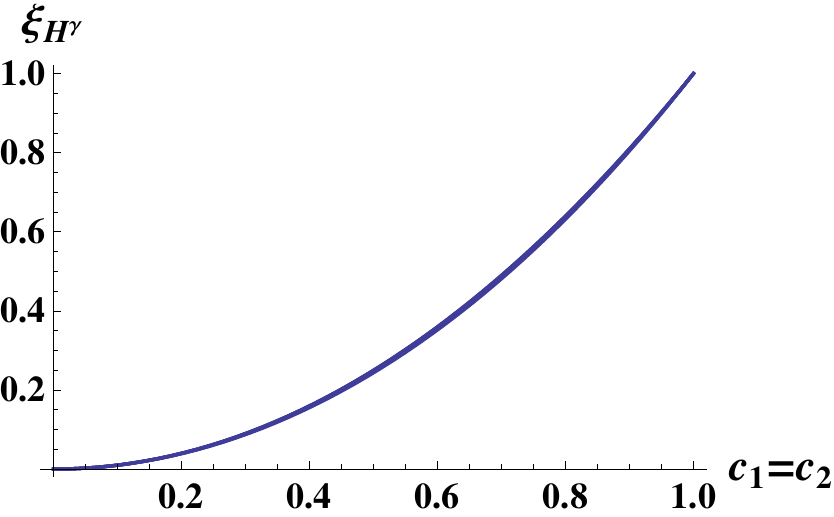}\hspace{1cm}
\includegraphics[width=0.4\textwidth]{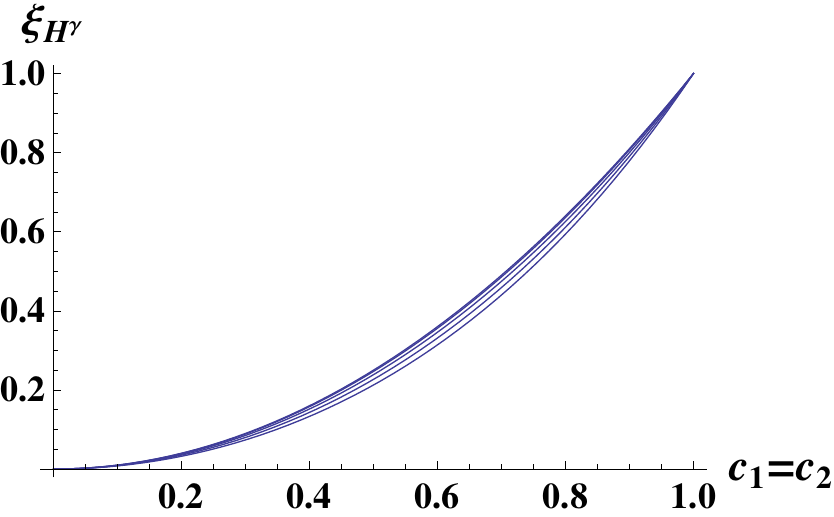}}
\caption{Correlation coefficients $ \xi_{H_\perp^\gamma}(c_1) $ for coincident
$c_1=c_2$ for several
$\gamma \in (1.5,2.8)$ (left) and $\gamma \in (2.5,4)$ (right). }
\label{fig:d1eqd2}
\end{figure}

Another special test case that can be treated analytically is the case of
equal correlation between normal components and correlation of
longitudinal components.   In this case, for arbitrary $\gamma$
\begin{equation}
\xi_{H_\perp^\gamma}(c_2=c_1) =
\frac{1 - _{2}\!\!F_1\left[-\frac{\gamma}{2},-\frac{\gamma}{2},1,c_1^2\right] }
{1 - \Gamma[1+\gamma]\Gamma\left[1+\gamma/2\right]^{-2} }
\label{eq:c1=c2}
\end{equation}
where $ _{2}F_1\left[-\frac{\gamma}{2}\right]$ is the hypergeometric
function \citep[see][]{aaa}.
As the Figure~\ref{fig:d1eqd2} shows,
this result is also nearly $\gamma$ independent for $\gamma \in (1.5,3)$, and
with only minor
differences showing only for $\gamma > 3$

\subsubsection{Conjecture of $\gamma$ independence}

\begin{figure}[ht]
\centerline{\includegraphics[width=0.4\textwidth]{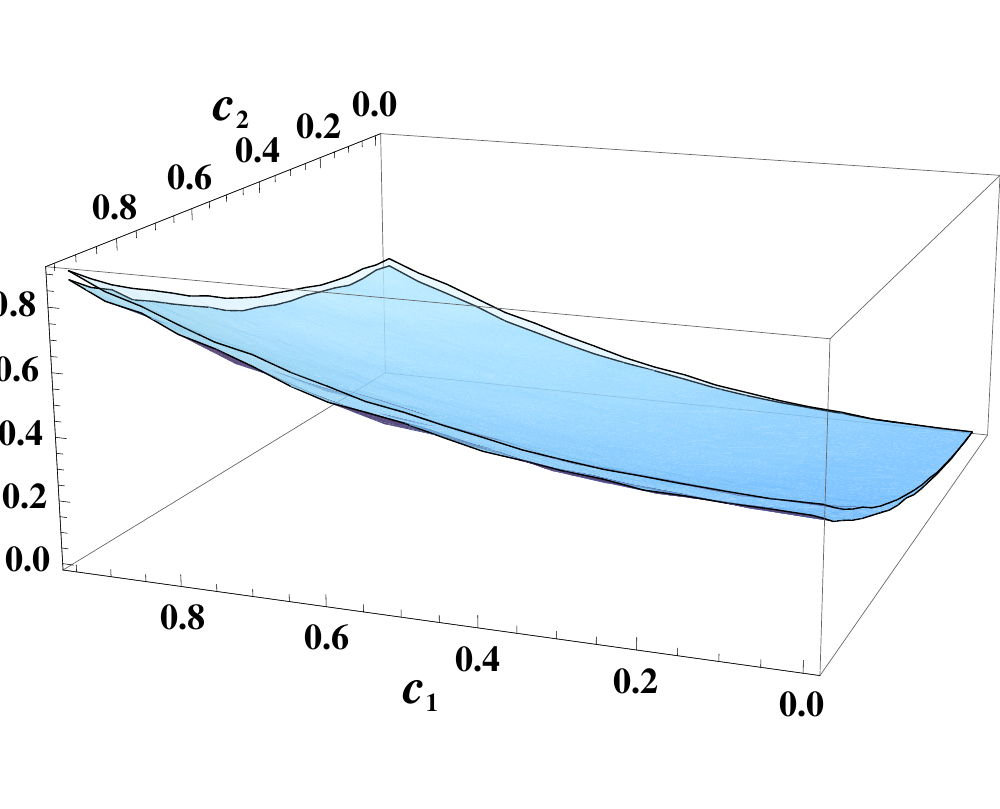}\hspace{
1cm}
\includegraphics[width=0.4\textwidth]{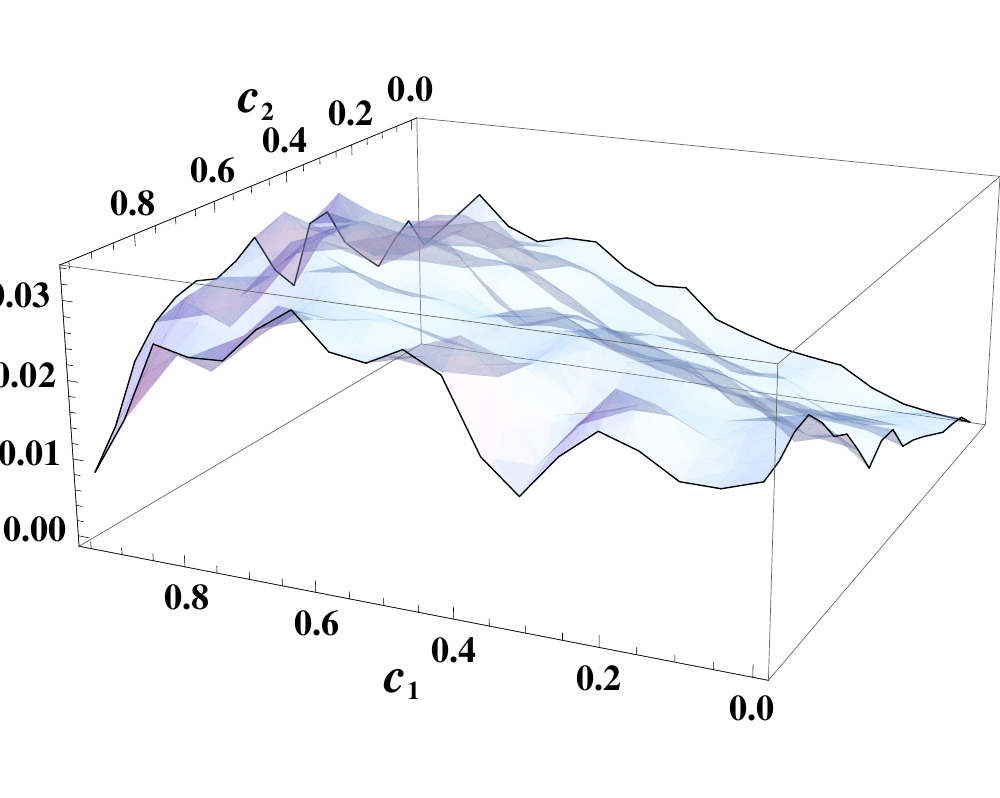}}
\caption{Left: Example comparison of the correlation coefficients $
\xi_{H_\perp^\gamma} $ for $\gamma=2$ (top surface)
 and $\gamma=3.5$ (bottom surface).
Right: The difference between two surfaces in the left panel. The systematic
separation between those two quite distinct
cases is below 5\%. Random features at 1-2\% level are due to limitations of
numerical integration.}
\label{fig:num_sol}
\end{figure}

Based on the test cases above we conjecture that the {\it normalized}
correlation function of the synchrotron
is insensitive to the $\gamma$-parameter in the range of our interest.
To test this conjecture we compute numerically $\xi_{H_\perp^\gamma}$ for
non-integer $\gamma$'s.
The results presented in Figure~\ref{fig:num_sol} confirm
insensitivity to $\gamma$ in all the range of interest. The maximal deviation in
correlations are limited to 3\% for $\gamma$
in the wide range $\gamma \in (1.2,3)$, while staying within 10\% for
$\gamma=4$.

Consequently, we conclude that the isotropic correlation of 3D synchrotron
emissivity
can be well approximated by a simple quadratic result of $\gamma=2$
\begin{eqnarray}
\tilde\xi_{H_\perp^\gamma} &\approx&  \frac{1}{2} \left( {c_1}^2 + {c_2}^2
\right)
\label{eq:xi_approx}
\\
\tilde D_{H_\perp^\gamma} & \approx & \left( d_1 + d_2 \right) - \frac{1}{4}
\left(d_1^2 + d_2^2 \right)
\label{eq:D_approx}
\end{eqnarray}
In other words, while the intensity of emission changes appreciably, the
correlation properties of the normalized intensities depend
on $\gamma$ only weakly.
It is important to note that while the mapping from the magnetic field to
synchrotron emissivity statistics
is essentially non-linear in terms of correlation function, it is linear at
small scales if the structure functions are used.
That makes a structure function the preferable statistics to apply.  The effects
of $\gamma$ manifests itself in an amplitude
shift of the small scale scaling, an effect difficult to observer without very
wide dynamical range.

\subsection{Example of statistically anisotropic signal}

While correlation properties of synchrotron emission in statistically isotropic
magnetic field are described with two functions, in general, full
description requires three function $D_{xx}$, $D_{yy}$ and $D_{xy}$. Study of
emission from axisymmetric turbulence belongs to this general case.

Here we shall consider the
synchrotron emission in the limit when the anisotropy can be expected to play
the largest role. This is the limit when a) the symmetry axis is orthogonal
to the line-of-sight and b) the field has no variations in the direction of the
symmetry axis.  If we chose x-direction to coincide with the axis of symmetry
(z-direction is along the line of sight) in this limit
\begin{eqnarray}
\sigma_{xx} &=& 0, \quad \sigma_{xy} = 0,  \quad\sigma_{yy} = \sigma \nonumber
\\
D_{xx} &=& 0, \quad D_{xy} = 0, \quad D_{yy} = \sigma^2 d ~,
\end{eqnarray}
i.e. the correlations are described just by a single function $d$.
Such a case has physical significance when the axis of symmetry is associated
with
a regular component of magnetic field, so realistic model should include
non-zero mean field in x-direction, $\langle H_x \rangle \ne 0$.
Perpendicular to the line of sight component of the magnetic field now has a
random, $y$ and regular, $x$ components, $H_\perp^2 = H_y^2 + \langle H_x
\rangle^2 $.

\begin{figure}[h]
\centerline{\includegraphics[width=0.4\textwidth]{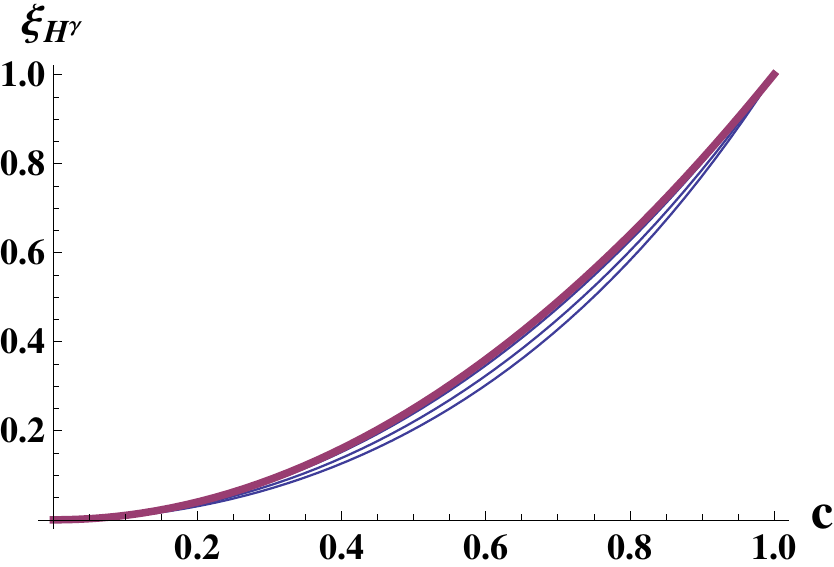} \hspace{1cm}
\includegraphics[width=0.4\textwidth]{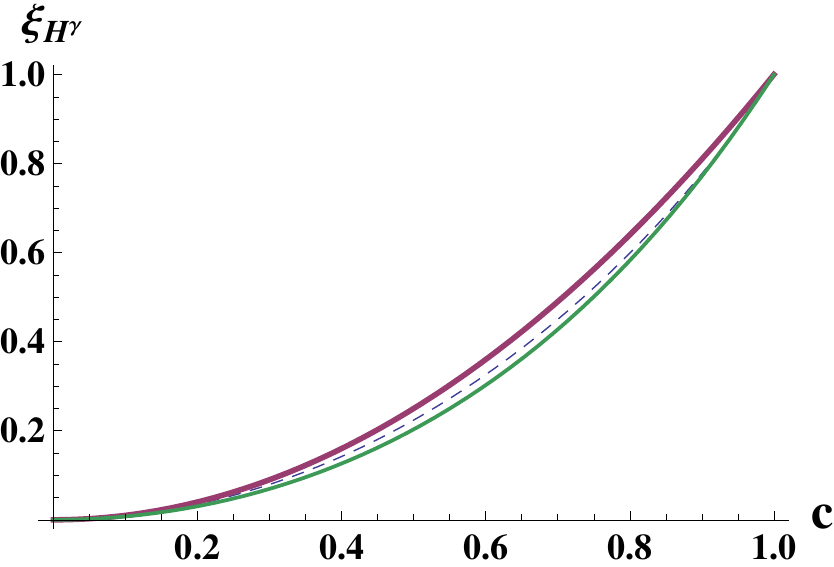}}
\caption{Left: correlation coefficients $ \xi_{H_\perp^4}(c) $ for $\gamma=4$
model in strongly
anisotropic turbulence with varying strength of the magnitude of
the regular component of the magnetic field (dashed curves starting for the
lowest one at $A=0$).
The correlation approaches the
$\gamma=2$ (solid) curve as $\langle H_x \rangle$ increases beyond
the rms of the random component $\langle H_y^2 \rangle^{1/2}$. Right:
$\gamma$-dependence in the limit of anisotropic turbulence with absent mean
field $A=0$.
The upper curve is $\gamma=2$ case, same as on the right, the dashed curve
corresponds to $\gamma=1$ and the bottom solid
one to $\gamma=4$.}
\label{fig:ani4A}
\end{figure}

The 3D emissivity of the synchrotron is correlated according to
\begin{equation}
\xi_{H_\perp^\gamma} = \left\langle
\left( H_{y}^2({\bf r_1}) + \langle H_x \rangle^2 \right)^{\gamma/2}
\left( H_{y}^2({\bf r_2}) + \langle H_x \rangle^2 \right)^{\gamma/2}
\right\rangle -
\left\langle \left( H_{y}^2 + \langle H_x \rangle^2 \right)^\gamma
\right\rangle
\end{equation}
which is readily computed
for $\gamma=2$ and $\gamma=4$
\begin{eqnarray}
 \tilde \xi_{H_\perp^2} &=& c^2  ~, \quad\quad\quad\quad\quad\quad\quad\quad
 \quad\quad\quad\quad\quad\quad\quad\quad
\gamma=2
\label{eq:anialpha=0}\\
\tilde \xi_{H_\perp^4} &=& c^2 \left( 1 - \frac{3 (\sigma^4 -
c^2)}{\langle
H_x \rangle^4 + 6 \langle H_x \rangle^2 \sigma^2 + 12 \sigma^4} \right)
~,\quad\quad \gamma=4
 \label{eq:anialpha=2}
\end{eqnarray}
where $c$ is a normalized correlation function of the magnetic field
fluctuations
perpendicular to the regular component of magnetic field.
We find that the correlation properties of the synchrotron do not depend at all
on the amplitude of the regular magnetic component if $\gamma=2$.
The left panel in Figure~\ref{fig:ani4A} demonstrates for $\gamma=4$ that the
difference from simple $\gamma=2$ result
decreases quickly with the strength of the regular field.
Between these two cases, the correlation differ by no more that 2\% along all
the range
of values when $ \langle H_x \rangle > 3 \sigma $. Similarly, for all
intermediate values of
$\gamma$, $\gamma=2$ formula (\ref{eq:anialpha=0}) provide the limiting
behaviour when
regular magnetic component is increased.

Thus, we conclude that $\gamma$ dependence of normalized correlations is very
weak for anisotropic turbulence in the presence of the notable regular magnetic
field component. With good precision the result can be approximated by
$\gamma=2$ formula $ \xi_{H_\perp^2} = \langle H_y({\bf r_1}) H_y ({\bf
r_2})\rangle^2 $.

The strongest sensitivity to $\gamma$ is  expected when the regular
component is small, $ \langle H_x \rangle \ll \sigma $. While the case
of strictly  one-dimensional $\langle H_x \rangle = 0$, $\langle H_x \rangle^2 =
0$ turbulent magnetic field is probably of little astrophysical importance, it
provides a useful, mathematically tractable limit, of strongly anisotropic
turbulence. In this limit
\begin{eqnarray}
\left\langle H_\perp^\gamma \right\rangle &=& \frac{2^{\gamma/2} \sigma^\gamma
\Gamma\left[\frac{1+\gamma}{2}\right]}{\sqrt{\pi}}
~,\quad \left\langle (H_\perp^\gamma)^2 \right\rangle = \frac{2^\gamma
\sigma^{2\gamma}
\Gamma\left[\frac{1}{2}+\gamma\right]
 }{\sqrt{\pi}}
 \\
\tilde \xi_{H_\perp^\gamma} &=& \frac{
{}_{2}F_1\left[-\frac{\gamma}{2},-\frac{\gamma}{2},\frac{1}{2},c^2\right] -1  }
{\sqrt{\pi}\Gamma\left[\frac{1}{2}+\gamma\right]\Gamma\left[\frac{1+\gamma}{2}
\right] ^ { -2 } -1 } \\
\tilde D_{H_\perp^\gamma} &=&
\frac{\sqrt{\pi}\Gamma\left[\frac{1}{2}+\gamma\right]-
\Gamma\left[\frac{1+\gamma}{2}\right]^{2}
{}_{2}F_1\left[-\frac{\gamma}{2},-\frac{\gamma}{2},\frac{1}{2},c^2\right]}
{\sqrt{\pi}\Gamma\left[\frac{1}{2}+\gamma\right]-\Gamma\left[\frac{1+\gamma}{2}
\right]^{2}}
\end{eqnarray}
The right panel in Figure~\ref{fig:ani4A} shows how mapping from the magnetic
field to
emissivity correlations behave in this limit. The $\gamma=2$
amplitude-independent
curve provides the upper bound for both $\gamma < 2 $ and $\gamma > 2$ models.

\subsection{ 2D correlations of synchrotron intensity in isotropic turbulence
and $\gamma$ dependence }

Integration of the synchrotron emissivity along the line of sight provides the
observer with the intensity of the emission.
To demonstrate that it does not modify our conjecture of weak sensitivity to
$\gamma$, we perform a detailed study
of $\gamma=2$ and $\gamma=4$ isotropic models.

Line-of-sight projection inevitably includes correlation between widely separated
regions of space. Thus one needs a
statistical model of the magnetic field that is self-consistent at large
separations. In physical environment the scaling
of the turbulence regime can not extend to arbitrary large scales and is
expected to saturate at the scales associated
with energy injection $r_I$ or the size of the turbulent region. We shall
consider a simple prescription for the structure
function of normal components
\begin{equation}
D_{NN}(r) = D(\infty) \left(\frac{r^2}{r^2 + r_I^2}\right)^{m/2}
\end{equation}
which follows the power-law scaling $ \propto r^m $ at small scales and
saturates at large separations at
the value equal twice the variance of the field component $D(\infty)=2
\left\langle H_i^2 \right\rangle$.
To satisfy the solinoidality condition of Eq.~(\ref{eq:solenoid_condition}) the
longitudinal structure function
is
\begin{equation}
D_{LL}(r) = \frac{D(\infty)}{1+m/2} \left(\frac{r}{r_I}\right)^m
{}_2F_1\left(1+\frac{m}{2},\frac{m}{2},2+\frac{m}{2},-\left(\frac{r}{r_I}
\right)^2 \right)
\end{equation}

\begin{figure}[ht]
\centerline{\includegraphics[width=0.4\textwidth]{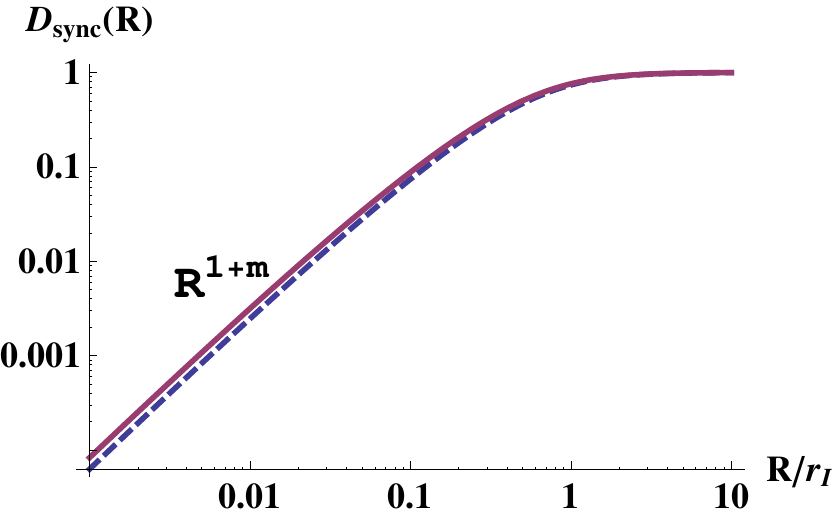}\hspace{1cm}
\includegraphics[width=0.4\textwidth]{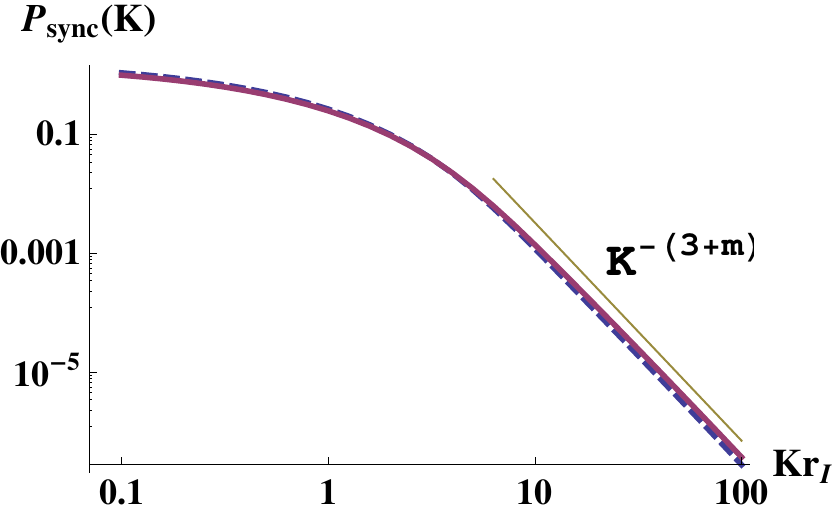}}
\caption{The structure functions of synchrotron intensity (left) and the power
spectrum (right) in the model of isotropic Kolmogorov magnetized turbulence
with $\gamma=2$ (dashed) and $\gamma=4$ indexes. The curves correspond to
$m=2/3$.}
\label{fig:2Dsyncg24}
\end{figure}

The left panel in Figure~\ref{fig:2Dsyncg24}
shows the resulting structure functions of synchrotron intensity on the sky in
this isotropic model for
$\gamma=2$ and $\gamma=4$.  While equal at large separations,  two cases show
small functional difference in the transitional
regime where effects of non-linear mapping between the magnetic field and
synchrotron manifest themselves.
As the result, the small-scale amplitudes, where the synchrotron structure
function essentially linearly reflects the
magnetic field structure function, differ for $\gamma=2$ and $\gamma=4$ by the
constant factor $\sim 6/5$, in accordance with
Eq.~(\ref{eq:Dalpha=2}). Such constant factor, however, is not readily
observable if synchrotron correlations are mapped over small scales only.
It is essential to track the behaviour beyond the scaling regime through
non-linear transition to large scales.
This encounters the following issue - the nonlinear, $\gamma$ sensitive,
response of the synchrotron takes place where
$d \propto D_{NN}/D(\infty) \sim 1$, i.e close to energy injection scale $r_I$.
Thus, to map it with accuracy sufficient to separate the
different $\gamma$ cases, would require an accurate modeling of the largest
scales of the turbulent cascade.

The right panel in Figure~\ref{fig:2Dsyncg24} demonstrates the power spectrum
that corresponds to the structure
functions in the left panel.  The power spectrum for $\gamma=2$ has been studied
in detail in \citet{1998A&AT...17..281C}. The nonlinear mapping
from the magnetic field to synchrotron emission affects the spectrum at long waves
$K r_I < 10$, proportional to the scale of turbulence
saturation. The short scale asymptotics $\propto K^{-(3+m)}$ follows the
linearized relation between magnetic field and synchrotron
statistics, with only the amplitude but not the slope being weakly dependent on
$\gamma$.

\subsection{Correlation function of intensity for arbitrary $\gamma$}
The investigation in this section allows us to approximate the
emissivity correlations by the  ansatz
\begin{eqnarray}
\xi_{H_\perp^\gamma}({\bf r}) \approx  {\cal P}(\gamma) \xi_{H_\perp^2}({\bf
r}), \quad\quad
D_{H_\perp^\gamma}({\bf r}) \approx {\cal A}(\gamma) {\cal P}(\gamma)
 D_{H_\perp^2}({\bf r}),
\end{eqnarray}
where the strongly dependent on $\gamma$ amplitude ${\cal P}(\gamma) \equiv
\frac{\left\langle (H_\perp^\gamma)^2 \right\rangle - \left\langle
H_\perp^\gamma \right\rangle^2}{\left\langle H_\perp^4 \right\rangle -
\left\langle H_\perp^2 \right\rangle^2}$ is factorized from
the scaling and angular dependences described by $\gamma=2$ term.
For the structure function, adjusting the amplitude in a weakly
$\gamma$-dependent fashion with ${\cal A}(\gamma) \sim 1$ makes the
approximation even more accurate at small scales.

This ansatz translates to the observable structure function of the synchrotron
radiation
\begin{eqnarray}
D_{sync, \gamma}({\bf R}) \approx {\cal A}(\gamma) {\cal P}(\gamma)
D_{sync, \gamma=2}({\bf R})
\end{eqnarray}
according to which the scaling and angular dependence of the
synchrotron structure function can be understood from studying $\gamma=2$ case.
For isotropic magnetic fields the above expression can be rewritten using Eq.
(\ref{disper}) 
\begin{eqnarray}
D_{sync, \gamma}({\bf R}) \approx {\cal A}(\gamma)
2^{\gamma-2} H^{2\gamma-4} \left(
\Gamma\left[1+\gamma\right]
-\Gamma\left[1+\onehalf\gamma\right]^2 \right)
D_{sync, \gamma=2}({\bf R})
\end{eqnarray}

In case of axisymmetric turbulence one can choose the $x$ coordinate to be
aligned with the sky projection of the symmetry axis. In this frame 
covariance of the components of the magnetic field is diagonal,
$\sigma_{xy}=0$ and the ratio of the variances $P(\gamma)$ can be expressed
via $\sigma^2=\sigma_{xx}+\sigma_{yy}$ and $\epsilon \equiv
\frac{\sigma_{xx}-\sigma_{yy}}{ \sigma_{xx} + \sigma_{yy}}$ to give
\begin{eqnarray}
D_{sync, \gamma}({\bf R}) &\approx& {\cal A}(\gamma)
\sigma^{2\gamma-4} D_{sync, \gamma=2}({\bf R}) \times  \\
&\times& \frac{\left(1 - \epsilon^2\right)^{\gamma}}{1+\epsilon^2}
\left[\left(1-\epsilon^2\right)^{\onehalf} 
\Gamma(1+\gamma) 
\mbox{}_{2}F_1 \left(\frac{1}{2}+\frac{\gamma}{2},1+\frac{\gamma}{2},1,
\epsilon^2\right) 
-\Gamma\left(1+\frac{\gamma}{2}\right)^2
\mbox{}_{2}F_1 \left(\frac{1}{2}+\frac{\gamma}{4},1+\frac{\gamma}{4},1,
\epsilon^2\right)^2
\right] \nonumber
\label{impor}
\end{eqnarray}

Eq. (\ref{impor}) relates the structure function of synchrotron intensity for arbitrary index of relativistic
electrons, i.e. for arbitrary $\gamma$, with the 
structure function for synchrotron intensity fluctuations corresponding to $\gamma=2$. The additional factors
do not depend on the distance between points for which the correlation is thought, but uniformly change the amplitude of the structure function. Both $\gamma$ and the $\epsilon$ can be obtained independently as (see Discussion). Further we simplify out treatment focusing on the case of $\gamma=2$, while Eq. (\ref{impor}) allows us to generalize the results obtained for $\gamma=2$ for an arbitrary index $\gamma$.

\section{Anisotropy of the synchrotron intensity statistics}

This section is focused on describing synchrotron fluctuations arising from a general model of axisymmetric anisotropic magnetic turbulence\footnote{A particular model of MHD turbulence that corresponds to numerical
simulations is considered in the next section.}. On very general grounds one can expect mean magnetic field to influence
statistics of magnetic turbulence. 

\subsection{Power spectrum of the axisymmetric MHD turbulence}

The axisymmetric tensors presented in \S 2 describe structure functions of
turbulence. In terms of the spectral representation
 of axisymmetric
turbulence a general form of the index-symmetric correlation tensor of the
solenoidal
vector field in the presence of a preferred direction ${\bf \hat \lambda}$ is
given by two scalar power spectra \citep{1997PhRvE..56.2875O}
\begin{eqnarray}
\left\langle H_i H_j \right\rangle = \frac{1}{(2\pi)^3} \int d^3{\bf k}
\; e^{i {\bf k}\cdot{\bf r}}
 \left[ E({\bf k}) \left( \delta_{ij} - \hat k_i \hat k_j
\right)
+ F({\bf k}) \frac{ ({\bf \hat k} \cdot {\bf \hat \lambda})^2 \hat k_i \hat
k_j + \hat \lambda_i \hat \lambda_j - ({\bf \hat k} \cdot {\bf \hat \lambda})
(\hat k_i \hat \lambda_j+\hat k_j \hat \lambda_i)}{1 - ({\bf \hat k} \cdot {\bf
\hat \lambda})^2}\right] ~.
\label{eq:axisym_turb}
\end{eqnarray}
In the isotropic limit only $E$ spectrum is present and is function only of
the wave number, i.e. $E(k)$.
For axisymmetric turbulence, $E({\bf k})$ and $F({\bf k})$ functions depend
on the wave number and the angle $\mu = {\bf \hat k} \cdot {\bf \hat \lambda}$
between the wave vector and the symmetry axis.
The form of Eq.~(\ref{eq:axisym_turb}) is fully general if the statistics has a
mirror symmetry in addition to
the axial one
\footnote{In the most general form two other pseudo-scalar spectra, $C$ and $H$
in
the notation of \citet{1997PhRvE..56.2875O}, that describe correlations between
toroidal and
poloidal components of the magnetic field appear. Both terms are traceless and
do not contribute to the rms energy
of the magnetic field and are absent for axisymmetric turbulence with mirror
symmetry.
One of them, $H$, describes helical correlation resulting in index antisymmetric
part of the correlation tensor
which is irrelevant for synchrotron studies. We shall not consider possible
contribution of the remaining $C$ term
in this paper.
}.
The normalization of tensors is chosen such as to have angle-independent traces.

What combinations of $E$ and $F$ spectra correspond to independent modes is
determined by the physical mechanisms of wave excitation and propagation.
In the context of MHD, where the axis of symmetry is associated with
the direction (perhaps local) of the mean magnetic field,
correlations due to Alfv\'enic modes correspond to
the linear combination of ``$E$-type'' and ``$F$-type'' spectra with
$F(k)=-E(k)$, while the slow and fast modes are described by the ``$F$-type''
correlation tensors \citep[][see also below]{2004ApJ...614..757Y}.

When  anisotropy is global, the scaling of power with wavenumber $k$ and
the angular dependence on $\mu$  factorizes. Limiting ourselves to the power-law
scaling we shall consider
\begin{equation}
E(k, \mu) = A_E k^{-3-m_E} \widehat{E} \left( \mu \right)~, \quad
F(k, \mu ) = A_F k^{-3-m_F} \widehat{F} \left( \mu \right)
\label{eq:globalasymspectrum}
\end{equation}
where $m=2/3$ would correspond to Kolmogorov scaling.
For example, in Alfv\'enic turbulence the fluctuations of the field are
suppressed in the direction of the mean magnetic field, so one expects
$\widehat{E} \left( \mu \right)$ to disfavor $\mu= \pm 1$ direction. We
normalize
the angular functions to have unit monopoles in Legendre expansion,
$\widehat{E}(\mu) = \sum_{l=0}^\infty \widehat{E}_l P_l(\mu)$, etc.,
\begin{equation}
\widehat{E}_0 \equiv \onehalf \int_{-1}^1 d \mu \widehat{E} \left( \mu \right) =
1,
\quad
\widehat{F}_0 \equiv \onehalf \int_{-1}^1 d \mu \widehat{F} \left( \mu \right) =
1
\end{equation}

In the presence of anisotropy the magnetic field components parallel
and perpendicular to the symmetry axis have unequal
variances, $\sigma^2_\parallel \equiv \langle ({\bf H} \cdot {\bf \hat
\lambda})^2 \rangle \ne
\onehalf \sigma^2_\perp \equiv \onehalf \langle({\bf H} \times {\bf \hat
\lambda})^2 \rangle$.
 Since
$\langle ({\bf H} \cdot {\bf \hat
\lambda}) ({\bf H} \times {\bf \hat
\lambda}) \rangle = 0 $ due to azimuthal symmetry, $\sigma^2_\parallel$ and
$\onehalf \sigma^2_\perp$ are the eigenvalues of the variance tensor
$\sigma_{ij}=\langle H_i({\bf x}) H_j({\bf x}) \rangle $, and are given by
\begin{eqnarray}
\sigma^2_\parallel &=& \frac{P_E}{2} \int_{-1}^1 d\mu \;
(1-\mu^2) \widehat{E}(\mu) +
\frac{P_F}{2} \int_{-1}^1 d \mu \; (1-\mu^2)  \widehat{F}(\mu) \\
\sigma^2_\perp &=& \frac{P_E}{2}  \int_{-1}^1 d\mu \;
(1 + \mu^2) \widehat{E}(\mu) + \frac{P_F}{2} \int_{-1}^1 d \mu \; \mu^2
\widehat{F}(\mu)
\end{eqnarray}
where $P_{E,F} = 4 \pi A_{E,F} \int k^2 dk k^{-3-m_{E,F}} $ are measures
of the
power in, respectively, $E$ and $F$ parts of the spectrum. Note that individual
amplitudes can, in principle, be negative, while the conditions
$P_E + P_F \ge 0 $ and $4 P_E + P_F \ge 0$ must hold.

The ratio of the two variances $r_\sigma$ is constant in the case of measurement
performed in the global system of reference
\begin{equation}
r_\sigma \equiv \frac{2 \sigma^2_{\parallel}}{\sigma^2_{\perp}}
= \frac{P_E + P_F - \frac{1}{5}\left(P_E \widehat{E}_2 + P_F \widehat{F}_2
\right)}{P_E + \onequarter P_F + \frac{1}{10} \left(P_E \widehat{E}_2 +
P_F \widehat{F}_2 \right)} \quad .
\label{epsilon}
\end{equation}
The parameter $r_\sigma=1$ in the isotropic case when  $P_F = 0$ and
$\widehat{E}(\mu)=1$. Values $r_\sigma < 1$ indicate the suppression while $r_\sigma >
1$ the enhancement of the variance in the component parallel to the symmetry
axis.
Notable anisotropic limit of interest is that of Alfv\'enic turbulence,
where $P_F = - P_E$ and the component of the
magnetic field along the mean is not perturbed, hence $\sigma_\parallel^2=0$
and $r_\sigma=0$.
$E$-type perturbations
with completely suppressed waves along the mean magnetic field,
$P_F=0$, $\widehat{E}(\mu) \propto \delta(\mu)$ lead to  $r_\sigma=2$.
On the other hand, the essentially anisotropic
$F$-type modes give $r_\sigma=4$ even when the spectral function $F$ is
angle-independent.


\subsection{Transition from local to global axisymmetry}
\label{sec:L_wandering}
In many physical situations, the preferred direction is local in nature and
is only approximately maintained over the extended regions. In particular,
in MHD turbulence $\hat \lambda$ is associated with the direction of
the mean magnetic field which may and usually does change over large scales.
In this paper we are
interested in the possible anisotropy of the synchrotron signal due to
existence of a preferred direction over extended
regions of space, i.e a global effect. Such anisotropy is expected when
the direction of the magnetic field has a long-range coherence, i.e
$\hat\lambda$ can be thought of as fluctuating around the mean global direction
$\hat\lambda_0$.

One can estimate the effect of $\hat\lambda$ ``wandering''
by averaging the tensor form of the correlations given  by
Eq.~(\ref{eq:axisym_turb}) over the distribution of $\hat\lambda$ around the
mean $\hat\lambda_0$ (see wandering induced by Alfvenic turbulence in Lazarian \& Vishniac 2000).
Qualitatively the outcome is clear if the distribution
of polar angle around $\hat\lambda_0$ is isotropic. Then, at
every level of the variance of $\hat \lambda$, the axisymmetric nature of the
statistics is maintained, so the correlation tensor must still possess the form
(\ref{eq:axisym_turb}) with global direction $\hat\lambda_0$ defining the
symmetry axis.  What changes when we average over local $\hat\lambda$ directions
is
the distribution of power between $E$ and $F$ components, with power shifted to
isotropic $E$ term as the variance of $\hat\lambda$ increases.
\begin{eqnarray}
\lefteqn{\overline{\left\langle H_i({\bf k}) H_j^*({\bf k})\right\rangle} =
\overline{E(k,{\bf \hat k} \cdot{\bf \hat\lambda_0})}
\left( \delta_{ij} - \hat k_i \hat k_j \right) +} \nonumber \\
&&+\overline{F(k,{\bf \hat k} \cdot{\bf\hat \lambda_0})} \left[
W_I ({\bf \hat k} \cdot{\bf \hat\lambda_0},\sigma_\lambda)
\left( \delta_{ij} - \hat k_i \hat k_j \right)
+ W_L({\bf \hat k} \cdot{\bf \hat\lambda_0},\sigma_\lambda)
 \frac{ ({\bf \hat k} \cdot {\bf \hat \lambda_0})^2
\hat k_i \hat k_j + \hat \lambda_{0i} \hat \lambda_{0j} -
({\bf \hat k} \cdot {\bf \hat \lambda_0})
(\hat k_i \hat \lambda_{0j}+\hat k_j \hat \lambda_{0i})}
{1 - ({\bf \hat k} \cdot {\bf \hat \lambda_0})^2} \right]
\label{eq:l_average}
\end{eqnarray}
where overline indicates the average over the cosine $\chi \equiv \hat \lambda
\cdot \hat \lambda_0$. $W_I$ and $W_L$ are weight functions that depend on
the measure of the average $\hat\lambda$ wandering $\sigma_\lambda$ so that
$ W_L({\bf \hat k} \cdot{\bf \hat\lambda_0},0)=1$ and
$W_I({\bf \hat k} \cdot{\bf \hat\lambda_0},0)=W_L({\bf \hat k} \cdot{\bf
\hat\lambda_0},\infty)=0$

While details of the averaging over local directions of $\hat \lambda$ depend on
exact form of the power spectra, main effect can be demonstrated by just
averaging the basic tensor forms.   In particular, the averaged $F$-tensor
(assigning $1-({\bf k}\cdot\hat\lambda)^2$ norm to the spectral function for
convenience)
becomes
\begin{eqnarray}
\label{eq:F_l_average}
\lefteqn{\overline{ ({\bf \hat k} \cdot {\bf \hat \lambda})^2 \hat k_i \hat
k_j + \hat \lambda_i \hat \lambda_j - ({\bf \hat k} \cdot {\bf \hat \lambda})
(\hat k_i \hat \lambda_j+\hat k_j \hat \lambda_i)}=} \\
&&= \onehalf \left(1 - \overline{\chi^2} \right)
\left(\delta_{ij}-{\hat k_i}{\hat k_j}\right)
+\onehalf \left( 3 \overline{\chi^2} - 1\right)
\left(({\bf \hat k} \cdot {\bf \hat \lambda}_0)^2 \hat k_i \hat
k_j + \hat \lambda_{0i} \hat \lambda_{0j}
 - ({\bf \hat k} \cdot {\bf \hat \lambda}_0)
(\hat k_i \hat \lambda_{0j}+\hat k_j \hat \lambda_{0i}) \right) ~.
\nonumber
\end{eqnarray}
We see that wandering of the local $\lambda$,
i.e. $\overline{\chi^2} \ne 1$, leads
to the appearance of the isotropic tensor part and the suppression of the
original $F$-type part.
Although the exact weight functions between the parts
depend on details, the structure of the transition described by
Eq.~{\ref{eq:F_l_average}}
is general -- the original tensor structure (of any form) that one starts with
when $\hat\lambda=\hat \lambda_0$ disappears as $\hat \lambda$ gets distributed
more and more isotropically, $\overline{\chi^2} \to
\onethird$, being replaced by the isotropic tensor.

\begin{figure}[ht]
\centerline{\includegraphics[width=0.5\textwidth]{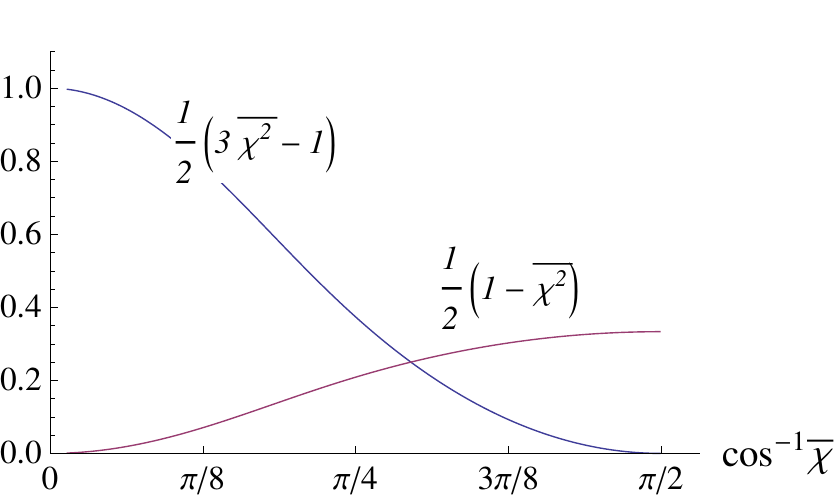}}
\caption{Example of the transition weights between anisotropic and isotropic
correlation tensor structures as functions of the mean angle between
a local and the global symmetry directions,
$\cos^{-1}\chi=\cos^{-1}\left(\overline{\hat\lambda \cdot
\hat\lambda_0}\right)$.}
\label{fig:L-wandering}
\end{figure}

To estimate the characteristic angle of the transition to isotropic tensor
structure we plot in  Figure~\ref{fig:L-wandering}  our model
weight functions $ W_I(\sigma_\lambda) =\onehalf\left(1 -
\overline{\chi^2}\right)$ and
$W_L(\sigma_\lambda)=\onehalf \left( 3 \overline{\chi^2}-1\right)$ assuming
a simple distribution $P(\chi) =
\frac{1}{\sigma_\lambda\left(1-e^{2/\sigma_\lambda}\right)}\exp\left[\frac{
\chi-1 } { \sigma_\lambda } \right ] d\chi$. With this parameterization,
$\sigma_\lambda$ matches very closely
the mean deviation of cosine $\chi$ from unity as long as
$\sigma_\lambda \lesssim 0.4 $,
 We find
that one can approximately use the local tensor structure of the correlation
for global studies when the mean deviation of $\hat \lambda$ from $\hat
\lambda_0$ is less than $\pi/8$. If the mean fluctuation of $\hat \lambda$
exceeds $\approx 3\pi/8$, one can use the isotropic tensor. In the intermediate
range
$\pi/8 \lesssim \cos^{-1}\overline{\chi} \lesssim 3\pi/8$ one has a mix of the
local and isotropic contributions. Two contributions are in similar proportions
when $\cos^{-1}\overline{\chi} \approx \pi/4$.

The averaging in Eq. (\ref{eq:l_average}) is an ensemble averaging for three dimensional correlations.
It worth noting that the averaging arising from field wandering in Eq. (\ref{eq:l_average}) acts
differently on the different parts of the correlation tensor.

\subsection{Structure functions at small scales}

We have argued that the correlation properties of the synchrotron emission at
small scales are well
approximated by the linearized limit of the $\gamma=2$ model.
 This allows one to study scaling properties of emission for $\gamma=2$ and then
generalize those for
any astrophysically relevant $\gamma$.
The full expression for the structure functional
for $\gamma=2$ (see Appendix~\ref{sec:g2functions})
\begin{eqnarray}
\lefteqn{D_{sync}({\bf R})  =
\left\langle \left [\int dz_1 H_\perp^2\left({\bf X_1},z_1\right)
-\int dz_2 H_\perp^2\left({\bf X_2},z_2\right)\right]^2 \right\rangle =}\\
&=&  \int dz_+ \! \int dz \left\{ 4 \sigma_{ij} \left[ D_{ji}\left({\bf X}
,z\right) - D_{ji}\left(0 ,z\right) \right]
- \left[ D_{ij}\left(X,z\right) D_{ij}\left(X,z\right)
- D_{ji}\left(0,z\right) D_{ij}\left(0,z\right)\right] \right\} ~.
\label{eq:Dsync_full}
\nonumber
\end{eqnarray}
in the small-scale limit (i.e. when the structure functions are much less than dispersions) 
provides a linearized expression
\begin{equation}
D_{sync}({\bf R})
\sim  4 \sigma_{ij}  \int dz \left[ D_{ji}\left({\bf R} ,z\right) -
D_{ji}\left(0 ,z\right) \right] ~~.
\label{eq:Dsync_approx}
\end{equation}
that involves  $x,y$ ($i,j=1,2$) components of the magnetic field correlation tensor.

For the following discussion we introduce the coordinate frame that has z-axis
along
the line-of-sight and y-axis perpendicular to the plane spanned by the
line-of-sight and the symmetry axis $\hat \lambda$.
The x-axis is then along the projection of $\hat \lambda$
onto the image $(x,y)$ plane, denoted by the unit 2D vector $\hat \Lambda$.
The vector $\hat {\bf \lambda}$ now has components $\hat \lambda_i =
(\sin\theta, 0, \cos\theta)$.
The polar angle of the 2D separation ${\bf R}$ is denoted by $\phi$
while $\psi$ is the polar angle of the 2D wave vector ${\bf K}$.
Both have the meaning of angles between the
correspondent vectors and $\hat \Lambda$.

The 2D part of the dispersion tensor $\sigma_{ij}$ is diagonal in the
chosen frame
\begin{equation}
\sigma_{xx} = \sigma_\parallel^2 \sin^2\theta  + \onehalf \sigma_\perp^2
\cos^2\theta ~, \quad
\sigma_{yy} = \onehalf \sigma_\perp^2 ~, \quad \sigma_{xy} = 0 \quad .
\end{equation}
Its anisotropy can be represented by the parameter
\begin{equation}
\epsilon \equiv \frac{\sigma_{xx} - \sigma_{yy}}{\sigma_{xx}+\sigma_{yy}} =
\frac{\sin^2\theta \left(\sigma_\parallel^2 - \onehalf \sigma_\perp^2 \right)}
{\sin^2\theta \left(\sigma_\parallel^2 - \onehalf \sigma_\perp^2 \right) +
\sigma_\perp^2}
=\frac{\sin^2\theta \left(r_\sigma-1\right)}{\sin^2\theta
\left(r_\sigma-1\right)+2}
\label{eq:epsilon}
\end{equation}
which depends both on the properties of 3D turbulence and the projection angle
$\theta$. The parameter $\epsilon$ play a key role in
interpretation of the synchrotron signal.

The integration over the line of sight is equivalent to
setting $k_z=0$ in the spectral domain.
The z-axis projected correlation function of the field becomes
\begin{eqnarray}
\int dz \langle H_i H_j \rangle \propto \frac{1}{(2\pi)^2} \int d^2 K
e^{i {\bf K \cdot R }}
\left[\vphantom{\frac{ \cos^2\psi \hat K_i \hat K_j +
\hat \Lambda_i \hat \Lambda_j - \cos\psi
(\hat K_i \hat \Lambda_j+\hat K_j \hat \Lambda_i)}{1 - \cos^2\psi \sin^2\theta}}
E({\bf K}) \left(\delta_{ij} - {\hat K_i} {\hat  K_j} \right) +   
\sin^2\!\theta \;  F({\bf K}) \; \frac{ \cos^2\!\psi\hat K_i \hat
K_j + \hat \Lambda_i \hat \Lambda_j - \cos\psi
(\hat K_i \hat \Lambda_j+\hat K_j \hat \Lambda_i)}{1 - \cos^2\!\psi
\sin^2\!\theta} \right]\quad .
\label{eq:Hij2D}
\end{eqnarray}
where, as we discussed earlier $E({\bf K})$ and $F({\bf K})$ functions describe
the underlying statistics of turbulence.

The 2D power spectra $E({\bf K})=E(k_z=0)$ and $F({\bf
K})=F(k_z=0)$\footnote{We shall use the same letter notation for 3D and 2D
spectra, which, hopefully, should not lead to the confusion.} depend,
parametrically, on the angle $\theta$ between the symmetry axis and the line of
sight. Notably, the contribution from the $F$-modes vanishes if the axis of
symmetry is along the line-of-sight, $\sin\theta=0$.

More interesting is the dependence on the positional angle $\psi$. One can deal with 
the harmonic decomposition of 2D spectra  which is defined as
$E_n(K) = \frac{1}{2\pi} \int_0^{2\pi}\! d \psi \; e^{- i n \psi} E(K,\psi)
$, and analogously for $F_n(K)$.

Below we obtain expressions for multipoles of the decomposition of 2D structure functions of 
magnetic field. These multipoles can be studied observationally. In what follows we transform back to the correlation function domain we express the spectral
tensors by the differential operators. We consider separately the additive $E$ and $F$ contributions to the
structure functions.

\subsubsection{Multipole expansion for E-term}
Correlation function that arises from $E$-type spectral tensor can be expressed
as
\begin{equation}
\int dz \langle H_i H_j \rangle \propto
\left( \partial_i \partial_j - \delta_{ij} \Delta \right) \Phi_E(R,\phi)~,
\end{equation}
with
\begin{equation}
\Phi_E(R,\phi) \equiv \sum_{n=-\infty}^\infty i^n  e^{i n \phi}
\int  K^{-1} dK J_n(KR) E_n(K)
\label{eq:2Dcorr}
\end{equation}
The synchrotron structure function has a similar form
\begin{equation}
 D_{sync}({\bf R}) \propto 8 \sigma_{ij} \left( \partial_i \partial_j -
\delta_{ij}  \Delta \right)
\Phi_{reg}(R,\phi)
\end{equation}
but with the regularized $ \Phi_{reg}(R,\phi) = \Phi(0,\phi) - \Phi(R,\phi) $
\footnote{We have left $\phi$ dependence in $\Phi(0,\phi)$ to signify that
regularization may affect not
only the monopole but also the higher harmonics.}.
It becomes
\begin{equation}
D_{sync}(R,\phi) \propto - (\sigma_{xx}+\sigma_{yy}) \Delta
\Phi_{reg}(R,\phi) +
(\sigma_{xx}-\sigma_{yy}) \left(\partial_x^2  - \partial_y^2 \right)
\Phi_{reg}(R,\phi)~.
\end{equation}
Both terms have concise multipole expansions
\begin{eqnarray}
\Delta \Phi_{reg}(R,\phi)  &=& \sum_{n=-\infty}^\infty i^n  e^{i n \phi}
\int_0^\infty \! K dK \left[ J_n(KR) - J_n(0) \right] E_n\\
\left(\partial_x^2  - \partial_y^2 \right) \Phi_{reg}(R,\phi) &=&
\sum_{n=-\infty}^\infty i^n e^{i n \phi} \int_0^\infty \! K dK
\left[ J_n(KR) - J_n(0) \right] \frac{ E_{n+2} +  E_{n-2} }{2}
\end{eqnarray}
that lead to the expression for the multipoles of the normalized
synchrotron structure  function
\begin{eqnarray}
 \lefteqn{\tilde D_n(R) \equiv \frac{1}{2\pi} \int_0^{2\pi} d \phi e^{-i n \phi}
\tilde D_{sync}(R,\phi) =} \nonumber \\
&=& i^n \int_0^\infty \! K dK \left[ J_n(0) - J_n(KR) \right]
\left[ E_n - \onehalf \epsilon \left(E_{n+2} +  E_{n-2}\right) \right] ~.
\label{eq:Dn_Eterm}
\end{eqnarray}
The first term directly maps the multipole terms of the power spectrum to
correspondent multipole terms in the
projected correlation function, while the second one describes the multipole
coupling in the presence of anisotropy.
In particular, the spectral quadrupole feeds into the monopole of the structure
function
\begin{equation}
\tilde D_0(R) =  \int_0^\infty \! K dK \left[ 1- J_0(KR) \right]
\left[ E_0 -\onehalf \epsilon \left(E_{2} +  E_{-2} \right) \right] ~ ,
\end{equation}
while the quadrupole dependence of the correlation is affected by the monopole
and the octupole of the spectrum
\begin{equation}
\tilde D_{\pm 2}(R) = \int_0^\infty \! K dK J_2(KR) \left[ E_{\pm 2} -
\onehalf \epsilon \left( E_{0} +  E_{\pm4} \right) \right] ~ .
\end{equation}

With factorization Eq.~(\ref{eq:globalasymspectrum}) in place, the multipole
terms of the
synchrotron intensity in small scale regime and $\gamma=2$ are
\begin{eqnarray}
\tilde D_n(R) & \approx &  A_E C_n(m)
\left( \widehat{E}_n - \onehalf \epsilon
\left(\widehat{E}_{n+2} +  \widehat{E}_{n-2}\right) \right) R^{1+m} ~,
\label{eq:DnRE}\\
&& C_n(m) =
-\frac{i^n
\Gamma\left[\frac{1}{2}(|n|-m-1)\right]}{2^{2+m}\Gamma\left[\frac{1}{2}
(|n|+m+3)\right]} \quad .
\label{eq:Cnms}
\end{eqnarray}
The expression for the coefficients is valid for $m < 1$.
For the Kolmogorov index, $m=2/3$,
 $C_0(2/3) \approx 1.12$, $C_2(2/3) \approx 0.51$, $C_4(2/3) \approx -0.03 $,
 decreasing to under
1\% of the monopole value for higher $n$. Lower $m$ somewhat enhances the higher
multipoles, e.g. for $m=1/2$  $C_0(1/2) \approx 0.93$, $C_2(1/2) \approx 0.40$,
$C_4(1/2) \approx -0.04$, although they remain small beyond $n=4$.

Eq. (\ref{eq:Cnms}) relates the multipoles of correlation function that can be measured
to the multipoles of the underlying power spectrum that characterizes  axisymmetric turbulence. The
dependence on the spectral index of turbulence enters both through $R^{1+m}$ and through $C_n(m)$ terms given
by Eq. (\ref{eq:Cnms}).

\subsubsection{Multipole expansion for F-term}
Expression for the $F$-type correlation function is somewhat more complex due to
explicit angular dependence in the spectral tensor. While it is possible to
represent it as the fourth-order derivatives of a single scalar function 
\citep[see][]{1997PhRvE..56.2875O} here we shall use the direct coordinate approach.

From Eq.~(\ref{eq:Hij2D}), two relevant component correlations are
\begin{eqnarray}
\int dz \langle H_x H_x \rangle &\propto& \frac{ \sin^2\!\theta}{(2\pi)^2} \int
d^2 K e^{i {\bf K \cdot R }}
 \;  F({\bf K}) \; \frac{ \sin^4\!\psi}{1 - \cos^2\!\psi \sin^2\!\theta}
\nonumber \\
\int dz \langle H_y H_y \rangle &\propto& \frac{ \sin^2\!\theta}{(2\pi)^2} \int
d^2 K e^{i {\bf K \cdot R }}
 \;  F({\bf K}) \; \frac{ \cos^2\!\psi \sin^2\!\psi}{1 - \cos^2\!\psi
\sin^2\!\theta} \nonumber
\end{eqnarray}
Combining these expressions into Eq.~(\ref{eq:Dsync_approx}) for $\tilde
D_{sync}(R,\phi)$ and performing multipole expansion we find that
the multipole coefficients of the synchrotron structure function from
waves of $F$-type are (compare with Eq.~(\ref{eq:Dn_Eterm}))
\begin{eqnarray}
\tilde D_n(R) \sim i^n \sin^2\!\theta \int_0^\infty \! K dK
\left[ J_n(0) - J_n(KR) \right]
 \sum_{p=-\infty}^\infty
\left[F_p - \onehalf \epsilon \left(F_{p-2} + F_{p+2}\right)\right] G^F_{n-p}
\label{eq:Dn_F}
\end{eqnarray}
where harmonic coefficients of the angular function that describes the
trace, $xx+yy$, of $F$-tensor (divided by $\sin^2\theta$) is
\begin{equation}
G^{F}_p(\theta) = \frac{1}{2 \pi} \int_0^{2\pi} d \psi e^{-i p \psi}
\frac{\sin^2\!\psi}{1 - \cos^2\!\psi\sin^2\!\theta}
\end{equation}
This function describes how the spectrum contributes to different angular
multipoles in the observed synchrotron correlations. It is especially
important in many interesting cases when the spectrum $F({\bf K})$ is isotropic,
since then it determines the angular behaviour of $D_{sync}$ in its entirety.

Geometric functions that arise in our studies are presented, in particular
graphically, in Appendix~\ref{sec:Gfunctions}.
In two limiting cases, $G^F$ contains only low harmonics.
For $\theta=0~$\footnote{We reiterate that if the symmetry axis is along the
line-of-sight this signal is not observed due to amplitude suppression from
line-of-sight integration. However, this case is relevant for consideration of
the fast modes in high-$\beta$ MHD turbulence}, $G^{F}$ contributes the
monopole
$G_0^{F}(0)=\onehalf$ and the quadrupole $G_2^{F}(0)=-\onehalf G_0^{F}(0)$
terms only,
while in the opposite limit, $\theta=\onehalf \pi$,
there is just the monopole  $G_0^{F}(\pi/2) = 2 G_0^{F}(0)$.
Varying the symmetry orientation angle $\theta$ (see Figure in
Appendix~\ref{sec:Gfunctions}) changes
the power distribution between harmonics $G^{F}_p(\theta)$ in
a rather complex way due to singular behaviour at $\theta \to \onehalf \pi$.
At steep angles $\theta < \pi/3$ as symmetry axis approaches sky
direction, the whole train of high multipoles is generated, albeit at
decreasing amplitudes. This may or may not lead to observable consequences,
especially in the context of steep spectra that suppress high multipole
contributions even further.

The relation between multipoles of the observed intensity correlation function
and multipoles of the underlying spectrum in case of the $F$-term is
\begin{eqnarray}
\tilde D_n(R) \sim
A_F C_n(m) \sin^2\!\theta  \sum_{p=-\infty}^\infty \left[\widehat{F}_p -
\onehalf \epsilon \left(\widehat{F}_{p-2} + \widehat{F}_{p+2}\right)\right]
G^F_{n-p} R^{1+m} ~.
\label{eq:Dn_Fscaling}
\end{eqnarray}

\section{Synchrotron emission from MHD turbulence}
\label{sec:physical_models}

In the previous section we dealt with a general form of the tensor of the
axisymmetric solenoidal field. A realistic MHD turbulence corresponds to the
particular form of this tensor. In what follows we use the model of compressible
MHD turbulence that follows from numerical simulations and also can be justified
theoretically. Predictions obtained within this model can be used to test the
model and can provide insight into the role of compressible (fast and slow
modes) versus non-compressible (Alfv\'en modes) motions.

\subsection{Model of Alfv\'enic turbulence}
It has been well known that small amplitude magnetic perturbations can be
decomposed into Alfv\'en, slow and fast basic MHD modes. However, as the
conventional discussion involves the decomposition in respect to the mean field,
such a decompositions becomes questionable for any magnetic field perturbations
of appreciable amplitude. The use of local system of reference 
\citep[see][]{1999ApJ...517..700L,2000ApJ...539..273C}
makes the procedure justifiable. The {\it statistical} decomposition of the modes
has been performed in \citet{2002PhRvL..88x5001C,2003MNRAS.345..325C, 2010ApJ...720..742K}.
It was proven that the modes provide cascades of their own and the drainage of
energy between the cascades is limited.

Alfv\'enic modes are the most important ones with the most of the energy residing
in them and they are less subject to damping in astrophysical plasmas 
\citep[see][]{2002PhRvL..88x5001C}. They have the statistical properties which
are similar to the Alfv\'enic modes obtained in incompressible simulations \citep{2003MNRAS.345..325C}.
This motivates us to study the Alfv\'enic turbulence as our primary example.

The spectrum of Alfv\'enic modes was obtained in \citet{2002ApJ...564..291C} for the
driving turbulence at large scale with $V_L$ equal to the Alfv\'en velocity:
\begin{equation}
E({\bf k}) \propto k^{-10/3} L^{-1/3} \exp\left[-
\frac{ k_{\|}}{L^{-1/3}k_{\bot}^{2/3}}\right]~,
\label{Cho}
\end{equation}
where $L$ is the injection scale of the turbulence.
The exponent in Eq.~(\ref{Cho}) reflects the so-called {\it critical balance}
condition between the parallel and perpendicular modes. It shows that most of
the energy resides around the wavenumbers where the turnover of the turbulent
eddies is equal to the period of the Alfv\'en wave.

Note that the driving at $V_L=V_A$ corresponds to model of GS95 turbulence. In
the case of weaker driving at the outer scale, the turbulence
becomes sub-Alfv\'enic at all scales and the relations between $k_{\|}$ and
$k_{\bot}$ from LV99 should be used. Namely, the critical balance condition
provides a modified relation between the parallel and perpendicular modes at
sufficiently small scales
\begin{equation}
k_{\|}\sim L^{-1}(k_\bot L)^{2/3} M_A^{4/3}
\label{LV}
\end{equation}
where $M_A$ is the Alfv\'en Mach number, which is $V_L/V_A$. Therefore one
expects that the description of the turbulence for $M_A<1$ involves the ratio
of the left and right sides of the Eq.~(\ref{LV}) in the exponent
of Eq.~(\ref{Cho}). This would provide the description of turbulence in the {\it
local} system of reference.

The tensor structure of Alfv\'en power spectrum is derived from the condition
that displacement and velocity of plasma in an Alfv\'en wave ${\bf k}$ are
orthogonal to the plane spanned by the magnetic field and  ${\bf k}$,
${\bf v}_A \propto {\bf \hat k} \times \hat \lambda$. Magnetic field response
is given by the condition that
fluctuating field is frozen, 
\begin{equation}
H({\bf k}) \propto {\bf k}
\times \left( {\bf v} \times {\bf \hat \lambda}\right)/ \omega({\bf k}) \quad ,
\end{equation}
where for the Alfv\'enic waves the frequency  $\omega \propto {\bf k} \cdot \hat \lambda $ .
Simple algebraic manipulations then show
\begin{equation}
\left\langle H_i({\bf k}) H_j^*({\bf k})\right\rangle_A \propto
\left(\delta_{ij} - \hat k_i \hat k_j \right)
- \frac{ ({\bf \hat k} \cdot {\bf \hat \lambda})^2 \hat k_i \hat
k_j + \hat \lambda_i \hat \lambda_j - ({\bf \hat k} \cdot {\bf \hat \lambda})
(\hat k_i \hat \lambda_j+\hat k_j \hat \lambda_i)}{1 - ({\bf \hat k} \cdot {\bf
\hat \lambda})^2}
\label{eq:Alfven_tensor_loc}
\end{equation}
which in terms of the general $E$ -- $F$ decomposition (\ref{eq:axisym_turb})
corresponds to $F({\bf k}) = - E({\bf k})$. This tensor reflects the absence of
the perturbations in the magnetic field component parallel to the mean field
$\hat \lambda$. 

In the global system of reference, however, the anisotropy is scale independent
and determined by the anisotropy at the outer scale. In this system of
reference the tensor of magnetic perturbations can be given by
\begin{equation}
E({\bf k}) \propto k^{-11/3} \exp\left[-M_a^{-4/3}
\frac{\left|\bf{ \hat k \cdot \hat \lambda_0}\right|}{\left(1-(\bf{ \hat k \cdot
\hat \lambda_0})^2\right)^{2/3}}\right]~,
\label{eq:LCV_Pk}
\end{equation}
arises from extending to global scales and
$M_a \ne 1$ the phenomenological, locally anisotropic, model of \citet{2002ApJ...564..291C}.
The axis of symmetry is associated with the direction of
the mean magnetic field $\hat \lambda_0$. The spectrum reflects the
consideration
that in critically balanced turbulence
with $M_a < 1$ the modes with parallel to magnetic field wavenumber  $k_{||}$
that exceed the critical value $ \sim L^{-1/3} k_{\perp}^{2/3} M_a^{-4/3}$ are
suppressed. At the injection scale $L$, if $k_{||} \sim L$.

Following Section~\ref{sec:L_wandering}, the global tensor structure is a
weighted mix of the local tensor of Eq.~\ref{eq:Alfven_tensor_loc} and the
isotropic tensor form.
The $k_z=0$ section of the spectrum becomes
\begin{eqnarray}
\lefteqn{\left\langle H_i({\bf K}) H_j^*({\bf K})\right\rangle_A \propto
 K^{-11/3} \exp\left[-M_a^{-4/3}
\frac{\left|\cos\psi \right|
\sin\theta}{\left(1-\cos^2\psi\sin^2\theta\right)^{2/3}}\right] \times }
\nonumber \\
&&  \times \left[\left(\delta_{ij} - {\hat K_i} {\hat  K_j}
\right) W_I +  \left(\left(\delta_{ij} - {\hat K_i}
{\hat K_j} \right) -
 \sin^2\!\theta \;   \frac{ \cos^2\!\psi \hat K_i \hat
K_j + \hat \Lambda_i \hat \Lambda_j - \cos\psi
(\hat K_i \hat \Lambda_j+\hat K_j \hat \Lambda_i)}{1 - \cos^2\!\psi
\sin^2\!\theta} \right) W_L \right] \quad .
\label{eq:ourmodel}
\end{eqnarray}
Now $\hat \Lambda$ is the projection of the global symmetry axis $\hat
\lambda_0$ but for brevity we are omitting index '0' henceforth.
We note that if the symmetry axis is
strictly along the line of sight, $\sin\theta=0$, the anisotropy effects are not
observable in sky projection.
Maximal anisotropic effects are observed when the symmetry axis is perpendicular
to the line of sight, $\sin\theta=1$. At small Mach numbers, parallel to the
mean magnetic field modes are strongly suppressed and the turbulence is highly
anisotropic. The effect is retained in sky projection, but is mitigated by
$\sin\theta$.

The model Eq~(\ref{eq:ourmodel}) that we use to discuss  Alfv\'enic turbulence
 has two parameters - the Alfv\'enic Mach number $M_a$ and the mean deviation
angle $\overline{\chi}$ that gives the weights $W_L(\overline{\chi})$ and
$W_I(\overline{\chi})$. On the other hand, the level of magnetic field wandering
due to turbulence is itself related to the Alfv\'enic Mach number. High $M_a > 1
$ turbulence leads to significant bending of magnetic field lines and
isotropization of the observable statistics. Conversely,
details of the local mode structure of the correlation tensor can be observable
only in sub-Alfv\'enic, $M_a < 1$ regime. As the guiding model of the wandering
consistent with given $M_a$ we shall employ a rough prescription
that $M_a$ gives a tangent of the mean deviation of the local angle from the
global axis, $\overline{\chi} \approx 1/\sqrt{1+M_a^2}$.
The sub-Alfv\'enic regime but with
some isotropization may be relevant to synchrotron studies if the line-of-sight
integration averages over regions with different magnetic field directions.

\begin{figure}[ht]
\centerline{\includegraphics[width=0.4\textwidth]{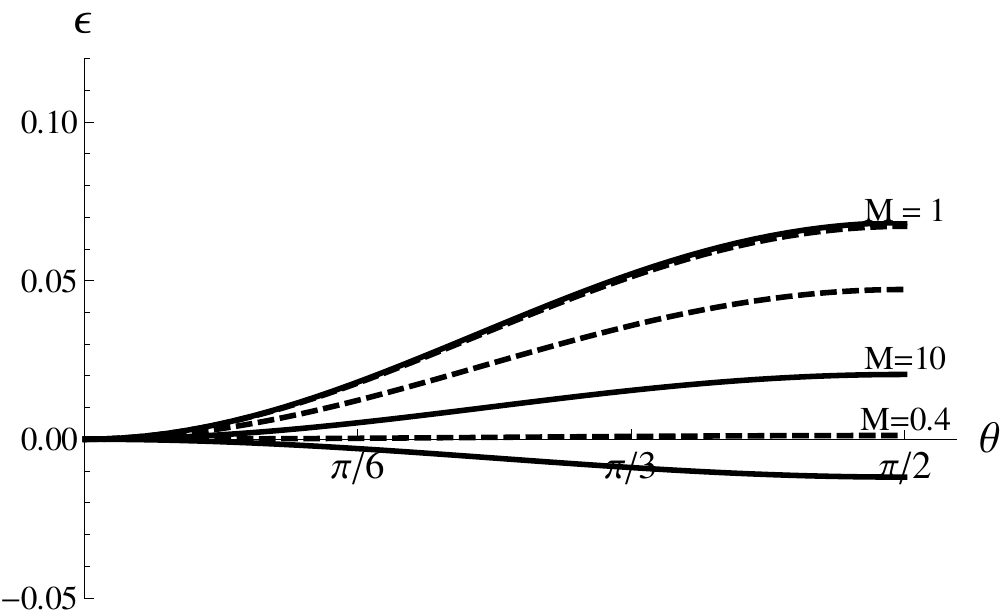}
\hspace{1cm}
\includegraphics[width=0.4\textwidth]{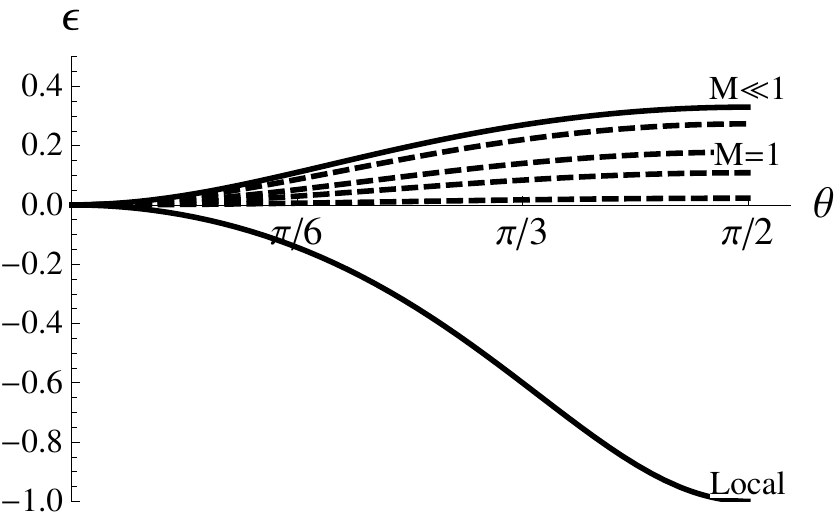}}
\caption{Asymmetry in the variance of the $H_x$ (parallel to sky projection of
the mean field) and perpendicular $H_y$,
components of the magnetic field for the
model of Alfv\'enic turbulence, Eq.~(\ref{eq:LCV_Pk}) as parameterized by
$\epsilon(\theta)=(\sigma_{xx}-\sigma_{yy})/(\sigma_{xx}+\sigma_{yy})$, as
the function of the projection angle $\theta$.
Left: change from the sub-Alfv\'enic ($M_a=0.1$ bottom solid)
to trans-Alfv\'enic ($M_a \approx 1$, top solid)
to super-Alfv\'enic ($M_a>1$, middle solid) behaviour.
Averaging over local direction of the magnetic field is
carried out in accordance with $M_a$.
Right: dependence on Mach number $M_a$ when field wandering is neglected.
The bottom solid curve in the negative sector is Mach independent result
($r_\sigma=0$) when local tensor correlation form is assumed to hold exactly.
$\epsilon > 0$ sector shows the dependence on Mach number if isotropic
$E$-tensor (but with anisotropic $E({\bf k})$) is used to model Alfv\'enic
correlations.
The dashed curves, in decreasing order of $\epsilon$, correspond to
$M_a=0.2,0.5,2$. The limiting behaviour with the highest
anisotropy $\epsilon$ ( top solid curve marked $M_a \ll 1$) is closely followed
already at $M_a \lesssim 0.2 $.
}
\label{fig:LCV_sigmas}
\end{figure}

Asymmetry in the variance in sky components of the magnetic field in this model
is shown in Figure~\ref{fig:LCV_sigmas}.
The parameter $\epsilon$, given by Eq.~(\ref{eq:epsilon}),
as expected, is larger by magnitude when the symmetry axis is aligned with the
sky plane, and is vanishing when it is parallel to the line of sight. Left
panel shows that $\epsilon$ magnitude is small to moderate when field wondering
is appropriately accounted for. $\epsilon$ is slightly negative for strongly
sub-Alfv\'enic turbulence, almost zero at $M_a \approx 0.4$, reaches maximum
positive values, $\epsilon \sim 0.1$
in trans-Alfv\'enic regime $M \sim 1-2 $ and then decreases to isotropic limit
$\epsilon \to 0$ for super Alfv\'enic $M_a$. However the result is somewhat
sensitive on exact procedure we model the transition.

The right panel provides a cautionary illustration. If one adopts
the exact local description Eq.~(\ref{eq:Alfven_tensor_loc})
for the Alfv\'enic turbulence to hold through to global scales,  $W_I
\approx 0$, $W_L \approx 1$, then we get negative and practically
$M_a$-independent $\epsilon$, since $r_\sigma \approx 0$. This means that the
perturbations in magnetic field component parallel to the mean field are
suppressed. On the other hand, if anisotropic power
spectrum is used with isotropized $E$-tensor form to describe the turbulence
the result is opposite - parallel component is presumed to fluctuate more than
the perpendicular one and $0 \le \epsilon < 1/3$, with increasing $M_a$
taking us to isotropic limit. Far from being just of academic interest, this
latter case appear in the description of a mix of Alfv\'en and Slow modes in
incompressible limit. We also see that averaging of the
statistics over local field direction may effectively bring us to this regime,
but probably not at the highest magnitude of the effect given by $r_\sigma=2$
and
$\epsilon \approx \sin^2\theta/(\sin^2\theta+2)$.

\begin{figure}[ht]
\centerline{\includegraphics[width=0.4\textwidth]{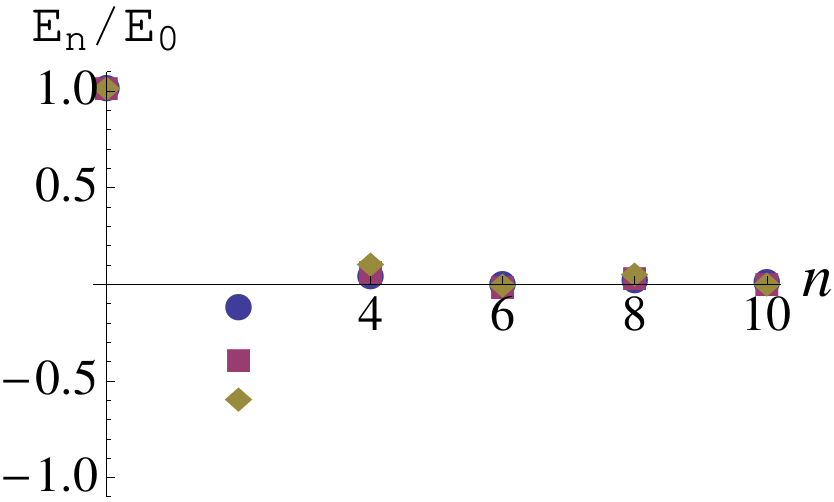}\hspace{1cm}
\includegraphics[width=0.4\textwidth]{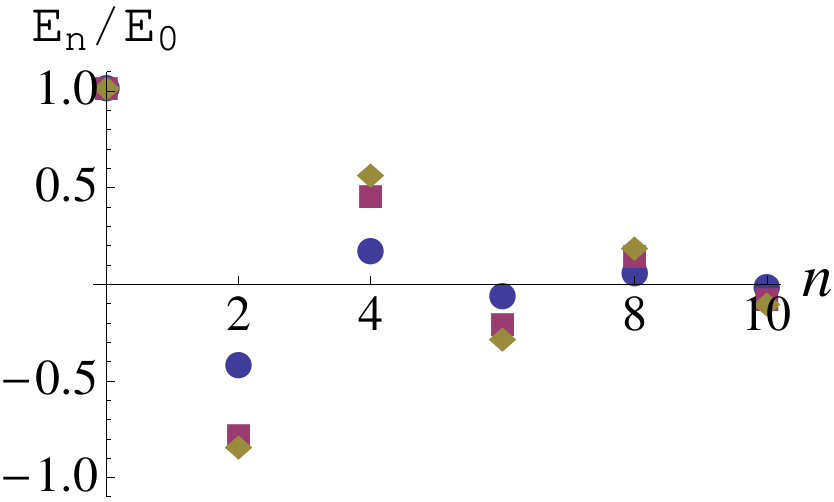}}
\caption{Multipole distribution of power in the units of the monopole
$\widehat{E}_n/\widehat{E}_0$ for $M_a=1$ (left) and $M_a=0.4$ (right)
for $\theta=\frac{1}{6}\pi,\frac{1}{3}\pi, \onehalf\pi $
(circles, squares and diamonds, respectively).}
\label{fig:LCV_Ens}
\end{figure}

In Figure~\ref{fig:LCV_Ens} the distribution of power in multipoles $E_n$
is shown for representative Alfv\'enic Mach numbers and several orientations
$\theta$.
The multipole expansion contains only even multipoles and the real nature of
coefficients reflects the presence of only the cosine terms in the
decomposition. The anisotropy of the spectrum is predominately a quadrupole one for
$M_a \gtrsim 1$. As $M_a$ decreases, the higher multipoles appear and in the
limit the spectrum contains all  $E_n$ of equal amplitudes with compensating
alternating signs. The presence of high multipoles is largely due to sharp
features in angular dependence of the spectrum which are artifacts of the model
\citep[see][ for technical remedies of this issue]{2002ApJ...564..291C}.

The structure function of the synchrotron fluctuations from Alfv\'enic modes is
a special combination of Eqs.~(\ref{eq:DnRE}) and (\ref{eq:Dn_Fscaling}) that
matches Eq.~(\ref{eq:ourmodel})
\begin{equation}
\tilde D_n(R) \approx   A_A C_n(2/3)
\left[ W_I \left( \widehat{E}_n - \onehalf \epsilon
\left(\widehat{E}_{n+2} +  \widehat{E}_{n-2}\right) \right)
+ W_L  \sum_{p=-\infty}^\infty \left[ \widehat{E}_p -
\onehalf\epsilon
\left(\widehat{E}_{p-2} +  \widehat{E}_{p+2} \right)\right] G^A_{n-p}
\right] R^{5/3} ~.
\label{eq:DnRAlfven}
\end{equation}
where we refer to Appendix~\ref{sec:Gfunctions} for the properties of
$G^A(\theta)$ function. In many cases Alfvenic modes are expected to dominate magnetic
field perturbations and therefore Eq. (\ref{eq:DnRAlfven}) provides a good representation of the
multipoles of the structure function of the synchrotron intensity for small 2D separations between
the lines of sight $|R|\ll L$, where $L$ is the injection scale of the turbulence\footnote{For the observed
correlation functions this means that the angle $\phi\ll L/D$, where $D$ is the distance to the object.}.

\subsection{Fast modes}
Fast modes are compressible modes which in high beta plasma, i.e. plasma with
gaseous pressure much larger than the magnetic one, transfer to sound waves,
while in low beta plasma propagate with Alfv\'en velocity due to compressions of
the magnetic field. These modes marginally interact with Alfv\'en modes
and were shown to form their own cascade. They were identified with an
isotropic spectral function in \citet{2002PhRvL..88x5001C,2003MNRAS.345..325C}.
The tensor form of the fast
mode turbulence was presented in \citet{2004ApJ...614..757Y}
\begin{eqnarray}
\left\langle H_i({\bf k}) H_j^*({\bf k})\right\rangle_F =
\frac{k^{-7/2} }{8\pi L^{1/2}} \;
\frac{({\bf \hat k} \cdot {\bf \hat \lambda})^2 \hat k_i \hat
k_j + \hat \lambda_i \hat \lambda_j - ({\bf \hat k} \cdot {\bf \hat \lambda})
(\hat k_i \hat \lambda_j+\hat k_j \hat \lambda_i)}{1 - ({\bf \hat k} \cdot
{\bf\hat \lambda})^2} \times
\left\{
\begin{array}{ll}
1 & (\mathrm{low}~\beta) \\
1 - ({\bf \hat k} \cdot {\bf\hat \lambda})^2 &
(\mathrm{high}~\beta)
\end{array}
\right.
\label{eq:Fspectrum}
\end{eqnarray}
The difference between fast modes in high-$\beta$ and low-$\beta$ plasmas comes
from different behaviour of the velocities in the medium. 
While by very definition of the fast modes the velocity at a wave
${\bf k}$ lies in the plane spanned by $\hat k$ and $\hat \lambda$, 
for $\beta \gg 1$
velocities are potential, while for  $\beta \ll 1$ they are 
orthogonal to the direction of the magnetic field $\hat \lambda$.
The frequency of fast modes $\omega_F({\bf k})$ has a weak additional dependence
on the direction of the wave vector if $\beta \sim 1$, the effect we do not
consider here.

In contrast to Alfv\'en modes, fast modes are of $F$-type. Despite the isotropic
power scaling, fast modes are not statistically
isotropic, with anisotropy built into the tensor structure of the spectrum.

\begin{figure}[ht]
\centerline{\includegraphics[width=0.4\textwidth]{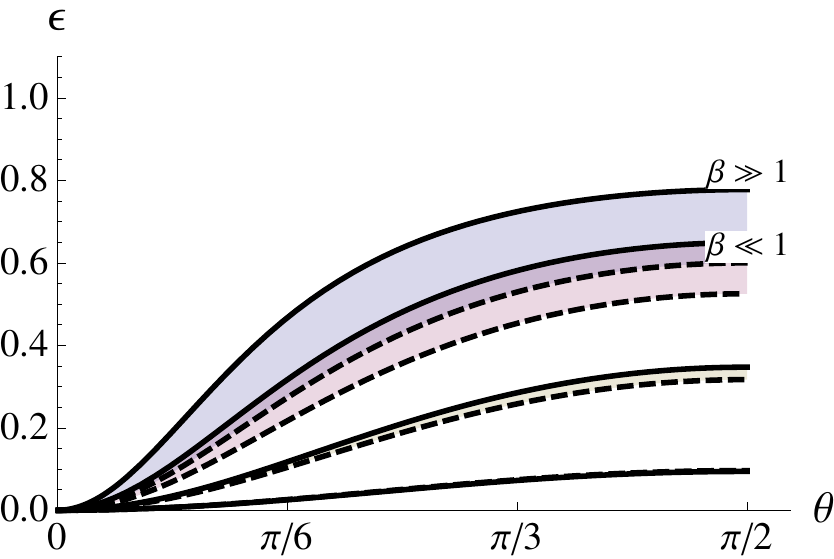}\hspace{1cm}
\includegraphics[width=0.4\textwidth]{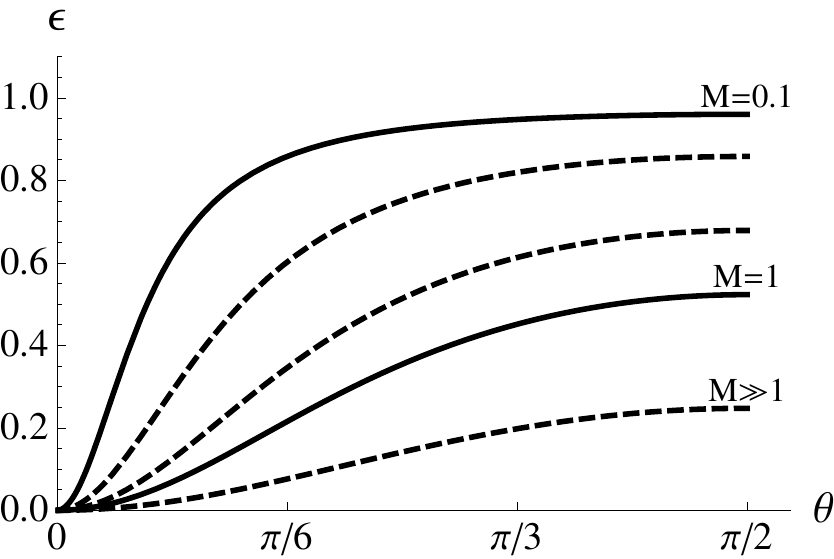}}
\caption{Left: $\epsilon(\theta)=(\sigma_{xx}-\sigma_{yy})/(\sigma_{xx}+\sigma_{
yy } )$ ,
as the function of the projection angle $\theta$ for the fast modes.
Solid lines: change in high-$\beta$ plasma from the local behaviour $W_L=1,
W_I=0$ (top) to the
isotropized limit $W_L=0$ (bottom) through intermediate cases that
corresponds to average angle of local $\lambda$ deviations of $\pi/8, \pi/4,
3\pi/8$ (from top to bottom).
Dashed curves - same for low-$\beta$ plasma. Shading covers the interval that
different $\beta$ will fill for the same $\cos^{-1}\overline{\chi}$.
Right: same for Slow modes. Isotropization of the slow modes is
treated as associated with the increase in the Alfv\'enic Mach number.
}
\label{fig:LCV_F}
\end{figure}

Asymmetry in the variance of the sky components of the magnetic field for fast
modes is plotted in Figure~\ref{fig:LCV_F}. Compared with that of the Alfv\'en
modes, fast modes always produce positive
anisotropy $\epsilon$. This effect is quite distinct from Alfv\'enic modes where
$\epsilon$ is much smaller, or even negative.
 Determining $\epsilon$ from fitting multipole
composition of synchrotron correlations
may give result that distinguishes  whether we are dealing with fast (positive
$\epsilon > 1/3$) modes.

Isotropic nature of the power scaling  leads to a very simple application
of Eqs.~(\ref{eq:Dn_Fscaling}) and (\ref{eq:DnRE}) to MHD fast modes. In
low $\beta$ case we have
$F({\bf K}) \propto K^{-7/2} $ which corresponds to $m=1/2$ and only the
monopole $\hat F_0$ present. This gives
\begin{equation}
\tilde D_n(R) \sim
A_F C_n(1/2) \left[W_I \left(\delta_{n0} - \epsilon \delta_{n2} \right) +
W_L \sin^2\!\theta \left( G^{F}_n(\theta) - \onehalf\epsilon
\left[ G^{F}_{n-2}(\theta) + G^{F}_{n+2}(\theta)\right]\right) \right]
\widehat{F}_0 \; R^{3/2} \quad \quad (
low~\beta
)
\label{eq:Dn_Fscaling_lowb}
\end{equation}

Additional angular dependence in the spectrum in high-$\beta$ medium can be
accounted for by just evaluating the $G^F_n$ coefficients at $\theta=0$
\begin{equation}
\tilde D_n(R) \sim
A_F C_n(1/2) \left[W_I \left(\delta_{n0} - \epsilon \delta_{n2} \right) +
W_L \sin^2\!\theta
 \left( G^{F}_n(0) - \onehalf\epsilon
\left[ G^{F}_{n-2}(0) + G^{F}_{n+2}(0)\right]\right) \right]
\widehat{F}_0 \;  R^{3/2} \quad \quad ( high~\beta )
\label{eq:Dn_Fscaling_highb}
\end{equation}
We note that in case of high $\beta$ the averaging over $\lambda$ wandering is
exact, so it is useful to write explicitly the set of all non-vanishing
multipoles the synchrotron structure has in this limit. Taking into account the
signs of $C_n(1/2)$ coefficients we have
\begin{eqnarray}
\tilde D_0(R) &\propto & \left( W_I + \left( 1 +
\onehalf \epsilon \right) \frac{W_L}{2} \sin^2\theta  \right)
R^{3/2} \nonumber \\
\tilde D_2(R) &\propto & -\left(\epsilon W_I + \left( 1 +
 \epsilon \right) \frac{W_L}{4} \sin^2\theta  \right)
R^{3/2}  \quad \quad ( high~\beta ) \nonumber \\
\tilde D_4(R) &\propto &  -\left( \epsilon \frac{W_L}{8} \sin^2\theta \right)
 R^{3/2}
\label{eq:Dn_Fmultipoles_highb}
\end{eqnarray}

The structure of the multipoles of correlation functions is different from that of multipoles arising from
Alfv\'en modes, which allows one to separate them, finding the measure of media compressibility.

\subsection{Slow modes}

Slow modes are usually subdominant but provide an interesting case of the
$F$-type modes but with the power spectrum that follows the Alfv\'enic
power spectrum. Displacements in slow modes lie in $\hat {\bf k}$ -- $\hat
\lambda$ plane and are orthogonal to the Fast mode ones.
In the low $\beta$
limit, the slow displacements are parallel to the magnetic field and do not
lead to  magnetic field perturbations.  So we focus at $\beta \gg 1$
case where the slow mode motion of plasma is perpendicular to the wave
vector, $v \propto \hat {\bf
k} \times \left( \hat {\bf k} \times \hat \lambda \right)$. The frequency
of the slow mode in this regime is similar to that of Alfv\'en ones, 
$\omega_S \propto {\bf k} \cdot \lambda$.
The frozen field condition gives the magnetic field correlation tensor
of the  $F$-type
\begin{eqnarray}
\left\langle H_i({\bf k}) H_j^*({\bf k})\right\rangle_S \sim E({\bf k}) \;
\frac{({\bf \hat k} \cdot {\bf \hat \lambda})^2 \hat k_i \hat
k_j + \hat \lambda_i \hat \lambda_j - ({\bf \hat k} \cdot {\bf \hat \lambda})
(\hat k_i \hat \lambda_j+\hat k_j \hat \lambda_i)}{1 - ({\bf \hat k} \cdot
{\bf\hat \lambda})^2}
\quad\quad\quad (\mathrm{high}~\beta~\mathrm{only}).
\label{eq:slowtensor}
\end{eqnarray}

Figure~\ref{fig:LCV_F} shows that the anisotropy in the sky component of the
magnetic field variance produced by Slow modes is similar to the Fast modes
demonstrating isotropization as $M_a$ increases. We stress, however, that
$\epsilon$ is determined by all the modes excited in the medium and cannot be
simply separated in individual contributions.  The expression for the
synchrotron structure function multipoles can be found by combination of the
previous results Eq~(\ref{eq:Dn_Fscaling}) and Eq.~(\ref{eq:LCV_Pk})
\begin{equation}
 \tilde D_n(R) \approx   A_S C_n(2/3)
\left[ W_I \left( \widehat{E}_n - \onehalf \epsilon
\left(\widehat{E}_{n+2} +  \widehat{E}_{n-2}\right) \right)
+ W_L \sin^2\theta \sum_{p=-\infty}^\infty
\left[\widehat{E}_p - \onehalf \epsilon \left( \widehat{E}_{p-2} +
\widehat{E}_{p+2} \right) \right]
G^{F}_{n-p} \right] R^{5/3} ~.
\label{eq:DnRslow}
\end{equation}

With slow modes being subdominant it is more relevant to study their
contribution to synchrotron 
in combination with other modes. The largest effect the slow modes have are in
$\beta \gg 1$ limit that describes,
in effect, an incompressible plasma. Slow and Alfv\'en modes in this limit are
two ``linear polarizations'' of a transverse displacement wave. In particular,
if in strong turbulence the powers in
slow and Alfv\'en modes are equal \citep{2003MNRAS.345..325C}, the waves are
unpolarized. The correlation
tensor that describes such waves is of a pure $E$-type
\begin{equation}
\left\langle H_i({\bf k}) H_j^*({\bf k})\right\rangle_{A+S} \propto
E({\bf k})  \left(
\delta_{ij} - \hat k_i \hat k_j \right)
\label{eq:Alfven+Slow_tensor}
\end{equation}
In Appendix~\ref{sec:3Dpower_to_corr} we discuss the multipole structure of the
3D correlation function that
corresponds to such spectra. The synchrotron correlation structure that
corresponds to Eq.~(\ref{eq:Alfven+Slow_tensor}) has the following multipoles
\begin{equation}
\tilde D_n(R) \approx   A_{AS} C_n(2/3)
 \left( \widehat{E}_n - \onehalf \epsilon
\left(\widehat{E}_{n+2} + \widehat{E}_{n-2}\right) \right)  R^{5/3} ~,
\end{equation}
which presents the structure different both from multipoles of Alfv\'en and fast modes.

\subsection{Angular structure of synchrotron correlations from MHD
modes}

Full observed synchrotron signal reflects the combination of Alfv\'en, Fast and
Slow modes in the proportion determined by the distribution of power between the
modes. The structure function of the resulting emission can be obtained
by linear combination of Eq.~(\ref{eq:DnRAlfven}),
Eq.~(\ref{eq:Dn_Fscaling_highb}) or Eq.~(\ref{eq:Dn_Fscaling_lowb}), and
Eq.~(\ref{eq:DnRslow}).
However the dependence on the underlying models is non-linear, since
the anisotropy of magnetic field variance $\epsilon$ is determined by all
the contributed modes at the same time.

First we observe that the Fast modes lead to different scaling
of the synchrotron correlations than the Alfv\'en or Slow modes, $\propto
R^{3/2}$ versus $\propto R^{5/3}$,
respectively. Here we suggest to focus on the measurement of the anisotropy of
$D(R,\phi)$ as independent from scaling signature of the contributions from
different modes. Measurement of the moments of $D_n(R)/D_0(R)$ may prove more
robust discriminant between the contribution modes in cases when power-law
scaling is either not exact or its accuracy is affected by observational
algorithms or theoretical approximations.

We first investigate the limit when only one of spectral contribution is
dominant, either Alfv\'en (Eq.~\ref{eq:DnRAlfven}), or Fast 
(Eqs~\ref{eq:Dn_Fscaling_highb}-\ref{eq:Dn_Fscaling_lowb}), modes,
thus assuming $\epsilon$ determined by the dominant mode only. Then, as an example of
a mix between  modes we discuss the case Eq.~(\ref{eq:Alfven+Slow_tensor}) of
strong incompressible turbulence  which
maintains both Slow and Alfv\'en modes.

\begin{figure}[ht]
\centerline{\includegraphics[width=0.45\textwidth]{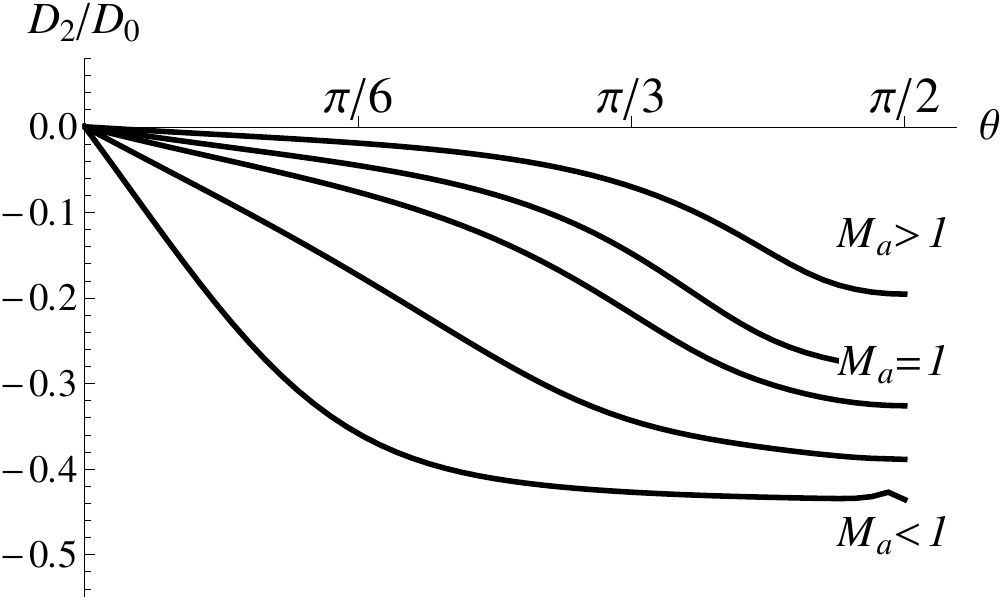}\hspace{1cm}
\includegraphics[width=0.45\textwidth]{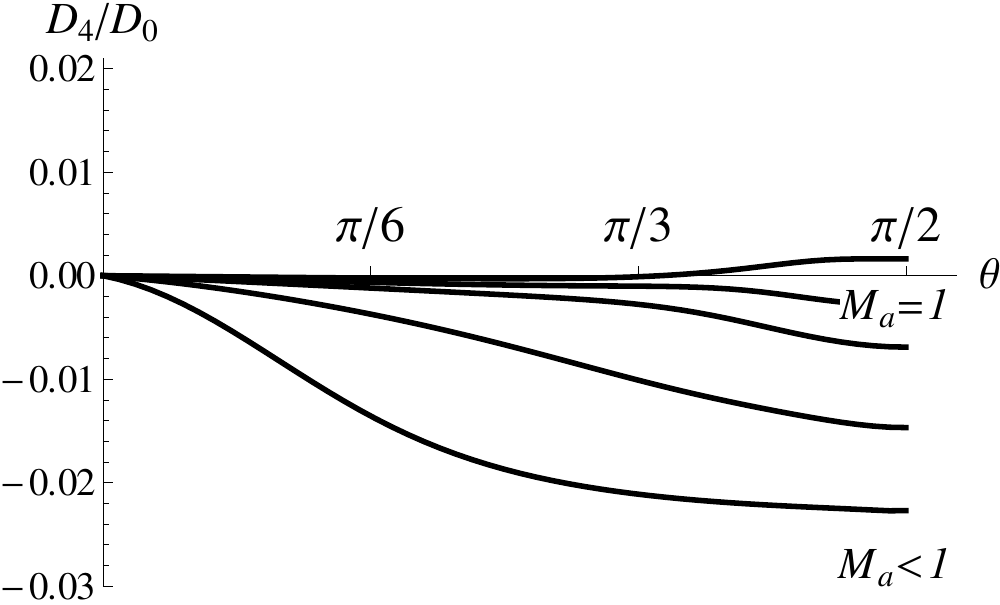}}
\caption{Left: Quadrupole of the synchrotron structure function in the units of
the monopole versus the mean magnetic field to the line-of-sight angle
$\theta$ for the Alfv\'en modes. Effect of Alfv\'enic Mach number,
including expected isotropization due to fluctuations of the local field
direction. The curves correspond to $M_a=0.1, 0.4, 0.7, 1, 2$
( from lower to upper ones). Right: same plot for the octupole.}
\label{fig:LCV_DnA}
\end{figure}

Figure~\ref{fig:LCV_DnA} shows  the details of the quadrupole and octupole 
contributions for Alfv\'enic turbulence.  Within our model of
Eq.~(\ref{eq:DnRAlfven}) the predicted
quadrupole depends on the Alfv\'enic Mach
number $M_a$ and follows one of the curves in the left panel of
Figure~\ref{fig:LCV_DnA}.

First observation from Figure~\ref{fig:LCV_DnA} is that the expected
quadrupole of the synchrotron correlations is {\it negative}. With structure
functions scaling as $\sim (D_0 + D_2 \cos\phi \ldots ) R^{1+m}$, negative
quadrupole means extended iso-correlation contours along the x-axis, the
sky projected direction of the magnetic field.  In case of Alfv\'enic
turbulence this corresponds to spectral suppression of the modes parallel to the
field having the dominant effect. It is easy to see from
Eq.~(\ref{eq:DnRAlfven}) that if the power distribution was 3D isotropic,
the local tensor structure of the Alfv\'en modes would have produced a {\it
positive} quadrupole in synchrotron, i.e the contours extended orthogonally to
the symmetry axis. Thus, observing negative synchrotron quadrupole in Alfv\'en
turbulence context is a strong argument for anisotropic underlying 3D cascade
that suppresses parallel modes.

For our calculations we have adopted
the model description that takes into account the fluctuations of the local
direction of the magnetic field at the level determined by the $M_a$.
Thus, changing $M_a$ involved both the field wondering effect and the change in
the power spectrum.
As the result of field wondering the anisotropy of
the observed synchrotron becomes suppressed when mean field is
roughly aligned with the line of sight, $\theta < \pi/6$, unless the turbulence
is strongly  sub-Alfv\'enic. If field axis is closer to be orthogonal to the
line of sight, the anisotropy is strong. Alfv\'enic turbulence leads
to the negative quadrupole which magnitude is limited to
$|D_2/D_0| < C_2(2/3)/C_0(2/3)=0.45$ as $M_a$ decreases, with values 
close to the limit $|D_2/D_0| \sim 0.4$ already exhibited at $M_a = 0.4$.
while $|D_2|/D_0 \sim 0.3$ for trans-Alfv\'enic $M_a \sim 1$.
  Measuring $D_2/D_0 \approx 0.4$ will be a
strong indication of the sub Alfv\'enic regime. 
Although diminishing with higher $M_a$,
quadrupole is still significant, $|D_2/D_0| \sim 0.1$ even for mildly
super-Alfv\'enic $M_a \sim $ few.

Right panel in Figure~\ref{fig:LCV_DnA} demonstrates that the octupole
contribution 
is quite small, at 2\% ($D_4/D_0 \sim 0.02$) level under the most favorable
parameters. Indeed  the steep spectra that
are produced in MHD turbulence, while reflecting the multipole decomposition of
the spectrum, demonstrate  suppression of the multipoles higher than the
quadrupole. This is the case for both Alfv\'enic and Fast modes and represents a
challenge to measure them.

\begin{figure}[ht]
\centerline{\includegraphics[width=0.45\textwidth]{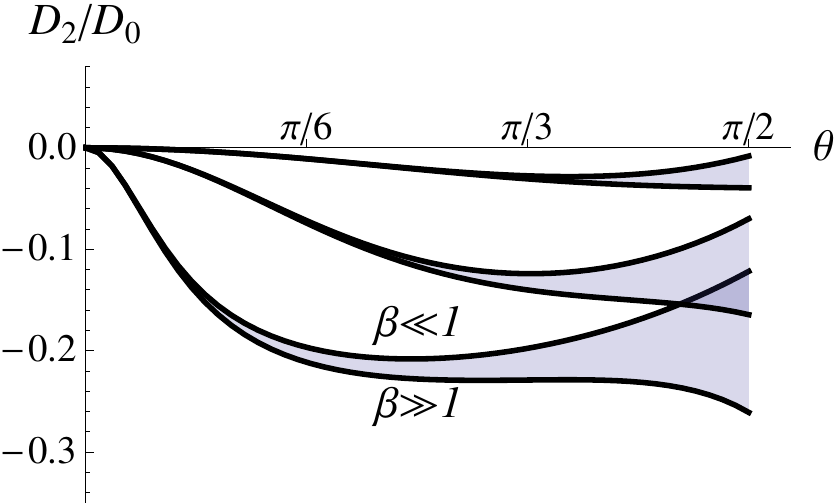}\hspace{1cm}
\includegraphics[width=0.45\textwidth]{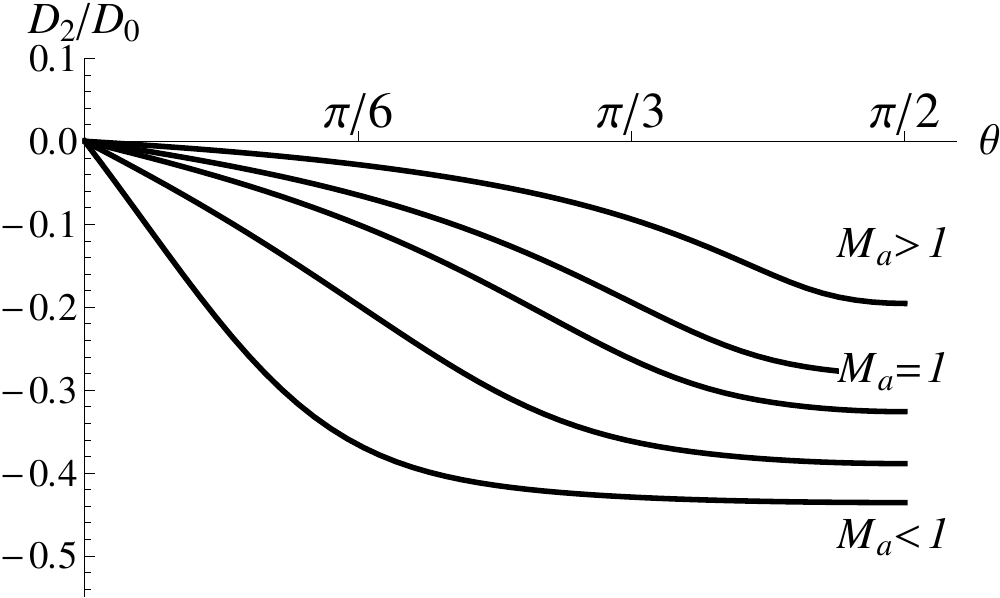}}
\caption{Quadrupole of the synchrotron structure function in the units of
the monopole versus the mean magnetic field to the line-of-sight angle
$\theta$. Left: Fast mode contribution shown in bands for
$\cos^{-1}\overline{\chi}
= \pi/16, \pi/6, \pi/3$, $D_2/D_0$, decreasing in magnitude with increased
field wandering. Each band corresponds to the range of different plasma
$\beta$, bounded by $\beta \gg 1$ and $\beta \ll 1$ limits.
Right: the same quantity for the mix of  Alfv\'en and Slow modes
in the regime of strong incompressible turbulence. The curves correspond to
$M_a=0.1, 0.4, 0.7, 1, 2$ ( from lower to upper ones).
}
\label{fig:D2D4}
\end{figure}

Fast modes have isotropic power spectrum and produce the anisotropy solely due
to the anisotropy built into  $F$-tensor,
Eq.~(\ref{eq:Fspectrum}), which is somewhat different in the cases of high and
low $\beta$ plasma.   Predictions of the quadrupole from the Fast modes are
given in the left panel of Figure~\ref{fig:D2D4}. 
There are two parameters,
$\beta$, and the mean deviation of the local field direction $\overline{\chi}$.
At each $\overline{\chi}$, for which three are shown, $\cos^{-1}\overline{\chi}
= \pi/16, \pi/6, \pi/3$, $D_2/D_0$ from Fast modes lie distinctly in a narrow
band limited by $\beta \gg 1$ and $\beta \ll 1$ predictions according to 
Eq.~(\ref{eq:Dn_Fscaling_highb}) and Eq.~(\ref{eq:Dn_Fscaling_lowb})
respectively. The Fast mode quadrupole anisotropy is rather uniform as the
function of the orientation angle at $\theta > \pi/6$ and at its maximum is
smaller in magnitude than that from Alfv\'enic modes. Fast mode signal
becomes nearly isotropic as field wandering increases,  being only at
5\% when local wandering is similar to $M_a=2$ of Alfv\'enic case.
The synchrotron quadrupole from Fast modes is never expected to exceed
$|D_2/D_0| \approx 0.25$. Thus, measuring higher quadrupole
will put the importance of the Fast modes  under question.

As the last exercise we are considering the mix of  Alfv\'en and Slow modes
in the regime of strong incompressible turbulence. This is the regime of pure
$E-type$ modes with Alfv\'en like power spectrum distribution and 3D isotropic
tensor spectral tensor as modeled in Eq.~(\ref{eq:Alfven+Slow_tensor}).
The results in the right panel of Figure~\ref{fig:D2D4} shows that the
synchrotron quadrupole structure in this model is very close to that of
Alfv\'enic turbulence. This confirms again that exact local structure of the
spectral tensor in Alfv\'enic turbulence is not decisive in determining the
observable quadruple of the synchrotron. It is primarily reflects the anisotropic
distribution of power in the cascade.

\section{Discussion}

Our paper is aimed to provide the quantitative analytical description of the
statistics of synchrotron fluctuations. In this section we discuss what our
results mean in terms of observations and what is the relation of this work with
previous related publications.

\subsection{Our advances in describing synchrotron intensity statistics}

Synchrotron fluctuations can be characterized statistically by the slope of
their spectrum and their amplitude. In this paper we have obtained expressions
for the synchrotron fluctuation amplitude and have shown that the spectral slope
of fluctuations depends weakly on the spectral index of relativistic electrons.
The latter presents a substantial advance compared to the earlier studies, which
were providing expressions for just one electron spectral index, namely,
$\gamma=2$ and were not able to estimate the errors arising from this
assumption. Our approximation for correlations at arbitrary $\gamma$, namely,
\begin{eqnarray}
D_{sync, \gamma}({\bf R}) &\approx& {\cal A}(\gamma)
\sigma^{2\gamma-4} D_{sync, \gamma=2}({\bf R}) \times 
\label{major1} \\
&\times& \frac{\left(1 - \epsilon^2\right)^{\gamma}}{1+\epsilon^2}
\left[\left(1-\epsilon^2\right)^{\onehalf} 
\Gamma(1+\gamma) 
\mbox{}_{2}F_1 \left(\frac{1}{2}+\frac{\gamma}{2},1+\frac{\gamma}{2},1,
\epsilon^2\right) 
-\Gamma\left(1+\frac{\gamma}{2}\right)^2
\mbox{}_{2}F_1 \left(\frac{1}{2}+\frac{\gamma}{4},1+\frac{\gamma}{4},1,
\epsilon^2\right)^2
\right] \nonumber
\end{eqnarray}
where $\sigma^2=\sigma_{xx}+\sigma_{yy}$ are magnetic field dispersions
and $\epsilon \equiv \frac{\sigma_{xx}-\sigma_{yy}}{ \sigma_{xx} + \sigma_{yy}}$
is the degree of anisotropy  (see more in \S3,  Eq.~(\ref{major1}) )
allows to express the statistics of fluctuations through the statistics
obtained for $\gamma=2$. This expression opens avenues for analysis of the synchrotron
intensities for a variety of astrophysical circumstances, for which $\gamma$ 
substantially deviates from 2. In other words, synchrotron intensity studies
for which possible statistical description was earlier limited to a very special
case of a single and unique spectral index of cosmic rays has been extended
to a wide range of spectral indexes which encompass astrophysically important
situations. Eq. (\ref{major1}) is our first major result.

Earlier studies were limited mostly to hypothetical isotropic magnetic
turbulence \citep[see][]{1998A&AT...17..281C}. 
This would provide a very special and rather unrealistic form for
$D_{sync,\gamma=2}$ in Eq. (\ref{major1}).
Thus the second part of our paper was devoted to obtaining adequate forms for
$D_{sync,\gamma=2}$.

Unlike other earlier studies \citep[see][]{LazarianShutenkov}, we avoided
dividing magnetic field into the regular
and turbulent components. Such a division is unrealistic in view of modern day
understanding of turbulence. Instead,
in the presence of mean magnetic field, we adopted the model of the axially symmetric magnetic turbulence, where
the mean field is directed along the direction of the axis. Both theory and simulations are suggestive
that this description should correspond well in the global frame of reference, related to the mean magnetic field 
\citep[see][]{1999ApJ...517..700L,2000ApJ...539..273C,2002ApJ...564..291C}.
To describe the axially symmetric magnetic turbulence we used the general expressions in \citet{1997PhRvE..56.2875O},
which generalize the classical results in \citet{1946RSPSA.186..480B}  and \citet{1950RSPTA.242..557C}.
With this, our expression provided a combination of two different tensors with two functions $E$ and $F$ in front of them. At this point one
can establish the existence of the anisotropy of synchrotron signal and its relation to the aforementioned functions.

To move one step further, we adopted a model of compressible MHD turbulence from 
\citet{2002PhRvL..88x5001C,2003MNRAS.345..325C} and
expressed the statistics of the Alfv\'en, slow and fast modes in terms of the combinations of $E$ and $F$ functions.
This allowed us to express the expected anisotropies of the synchrotron emission through the physically motivated
basic modes existing in MHD turbulence. Our approach allows to predict the synchrotron emission statistics if
the turbulence is defined and, alternatively, study fundamental properties of MHD turbulence analyzing spectra and
anisotropies of the synchrotron statistics. 

While we believe that the decomposition into modes in \citet{2002PhRvL..88x5001C,2003MNRAS.345..325C} is
well motivated theoretically and proven numerically, our approach in the present paper can be applied to any arbitrary model of
axisymmetric turbulence. For instance, it can be applied to test the predictions of the model of MHD turbulence consisting
from slab and 2D fluctuations\footnote{We believe that this model may represent only transient state of the actual MHD turbulence and
is not physically motivated in terms of cascades of turbulent energy.} 
\citep{1990JGR....9520673M,1993PhFl....5..257Z}.
We are far from assuming that astrophysical magnetized compressible turbulence is completely understood. 
For instance, recent numerical simulations in \citet{2008ApJ...688L..79F,2009ApJ...692..364F,2010A&A...512A..81F}
demonstrate the effects of turbulence injection on the resulting turbulence. In the presence
of multiple sources of energy injection, instabilities and damping processes the spectrum of magnetic turbulence inferred from
synchrotron observations may be rather complex. Our description accounts for turbulence compressibility and our analytical expressions
can be used for turbulence with several injection scales. This opens avenues for realistic magnetic turbulence in its complexity.

\subsection{Studying magnetic field direction}

Our results show that magnetic field direction can be studied through the analysis of the
anisotropy of synchrotron intensity correlation functions. This provides a synergetic way of
studying magnetic fields. 

Other approaches to studying magnetic field direction are known in the literature. Some of
them are based on direct measurements of synchrotron polarization (see Ginzburg 1981), 
polarization from aligned grains (see review by Lazarian 2007 and references therein),
or aligned atoms (see Yan \& Lazarian 2008). In addition,
a statistical determination of the mean magnetic field direction is 
possible with the statistical analysis of the Doppler shifted lines
\citep{2002ASPC..276..182L,2005ApJ...631..320E,Esquivel2011}, that has similarities with the one
we proposed here. In this technique the difference in velocity correlations parallel
and perpendicular to mean magnetic field also reveals the direction of magnetic
field. The aforementioned difference was recently used to find the direction of
magnetic field in interstellar gas  \citep{2008ApJ...680..420H}. 

All these techniques measure magnetic field in different media and have their limitations.
For instance, the technique. based on the analysis of spectral lines is not applicable to
hot rarefied halo gas with little emission. Studying aligned dust in the latter environment
is also extremely challenging. At the same time, this is the domain of the new technique.
Needless to say that obtaining the same direction of the magnetic field with different
techniques increases the confidence of the result, while the discrepancy in the directions
may testify on the existence of different regions with different conditions and different
direction of magnetic field along the line of sight.

\subsection{Towards a cookbook for synchrotron studies}

Turbulence in our galaxy is believed to span from hundreds of parsec to hundreds
of kilometers. This follows both from the analysis of electron density
fluctuations (Armstrong et al. 1994, Chepurnov \& Lazarian 2010), spectral lines
(Larson 1981, see Lazarian 2009 and ref. therein), synchrotron fluctuations (see
Cho \& Lazarian 2010 and ref. therein) etc. With synchrotron emission one
expects to measure fluctuations arising from magnetic fluctuations starting from
the largest scale to very small scales, where the physical limitations on the
small scale arise either from turbulence damping or Larmor radius of
relativistic electrons. The practical limitations on turbulence studies will
arise from both noise and the resolution of the data. As turbulence fluctuations
decrease in amplitude corresponding to the Kolmogorov power spectrum, i.e. with
the two dimensional measured spectrum $P_2\sim K^{-11/3}$, the small scale
fluctuations are more subjected to the effect of noise. 

The new generation of telescopes (e.g. SKA, LOFAR) present observers with lower
noise and higher resolution data which should enable them to reliably study
synchrotron fluctuations over a wider range of scales. This will open prospects
of detailed studies magnetic turbulence not only in Milky Way but also in
supernovae remnants, radio lobes and external galaxies. For this wide variety of
objects electron spectral index varies, but our present work shows that this is
not a problem. 

Our present paper opens a new dimension in studies of synchrotron maps. Apart
from getting the spectral index of magnetic turbulence, our analytical results
testify that by the analysis of synchrotron anisotropies one can obtain the direction
of magnetic field. Moreover, it can also deliver very unique piece of information, i.e. the
relative contribution of compressible versus incompressible motions. Indeed, our
study of the spatial multipoles of the synchrotron correlations shows that the
signatures of Alfvenic and fast-wave turbulence are different. These signatures can
be studied observationally to identify regions with different degrees of
compressibility. Does higher or lower compressions of magnetic field correlate
to higher star formation rate? What about compressibility and magnetic field
amplification? These can be the questions that one will be able to answer
analyzing synchrotron maps and correlating the results with other astrophysical
data.

\subsection{Statistics of synchrotron polarization}

While most of the present paper is devoted to studies of intensities, in Appendix E we provide the 
description of the fluctuations of other Stocks parameters and their combination. This provides
an important extension of the approach suggested in this paper. 

We noticed that the structure of the correlations of other Stocks parameters is similar to that
of intensity. On the basis of this we could conjecture that the dependence of the correlations
involving polarization will be similar to the correlations of intensity, i.e. with the dependence
on the cosmic ray spectral index affecting mostly the prefactor of obtained correlations. Numerical
studies of synchrotron fluctuations with cosmic rays corresponding to different $\gamma$ 
in Ensslin et al. (2010) are suggestive of this outcome. 

In any case, in Appendix E, in analogy with the rest of the paper, we provided calculations of correlations
of other Stocks parameters and their combinations. We observe that that combining intensity and polarization
statistics one can remove the degeneracies, for instance, arising from the unknown 3D direction of the
mean magnetic field, which determines the axis of anisotropy for magnetic turbulence. The bonus of 
combining the statistics of polarization and intensity could be determining of this 3D direction. This study should
be continued. 

Our study opens avenues of addressing subtle issues related to magnetic field structure. For instance, Waelkens et al. (2009) proposed to study tension force spectrum from fourth order statistics of polarized data. They, however,
used the isotropic model of turbulence, which does not correspond to what we now know about magnetic turbulence. Our approach allows such studies within a realistic model of magnetic turbulence. 

In Appendix~\ref{sec:zcomponents} we gave, for completeness, the expressions for
the sky-projected structure functions between the line-of-sight components
of the magnetic field. While not appearing in synchrotron intensity studies,
the line-of-sight part of the magnetic field is important in describing the
Faraday rotation of the polarization direction.

Having the statistical description of synchrotron fluctuations and the effects of Faraday rotation on the fluctuations
we open prospects of describing the statistics of synchrotron polarization data cubes which contain synchrotron polarization map for different frequencies (see Haverkorn  2010, Gaensler et al. 2011). Our approach opens ways
to provide detailed quantitative description of such data cubes. 

\subsection{Synchrotron foreground for CMB and high-redshift HI studies}

It is important to stress, that the description that we obtained is important for both studying galactic and
extragalactic magnetic fields and for removing the synchrotron fluctuations from
cosmological CMB fluctuations as well as from the fluctuations high redshifted
atomic hydrogen which can provide a unique insight into the processes at the
beginning of the Universe.

We expect our description to be important for studies of galactic synchrotron foregrounds.
In terms of weeding out the foreground fluctuations, a procedure of partial
filtering was developed in \citet{2010ApJ...720.1181C}. It was proven there that the
galactic foreground fluctuations, e.g. synchrotron fluctuations,
arise from galactic turbulence with the spectral index of fluctuations that
depends on the spectral index of the underlying turbulence and the geometry of
the emitting turbulent volume. This determinism allows one to predict the
fluctuations
of foreground emission at a given scale, provided that the foreground
fluctuations are
measured at a different scale. The procedure is applicable to both spatial
filtering
of CMB signal and high redshifted HI emission from the early Universe.  For
instance, if synchrotron is separated in low
resolution measurements, the procedure in \citet{2010ApJ...720.1181C} allows to
remove its contribution for high resolution measurements. The accuracy of the
removal increases if we know more about the expected power law of the spatial
fluctuations. In view of this the effect of $\gamma$ not being equal to the
value 2 assumed in the estimates was a concern. Our present work removes this
concern and allows to use the procedure in \citet{2010ApJ...720.1181C} with higher
confidence.

Our obtaining in Appendix E of the tensor of synchrotron polarization should also
help to weeding off the polarized galactic synchrotron foreground. This foreground
is known to be important for the CMB polarization studies, in particular, for
the studies of the illusive B-modes.

In general, the more we know about the synchrotron statistics, the more reliable is its removal. For
instance, if the filtered signal show the anisotropies expected from the synchrotron
emission, this would testify that the synchrotron foreground was not properly removed.
Additional, more sophisticated procedures may also be developed. For instance, the procedure of
foreground removal proposed in Cho \& Lazarian (2010) requires the best possible knowledge of
the foreground spectrum at one spatial resolution in order to remove the foreground at another resolution.
The a priori knowledge of what we expect in terms of the anisotropy of the synchrotron statistics help
to determine the foreground statistics with higher accuracy.

\subsection{Anisotropic part of synchrotron correlation tensor}

In this paper we provided a detailed analysis of the expected properties of the 
anisotropies of synchrotron fluctuations. These anisotropies were also of interest
for earlier studies, which were strongly limited in applicability because both of
the assumption about the assumed single value of electron spectral index and of the 
assumed models of magnetic turbulence. These earlier studies are of interest as the
approach presented in our paper allows to cure their deficiencies.

For instance, a technique for obtaining the size of galactic halo was proposed 
in \citet{LazarianChibisov}. The authors proposed to use HII regions as distance indicators
and calculate anisotropic part of synchrotron intensity correlations for the directions 
towards and away from the HII region. Choosing sufficiently long radio wavelengths for
which HII regions are opaque, one can probe magnetic fields on the scale of the
distance from the observer to the HII region. These correlation functions were to be
compared with the correlation functions obtained for both lines of sight avoiding HII
region or with correlation functions at the wavelengths for which the HII regions are transparent. 

It is possible to see that if the angular scale of
synchrotron fluctuations is $\theta$ and the distance to the HII region is $D$,
the physical size of the magnetic fluctuations is $\sim D\sin\theta$. If the
angular scale of synchrotron fluctuations is $\theta'$, then the ratio of the
halo size to the size of the distance to the HII region will be
$\sin\theta/\sin\theta'$. Our study of the weak dependence of synchrotron statistics 
expressions on $\gamma$ gives more confidence that similar studies using anisotropic part of
the correlation tensor can be done in practice.

\subsection{Effects of cosmic rays}

The fact that the expression for synchrotron intensity depends on the spectral
index of cosmic rays has been an impediment for the quantitative studies of the
statistics of synchrotron variations. Our study shows that cosmic ray spectral
variations change only the amplitude of the
synchrotron variations and do not change the angular dependences of the correlation
and structure functions of the synchrotron intensities (see Eq. (\ref{major1}). They
do not change the direction of the anisotropies either. This simplifies the interpretation of
the observations.     

Our study has been performed in the assumption of no magnetic field correlation
with the density of relativistic electrons. This assumption corresponds to
observation of much smoother distribution of relativistic electrons compared to
the distribution of magnetic fields in the Milky Way. Our formalism, however,
may be extended for the case of relativistic electrons correlated with magnetic
fields. We do not expect substantial changes in our present conclusions, e.g. in
the dependence of synchrotron intensity spectrum on $\gamma$.

\subsection{Complex geometry of observations}

In the paper we assumed that synchrotron emission is coming to the observer from
a single region. Even in this case, the statistics of fluctuations can change
from geometrical effects of observing galactic turbulence. For angles
$\phi$ between lines
of sight less than
$L_{turb}/D_{obj}$ where $L_{turb}$ is the turbulence injection scale and
$D_{obj}$ is the size the observed region size along the line of sight, the
spectral index of the angular spectrum of synchrotron fluctuations coincides 
with the spectral index of synchrotron emissivity fluctuations. For larger
angles the spectral index changes. In an idealized case of uniform turbulence
surrounding an observer, it gets universal corresponding to the correlation
function of intensity $\sim \phi^{-1}$. This effect was first discussed in
Shutenkov \& Lazarian (1990) and attributed to the converging lines of sight
geometry of observations. Later the change of the spectral index of synchrotron
fluctuations was confirmed with the analysis of the observational
data\footnote{It is worth mentioning that the change of the spectral index
reported in Cho \& Lazarian (2002) on the basis of the analysis of data sets for
high latitude synchrotron emission as well as for
synchrotron emission from the disk is consistent with the injection scale of 100
pc.} in Cho \& Lazarian (2002). This paper reveals the relation of the spectral
index of synchrotron emissivity with the spectral index of underlying magnetic
turbulence which shows that for $\phi<\phi_{max}=D_{obj}/L$ we get the
actual spectrum of the underlying magnetic turbulence.

For studies of synchrotron emission in a volume that does not surround the
observer, as this is the case for extragalactic studies, it is possible to show
that the direct relation of the spectral index of angular fluctuations and 
the underlying magnetic turbulence is preserved if
$\phi<\theta_{max}=\Delta_{obj}/L$, where $\Delta_{obj}$ is the extend of the
emitting volume along the line of sight. This is the regime that one is mostly
interested in.

In terms of anisotropy studies, both regimes of $\phi>\phi_{max}$ and
$\phi<\phi_{max}$ allow determination of the magnetic field direction.
This is true for both galactic and extragalactic studies. 

If the observations are done looking through multiple disconnected turbulent
volumes with uncorrelated direction of magnetic field, a
decrease of the degree of anisotropy is expected. Naturally, this is not the
best case for
studies of turbulence either. Provided that for all
regions along the line of sight $\phi$ is less than $\phi_{max}$, one is
expected to obtain the same power spectral index provided that for all volumes
we study, synchrotron fluctuations arise from turbulence in the inertial
range.

\subsection{Synergy of techniques for studying turbulence spectra}

In the section above we have discussed the advantages of using synchrotron fluctuations
for studies of the ISM. However, the technique of magnetic turbulence study that we propose in the present paper
should be viewed as a technique complementary to other existing techniques. We
believe that the most interesting questions can be answered if properties of turbulence
in different parts of the galaxy and in different ISM phases are compared. 

In addition to the magnetic field turbulence spectrum discussed in our paper it is
advantageous to study also spectra of density and velocity. If the study of the 
former is rather straightforward from the column
density maps  \citep[see][]{1999ptep.proc...48S}, studying velocity is far from trivial. In fact,
our studies in \citet{2005ApJ...631..320E,2007MNRAS.381.1733E}
revealed that a routinely adopted tool for turbulent velocity
studies, namely, velocity centroids\footnote{One may also use so-called
"modified velocity centroids" \citep{2003ApJ...592L..37L}, but those, while
increasing the range of applicability of centroids and increasing their accuracy
in tracing velocities, still cannot be applied for the observational studies of
high Mach number turbulence.} cannot reliably reveal velocity fluctuations for
supersonic turbulence. However, two new techniques termed Velocity Channel
Analysis (VCA) and Velocity Coordinate Spectrum (VCS) have been introduced
during the last decade \citep{2000ApJ...537..720L,2004ApJ...616..943L,2006ApJ...652.1348L,2008ApJ...686..350L}.
These techniques have been successfully tested and applied to observational data
\citep[][see also review in \citet{2009SSRv..143..357L} for the
references therein]{2001ApJ...555..130L,2006ApJ...653L.125P,2009ApJ...707L.153P,2009ApJ...693.1074C,2010ApJ...714.1398C}.
They allow to reliably retrieve data on velocity spectra
from the Doppler shifted emission and absorptions lines.

We would like to stress that the overall shape of the turbulence spectrum
and its spectral index can be very informative. The former provides information about
the sources and sinks of turbulence, the latter characterizing the cascading of
energy and, possibly, the role of plasma instabilities acting in collisionless media.
Thus advantages and insight arising from the ability of obtaining of magnetic field spectra and comparing its to spectra obtained in Doppler shifted lines are difficult to overestimate.

Spectrum is not the only measure of turbulence. New techniques to study sonic and Alfvenic Mach numbers (i.e.
$M_s$ and $M_A$, respectively) of turbulence have been developed recently.
They include so called Tsallis statistics
\citep[see][]{2004GeoRL..3116807B,2004JGRA..10912107B,2005PhyA..356..375B,2006PhyA..361..173B,2006ApJ...642..584B,
2006ApJ...644L..83B,2007ApJ...668.1246B,2009ApJ...691L..82B,2010ApJ...710..125E,2011arXiv1103.3299T},
bi-spectrum \citep[see][]{2009ApJ...693..250B,2010ApJ...708.1204B},
genus analysis \citep[see][]{2008ApJ...688.1021C}, etc. They allow us
to get a comprehensive picture of magnetic turbulence.  
Our analysis also provides a synergetic way of studying magnetic field compressibility.  
For instance, synchrotron fluctuations
may map extended turbulent volume of the galactic halo. The relation of this
turbulence taking place in tenuous plasma to the turbulence of atomic and
molecular species which can be revealed with the VCA and VCS techniques. This will be
very valuable for understanding of the complex interrelated dynamics of ionized and neutral  media in our galaxy.

In addition, the detailed work that we performed in this paper on describing of anisotropy of synchrotron fluctuations
arising from anisotropic magnetic turbulence will benefit other directions of turbulence studies. Similar anisotropies
arise, for instance in velocity fluctuations. These anisotropies were suggested by Lazarian, Pogosyan \& Esquivel (2002) as a means of studying the magnetic field direction and were shown to be present both in channel maps of
position-position-velocity (PPV) data cubes of synthetic data and in the statistics obtained with velocity centroids. Later, velocity centroids were studied in Esquivel \& Lazarian (2005, 2011) with synthetic data corresponding to numerical simulations with different Mach numbers, $M_s$ and $M_A$, and empirical dependences of the degree of anisotropy on these numbers was obtained. Applying the analytical approach developed in this paper it is possible to
separate contributions of the compressible and Alfvenic parts of the magnetic turbulence similar as we described above. This is going to provide new ways of detailed quantitative studies of interstellar turbulence using the wealth of spectroscopic surveys.

\section{Summary}

Our study has revealed the following important properties of the statistics of
synchrotron fluctuations:\\
1. In terms of functional dependence, the spectrum of spatial synchrotron
fluctuations marginally depends on the
spectral index of relativistic electrons, while the amplitude of the
fluctuations depends on the relativistic electron spectral
index. \\
2. Magnetic turbulence makes fluctuations of synchrotron radiation anisotropic
with the direction anisotropy given by the averaged mean magnetic field. As we
found that the quadrupole anisotropy is the most prominent, this provides a new
way of studying the
direction of magnetic field using just synchrotron intensity variations.\\
3. The relation between the isotropic and anisotropic parts of the synchrotron
intensity fluctuations provides a way to study the compressibility of magnetic
turbulence.

\acknowledgments

AL acknowledges the NSF grant AST 0808118, NASA grant NNX11AD32G and the support of the Center of
Magnetic Self-Organization (CMSO). Humboldt Award and related stimulating stay in Universities of Cologne and
Bochum is also acknowledged by AL. AL and DP thank Mark Baker for many useful
discussions and reviewing the formulas in Appendix~\ref{sec:3Dpower_to_corr}.

\appendix

\section{Table of notation}
\begin{deluxetable}{ccl}
\tablecaption{List of notations}
\startdata
\hline
&& \\
${\hat \lambda}$ && unit vector of direction of the statistical symmetry axis 
(mean magnetic field) \\
z-axis &&  is along the line of sight \\
y-axis &&  is perpendicular to both the line of sight and the 
symmetry axis $\hat \lambda$\\
x-axis &&  is perpendicular to the line of sight and parallel to the sky
projection of the symmetry axis $\hat \Lambda$\\
${\bf x}_1, {\bf x}_2 \ldots$ && 3D position vectors \\
${\bf r},~r$ && 3D separation vector and its magnitude, ${\bf r} = ({\bf
R},z)$\\
${\bf k},~k,~{\bf \hat k}$ && 3D wave vector and its magnitude, ${\bf k} = ({\bf K}, k_z)$, and the unit
vector ${\bf \hat k} = {\bf k}/k$
\\
$\mu$  && cosine of the angle between the 3D wave vector and the symmetry axis
$\mu = {\bf \hat k \cdot \hat \lambda}$ \\
&&\\
\hline
&&\\
$\hat \Lambda$ && 2D unit vector of the sky projection of the symmetry axis
$\hat \lambda$ \\
${\bf X}_1, {\bf X}_2 \ldots$ && 2D position vectors \\
${\bf R},~R$ && 2D sky projection of the separation vector and its magnitude \\
${\bf K},~K,~{\bf \hat K}$ && 2D wave vector and its magnitude,  and the unit
vector ${\bf \hat K} = {\bf K}/K$\\
$\psi$ && polar angle between 2D wave vector ${\bf K}$ and the x-axis, ${\bf \hat K}=(\cos \psi, \sin \psi)$\\
$\phi$ && polar angle between 2D separation ${\bf R}$ and the x-axis \\
&& \\
\hline
&& \\
$I_{sync}({\bf X})$ && intensity of synchrotron emission \\
$\xi_{sync}({\bf R})$, $D_{sync}({\bf R})$  && correlation/structure function of synchrotron intensity \\
$\tilde \xi_{sync}({\bf R})$, $ \tilde D_{sync}({\bf R})$  && normalized correlation/structure
 function of synchrotron intensity \\
$ \tilde D_n({\bf R})$  &&  coefficient of the multipole decomposition of $ \tilde D_{sync}({\bf R})$  \\ 
&& \\
\hline 
&& \\
$H_\perp({\bf x})$ && magnitude of the component of the magnetic field orthogonal to the
line of sight. \\
$\xi_{H^\gamma_\perp}({\bf r})$, $D_{H^\gamma_\perp}({\bf r})$  && 3D correlation/structure function
 of $\gamma$'s power of $H_\perp({\bf r})$ \\
$\tilde \xi_{H^\gamma_\perp}({\bf r})$, $\tilde D_{H^\gamma_\perp}({\bf r})$  && normalized 3D correlation/structure function
 of $\gamma$'s power of $H_\perp({\bf r})$ \\ 
&& \\
\hline
&& \\
$E({\bf k})$, $F({\bf k})$ && 3D power spectra functions.
For axisymmetric turbulence $E({\bf k}) = E(k,\mu)$ and
$F({\bf k}) = F(k,\mu)$ \\
$E_l$, $F_l$ && multipole coefficient of Legendre expansion of $E(k,\mu)$ and $F(k,\mu)$ in $\mu$ \\
$\hat E(\mu)$, $\hat F(\mu)$ && angular dependence of $E(k,\mu)$ and $F(k,\mu)$ if scale dependence is factorized \\
$\hat E_l $, $\hat F_l$ && multipole coefficient of Legendre expansion of $\hat E(\mu)$ and $\hat F(\mu)$ \\
$\sigma_{||}$, $\sigma_\perp $&& variances of the magnetic field components parallel and perpendicular
to the symmetry axis (mean field) \\
$r_\sigma$ && ratio of $\sigma_{||}$ and $1/2 \sigma_\perp$ \\

$E({\bf K)}$, $ F({\bf K})$ && $  E({\bf k}) $ , $F({\bf k})$ with $k_z=0$, power spectra of the fluctuations integrated
along the line of sight\\
$E_n $, $F_n$, $\hat E_n $, $\hat F_n$&& coefficients of the harmonic expansion
 of $E({\bf K)}$ and $ F({\bf K})$ or their factorized version \\
$\sigma_{xx},\sigma_{yy}$ && variances of the $x$ and $y$ magnetic field components\\
$\epsilon$ && anisotropy measure of the variance of the sky components of the magnetic field, 
$\epsilon = \frac{\sigma_{xx}-\sigma_{yy}}{\sigma_{xx} + \sigma_{yy}}$ \\
$D_{ij}, \; D_{zz}$ && 2D sky and line-of-sight components of the z-integrated correlation (structure) tensor 
of the magnetic field\\
$D^+,\; D^-, \;D^\times$ && $D^+ = D_{xx}+D_{yy}$, $D^+ = D_{xx}-D_{yy}$, $D^\times = D_{xy}$ \\

&& \\

\hline
&& \\ 
${\hat \lambda_0}$ && 
unit vector of direction of the global symmetry axis in case of mean field wandering \\
$\bar \chi$ && average cosine of the angle between the local and the global symmetry directions\\
$W_I(\bar \chi)$ && weight of the isotropized spectral part \\
$W_L(\bar \chi)$ && weight of the local anisotropic spectral part\\
$G^A(\psi,\theta),  G^F(\theta) $ && geometric functions describing the 2D structure of the turbulent mode\\
$G^A_n(\theta),  G^F_n(\theta) $ && multipole decomposition of the geometric functions\\
$r_I$ &&  energy injection scale \\
$\beta$ && plasma constant \\
$M_a$ && Alfvenic Mach number \\
\enddata
\end{deluxetable}

\section{Axisymmetric statistics of vector fields}

\label{sec:3Dpower_to_corr}

\subsection{From power spectrum to correlation function}
Let us consider axisymmetric MHD turbulence that is described by a single power
$E-type$ power spectrum
\begin{equation}
E({\bf k}) = E(k, {\bf k} \cdot \hat {\bf \lambda}) \left( \delta_{ij} - \hat
k_i \hat k_j \right)
\end{equation}
This is not the most general case of axisymmetric
spectrum, since in general one can have four independent spectral functions.
However, this ansatz arise in the regime of strong incompressible MHD turbulence
where contribution of Alfv\'enic and slow modes is similar, as found in  numerical
simulations by \citet{2003MNRAS.345..325C}.

Let us consider what structure functions correspond to these models.
We start from the general expression for the correlation function
\begin{equation}
\left\langle H_i({\bf x_1}) H_j({\bf x_2}) \right\rangle = \int d^3 {\bf k} e^{i
{\bf k}
\cdot {\bf x}} E({\bf k}) \left(\delta_{ij} - \hat k_i \hat
k_j  \right)
\end{equation}
and rewrite it as
\begin{eqnarray}
\left\langle H_i({\bf x_1}) H_j({\bf x_2}) \right\rangle &=& \left[
\frac{\partial}{\partial r_i} \frac{\partial}{\partial r_j}
-\delta_{ij}  \Delta \right]
\int dk d\Omega_{\bf k} e^{i {\bf k} \cdot {\bf x}} E(k,{\bf k} \cdot
\hat {\bf \lambda}) \nonumber \\
&=& \left[
\frac{\partial}{\partial r_i} \frac{\partial}{\partial r_j}
-\delta_{ij}  \Delta  \right] \Phi(r,r \mu)
\end{eqnarray}

Scalar function $\Phi(r,r \mu)$ depend on $r$ and
$\alpha = {\bf r \cdot \hat \lambda} = r \mu $. Following \citet{1950RSPTA.242..557C},
it is convenient to introduce
the following differential operators
\begin{equation}
D_r = \left. \frac{1}{r} \frac{\partial}{\partial r} \right|_{\alpha=const}
= \frac{1}{r}\frac{\partial}{\partial r} - \frac{\mu}{r}
\frac{\partial}{\partial \mu},
\quad
D_\mu = \left. \frac{\partial}{\partial \alpha}\right|_{r=const} = \frac{1}{r}
\frac{\partial}{\partial \mu}
\end{equation}
Notice, that $D_r$ reflects partial derivative with respect to $r$ when
$\alpha$, rather than $\mu$ is kept constant.
$D_\mu$ is actually derivative with respect to $\alpha$ and has dimensions of
$1/r$.

In terms of these differential operators, the tensor of second partial
derivatives and the Laplacian of an axisymmetric function are
\begin{eqnarray}
\frac{\partial^2}{\partial r_i \partial r_j} \Phi(r, r\mu) &=&
 r_i r_j D_{rr} \Phi + \delta_{ij} D_r \Phi  + \hat \lambda_i \hat \lambda_j
D_{\mu\mu} \Phi + \left(r_i \hat \lambda_j + r_j \hat \lambda_i\right) D_{r\mu}
\Phi \\
\Delta \Phi(r, r\mu) &=& \left( r^2 D_{rr} + 3 D_r + D_{\mu\mu} + 2 r \mu D_r
D_\mu \right) \Phi
\label{eq:Laplacian_def}
\end{eqnarray}
and one obtains
\begin{equation}
\left\langle H_i({\bf x_1}) H_j({\bf x_2}) \right\rangle =
\left(r^2 D_{rr} \Phi \right) \hat r_i \hat r_j +
\left(D_r - \Delta \right) \Phi \delta_{ij} + D_{\mu\mu}
\Phi \hat \lambda_i \hat \lambda_j  + r D_{r\mu} \Phi \left(\hat r_i
\hat \lambda_j + \hat r_j \hat \lambda_i\right)
\end{equation}
that leads to the following identification of the correlation coefficients
\begin{equation}
A_\xi = r^2 D_{rr} \Phi \;, \quad B_\xi = \left(D_r - \Delta \right) \Phi
\;,\quad C_\xi = D_{\mu\mu} \Phi \;, \quad D_\xi= r D_{r\mu}
\Phi ~.
\end{equation}

\subsection{Multipole decomposition}

Decomposition of the plain wave and the spectrum in spherical harmonics
\begin{eqnarray}
e^{i {\bf k} \cdot {\bf x}} &=& 4 \pi \sum_l i^l j_l(k r) \sum_{m=-l}^l
Y_{lm}(\hat k) Y_{lm}^*(\hat r) \\
E(k,{\bf k} \cdot \hat {\bf \lambda}) &=& \sum_{lm} \frac{4 \pi}{2l+1} E_l(k)
Y_{lm}^*(\hat k) Y_{lm}(\hat r)
\end{eqnarray}
gives after integration over angles $d\Omega_{\bf k}$
\begin{equation}
\Phi(r,r\mu) = 4 \pi \sum_l i^l P_l(\mu) \int \! \! dk \; j_l(k r) \; E_l(k) ~.
\end{equation}
The multiple expansion of the correlation tensor coefficients follows
\begin{eqnarray}
\label{eq:Axi}
 A_\xi &=& 4 \pi \sum_{n=0}^\infty i^n P_n(\hat {\bf r} \cdot \hat {\bf
\lambda})
 \left[  -\tilde \Upsilon_n + (3 + 2 n) \tilde \Gamma_n - 2 (1+2 n) \tilde
\Gamma_{n+2}
+ (1+2 n )(3 + 2 n) \Lambda_{n+4} \right] \nonumber \\
&+& 2 \pi \sum_{n=0}^{\infty}  P_n(\hat {\bf r} \cdot \hat {\bf \lambda}) (1 + 2
n)
\sum_{l=n+4,2}^{\infty} i^l  \left[4 \tilde \Gamma_l - \left(
l(l+1)-n(n+1)\right) \Lambda_{l+2}  \right] \\
\label{eq:Bxi}
B_\xi &=& 4 \pi \sum_{n=0}^ \infty i^n P_n(\hat {\bf r} \cdot \hat {\bf
\lambda})
\left[ \tilde \Upsilon_n - \tilde \Gamma_n + (2 n + 1) \Lambda_{n+2} \right]
\nonumber \\
&-& 4 \pi \sum_{n=0}^\infty P_n(\hat {\bf r} \cdot \hat {\bf \lambda})
(1 + 2 n ) \sum_{l=n+4,2}^\infty i^l \Lambda_l \\
\label{eq:Cxi}
C_\xi &=& 2 \pi \sum_{n=0}^\infty (1+2n) P_{n}(\hat {\bf r} \cdot \hat {\bf
\lambda})
\!\!\sum_{l=n+2,2}^\infty \!\! i^l (l-n)(l+n+1) \Lambda_l \\
\label{eq:Dxi}
 D_\xi & = &
-2\pi \sum_{n=0}^\infty (1+2n)  P_{n} (\hat {\bf r} \cdot \hat {\bf \lambda})
\sum_{l=n+1,2}^\infty i^l \Bigg[ 2 \tilde \Gamma_l - \Bigl( (l+1)(l+2)-n(n+1)
\Bigr)\Lambda_{l+2}\Bigg]
\end{eqnarray}
where $\tilde \Upsilon_l(r)  = \int \! k^2 dk \; j_l(kr) E_l(k)$,
$\tilde \Gamma_l(r) = \int\! k^2 dk \; \frac{ j_{l+1}(kr)}{kr} E_l(k) $ and
$\Lambda_l(r) = \int\! k^2 dk\;  \frac{ j_{l}(kr)}{(kr)^2} E_l(k) $

It is instructive to write out explicitly low multipole  terms
wrt the angle with the symmetry axis and investigate their behaviour as $r \to
0$.
Leaving only the $r \to 0$ leading terms, for the monopole
\begin{eqnarray}
 A_0(r) &\sim& 4 \pi \int \! k^2 d k \left[-\left( j_0(kr) - 3 \frac{
j_1(kr)}{kr} \right) E_0
- 2 \frac{ j_3(kr)}{kr} E_2 + 3 \frac{ j_4(kr)}{(kr)^2} E_4 \right]\\
B_0(r) &\sim& 4 \pi \int \! k^2 d k  \left[ \left( j_0(kr) - \frac{ j_1(kr)}{kr}
\right) E_0
 + \frac{ j_2(kr)}{(kr)^2} E_2 \right] \\
C_0(r) &\sim& -12 \pi \int \! k^2 d k \left[\frac{ j_2(kr)}{(kr)^2} E_2 \right]
\quad ,
\end{eqnarray}
and for the next order
\begin{eqnarray}
 A_2 &\sim&  4 \pi \int \! k^2 d k \left[\left( j_2(kr) - 7 \frac{ j_3(kr)}{kr}
\right) E_2
+ 10 \frac{ j_5(kr)}{kr} E_4 - 35 \frac{ j_6(kr)}{(kr)^2} E_6 \right]\\
B_2 &\sim& -4 \pi \int \! k^2 d k  \left[ \left( j_2(kr) - \frac{ j_3(kr)}{kr}
\right) E_2
 + 5 \frac{ j_4(kr)}{(kr)^2} E_4 \right] \\
C_2 &\sim&  140 \pi \int \! k^2 d k \left[\frac{ j_4(kr)}{(kr)^2} E_4 \right]\\
D_1 &\sim& 12 \pi  \int \! k^2 d k
\left[\frac{ j_3(kr)}{kr} E_2 - 5  \frac{ j_4(kr)}{(kr)^2} E_4\right] \quad .
\end{eqnarray}

At zero separation $r \to 0$, $A_0(0)$ and $D_0(0)$ while $B_0(0)= 4 \pi \int \!
k^2 d k  \left[ \frac{2}{3} E_0
 + \frac{1}{15} E_2 \right]$ and $C_0(0)= -\frac{4 \pi}{5} \int \! k^2 d k E_2$
describe the variance and same-point correlation
of the components of the vector field
\begin{equation}
\left\langle H_i({\bf x_1}) H_j({\bf x_1}) \right\rangle = B(0) \delta_{ij} +
C(0) \lambda_i \lambda_j
\end{equation}
Hence, for axisymmetric field not only the variances of the components differ,
but there is zero separation correlation
between different components if no frame axis is aligned with the symmetry axis.

In case of structure functions one computes $B(0)-B(r), \ldots $, which leads to
regularization of the monopole terms $B_0$ and $C_0$.

\subsection{Derivation of the multipole decomposition}
Derivations of the expressions Eqs.~(\ref{eq:Axi}-\ref{eq:Dxi}) involves
evaluating derivatives of the $j_l(kr) P_l(\mu)$ basis functions and rearranging
the series in multipole ``l''.  We start with the following relations for the
the angular derivatives
\begin{eqnarray}
\mu \frac{\partial}{\partial \mu} P_l(\mu) &=& l P_l(\mu) +
\sum_{n=(0,1),2}^{l-2} (2n + 1) P_n(\mu)\\
\left(\mu \frac{\partial}{\partial \mu}\right)^2 P_l(\mu) &=& 
l^2 P_l(\mu) + \frac{1}{2}  \sum_{n=(0,1),2}^{l-2} \left[(l+2)(l-1) -
n(n+1)\right](2n + 1) P_n(\mu) \\
\frac{\partial^2}{\partial \mu^2} P_l(\mu) &=& \frac{1}{2}
\sum_{n=(0,1),2}^{l-2}
(l - n)(l + n + 1)(2n + 1) P_n(\mu)
\end{eqnarray}
where the notation $\sum_{n=(0,1),2}$ means that the sum over $n$ starts from
the first value in the parenthesis (zero) when $l$ is even and from the second one (unity) when $l$ is odd,
and proceeds with the step of two. 
The sums are zero for $l=0,1$. 

For the radial derivative we shall use
\begin{equation}
\frac{1}{k} \frac{\partial}{\partial r} j_l(k r) = \frac{l}{k r} j_l(k r) -
j_{l+1} (k r) \quad .
\end{equation}

Now we can write out the actions of several operators that we need. The
following expressions show some intermediate steps, and for the final conclusion
point to the structure that the operator creates in the full multipole
expansion after the integration over $k$ has been carried out
\begin{eqnarray}
\label{eq:dr}
D_r \left[ j_l(kr) P_l(\mu) \right] &=& P_l(\mu) \frac{1}{r}
\frac{\partial}{\partial r} j_l(kr) - \frac{j_l(kr)}{r^2} \mu
\frac{\partial}{\partial \mu} P_l(\mu)
= -k^2 \frac{j_{l+1}(kr)}{k r} P_l(\mu) - k^2 \frac{j_l(k r)}{k^2 r^2}
\sum_{n=(0,1),2}^{l-2} \!\!\!\! (2n+1) P_n(\mu) \nonumber \\
& \to & - \tilde \Gamma_l(r) P_l(\mu)
- \Lambda_l(r) \sum_{n=(0,1),2}^{l-2} (2n+1) P_n(\mu) \quad ,\\
\label{eq:muDrDmu}
r \mu D_\mu D_r \left[ j_l(kr) P_l(\mu) \right] &=& \left( \mu
\frac{\partial}{\partial \mu} \right)   D_r \left[ j_l(kr) P_l(\mu) \right]  \\
&\to& -  \tilde \Gamma_l(r) \left[l \;P_l(\mu) +
\!\!\!\sum_{n=(0,1),2}^{l-2} \!\!\!\! (2n+1) P_n(\mu) \right] - 
 \frac{1}{2}  \Lambda_l(r) \!\!\! \sum_{n=(0,1),2}^{l-2}  \!\!\!\!
\left[(l+1)(l-2) -
n(n+1)\right](2n + 1) P_n(\mu) \nonumber \\
\label{eq:Dmumu}
D_{\mu\mu} \left[ j_l(kr) P_l(\mu) \right] &=& k^2 \frac{j_l(kr)}{k^2
r^2}\frac{\partial^2}{\partial \mu^2} P_l(\mu) 
\to  \frac{1}{2} \Lambda_l(r)  \sum_{n=(0,1),2}^{l-2}
(l - n)(l + n + 1)(2n + 1) P_j(\mu) \quad ,\\
\label{eq:Laplacian}
\Delta \left( j_l(kr) P_l(\mu) \right) &=& - k^2 j_l(k r) P_l(\mu) 
\to - \tilde \Upsilon_l(r) P_l(\mu) \quad .
\end{eqnarray}

Multipole expansion  of the correlation coefficients now follows from
combination of the above results. In order to obtain the final expressions, we
make
two important steps. Firstly we change the order of summation $\sum_{l=0}^\infty
\sum_{n=(0,1),2}^{l-2} = \sum_{n=0}^\infty \sum_{l=n+2,2}^\infty$ and take
the label for the exterior sum  to be $n$ everywhere. Secondly we
collect together the lowest order terms of the same power of divergence in $k$
integrals.

Let us proceed in the order of difficulty 
\subsubsection{$C_\xi$}
$C_\xi$ is obtained directly from Eq.~(\ref{eq:Dmumu}).

\subsubsection{$B_\xi$}
$B_\xi$ is a combination of Eq.~(\ref{eq:dr}) and Eq.~({\ref{eq:Laplacian}).
Let us illustrate the reversal of the summation order on this example. Using
$n$ for the exterior index the direct combination of two equations and
the change of the summation order in the double sum gives
\begin{eqnarray}
B_\xi = 4 \pi \sum_{n=0}^\infty i^n P_n(\mu) \left( \tilde\Upsilon_n - \tilde
\Gamma_n  \right) 
- 4 \pi \sum_{n=0}^\infty (2n+1) P_n(\mu) \sum_{l=n+2,2}^\infty i^l \Lambda_l
\end{eqnarray}
It is important that $\Lambda_0$ contribution  is not
present since the integral in $\Lambda_n \propto
j_n(k r)/(kr)^2 \sim (kr)^{n-2}$ is potentially the most divergent one. $\tilde
\Upsilon_n \sim j_n(kr) \sim (k r)^n $ and $\tilde \Gamma \sim j_{n+1}/(kr)
\sim (kr)^n$ has the same level of potential divergence as $\Lambda_{n+2}$.
These considerations are important for the steep spectra that play role in
turbulence studies, where correlation functions are not formally determined at
long wavelengths, and transition to regularized structure function must be
performed. Such transition amounts to regularization of the integrals, mainly in
$\Lambda_n$.
Thus, we group the $l=n+2$ term from the double sum together with the first
term to get
\begin{eqnarray}
B_\xi = 4 \pi \sum_{n=0}^\infty i^n P_n(\mu) \left( \tilde\Upsilon_n - \tilde
\Gamma_n + (2n+1) \Lambda_{n+2} \right) 
- 4 \pi \sum_{n=0}^\infty (2n+1) P_n(\mu) \sum_{l=n+4,2}^\infty i^l \Lambda_l
\end{eqnarray} 

\subsubsection{$A_\xi$}

Using Eq.~(\ref{eq:Laplacian_def}) we represent the coefficient $A_\xi$ as 
$ A_\xi = \left( \Delta - 3 D_r - D_{\mu\mu} - 2 r \mu D_r D_\mu \right) \Phi $,
thus reducing it to the combination of Eqs.~(\ref{eq:dr}-\ref{eq:Laplacian}).
Let us collect separately the terms in the single sum and the terms in the
double sum
\begin{eqnarray}
A_\xi &=& 4 \pi \sum_{n=0}^\infty i^n  P_n(\mu) \left[
- \tilde \Upsilon_n + (3+2n) \tilde \Gamma_n 
\right] + \nonumber \\
&+& 4 \pi \sum_{n=0}^\infty (2n+1) P_n(\mu) \sum_{l=n+2,2}^\infty i^l
\left[2 \tilde \Gamma_l + \onehalf \left((l-1)(l-2)-n(n+1)\right) \Lambda_l   
\right] 
\end{eqnarray}
We again separate the first term in the double sum, noting that
there is an additional cancellation of $\Lambda_{n+2}$ contribution
\begin{eqnarray}
A_\xi &=& 4 \pi \sum_{n=0}^\infty i^n  P_n(\mu) \left[
- \tilde \Upsilon_n + (2n+3) \tilde \Gamma_n - 2 (2n+1) \tilde \Gamma_{n+2}
\right] + \nonumber \\
&+& 4 \pi \sum_{n=0}^\infty (2n+1) P_n(\mu) \sum_{l=n+4,2}^\infty i^l
\left[2 \tilde \Gamma_l + \onehalf \left((l-1)(l-2)-n(n+1)\right) \Lambda_l   
\right] 
\end{eqnarray}
and then shift $l \to l+2$ in the  $\Lambda$  part of the double sum
to select $\Lambda_{n+4}$ which has the same scaling at $kr \to 0$ as the
terms in the single sum
\begin{eqnarray}
A_\xi &=& 4 \pi \sum_{n=0}^\infty i^n  P_n(\mu) \left[
- \tilde \Upsilon_n + (2n+3) \tilde \Gamma_n - 2 (2n+1) \tilde \Gamma_{n+2}
+(2n+1)(2n+3) \Lambda_{n+4}
\right] + \nonumber \\
&+& 4 \pi \sum_{n=0}^\infty (2n+1) P_n(\mu) \sum_{l=n+4,2}^\infty i^l
\left[2 \tilde \Gamma_l + \onehalf \left((l+1)l-n(n+1)\right) \Lambda_{l+2}   
\right] \quad ,
\end{eqnarray}
which gives our original expression.

\subsubsection{$D_\xi$}

The last coefficient, $D_\xi$ is different from the ones that have just been
considered in
that its multipoles are linked to multipoles of the power spectrum of the
opposite parity. While even multipoles of $E_n$, the only expected
to be present, give rise to even multipoles of $A_\xi,B_\xi$ and $C_\xi$, they
create odd multipoles of the $D_\xi$. As a consequence, $D_\xi$ cannot be
assembled using just even  operators that we have studied. The
additional odd derivative that we need is just 
\begin{equation}
\frac{\partial}{\partial \mu} P_l(\mu) = \sum_{n=(1,0),2}^{l-1} P_n(\mu) 
\end{equation}
(here summation starts with unity if l is even, and zero if l is odd, $l=0$ term
is absent) which allows us to write
\begin{eqnarray}
\label{eq:DrDmu}
r D_\mu D_r \left( j_l(kr) P_l(\mu) \right) &=&  \frac{\partial}{\partial
\mu} \left[ D_r j_l(kr) P_l(\mu) \right]  =
- \tilde \Gamma_l(r)\frac{\partial}{\partial \mu}  P_l(\mu)
- \Lambda_l(r) \sum_{n=(0,1),2}^{l-2} (2n+1) \frac{\partial}{\partial \mu}
P_n(\mu) \nonumber \\
&\to& -  \tilde \Gamma_l(r) \sum_{n=(1,0),2}^{l-1} \!\!\!\! (2n+1) P_n(\mu) - 
 \frac{1}{2}  \Lambda_l(r) \!\!\! \sum_{n=(1,0),2}^{l-1}  \!\!\!\!
(l+n) (l-n-1) (2n + 1) P_n(\mu) 
\end{eqnarray}
Reverting the summation order again in the full multipole expansion, 
$\sum_{l=0}^\infty
\sum_{n=(1,0),2}^{l-1} = \sum_{n=0}^\infty \sum_{l=n+1,2}^\infty$, and shifting
the summation index in $\Lambda_l$ part by two, $l \to l+2$,
we obtain the  Eq.~(\ref{eq:Dxi}). 

\section{Correlation of the synchrotron intensity in axisymmetric magnetic field
for $\gamma=2$}
\label{sec:g2functions}

For $\gamma=2$ we have the following relations
\begin{eqnarray}
\left\langle H_{\perp}^2 \right\rangle &=&
\left\langle H_x^2+H_y^2 \right\rangle = \sigma_{xx} + \sigma_{yy} \\
\left\langle H_{\perp}^4 \right\rangle &=&
\left\langle (H_x^2+H_y^2)^2 \right\rangle = 3 \sigma_{xx}^2 + 3 \sigma_{yy}^2
 + 2 \sigma_{xx} \sigma_{yy} + 4 \sigma_{xy}^2 \\
\left\langle H_{\perp}^2 \left({\bf x}_1 \right) H_{\perp}^2
\left({\bf x}_2 \right) \right\rangle
&=&\left\langle (H_x^2+H_y^2)_1 (H_x^2+H_y^2)_2 \right\rangle  = \\
&=& \left\langle H_x^2\left({\bf x}_1 \right) H_x^2\left({\bf x}_2 \right)
\right\rangle
+ \left\langle H_x^2\left({\bf x}_1 \right) H_y^2\left({\bf x}_2
\right)\right\rangle
+ \left\langle H_y^2\left({\bf x}_1 \right) H_x^2\left({\bf x}_2
\right)\right\rangle
+ \left\langle H_y^2\left({\bf x}_1 \right) H_y^2\left({\bf x}_2
\right)\right\rangle
\nonumber \\
&=& \sigma_{xx}^2 + 2 \sigma_{xx} \sigma_{yy} + \sigma_{yy}^2 \nonumber \\
&+& 2 \left\langle H_x\left({\bf x}_1 \right) H_x\left({\bf x}_2
\right)\right\rangle^2
 + 2 \left\langle H_x\left({\bf x}_1 \right) H_y\left({\bf x}_2
\right)\right\rangle^2
+ 2 \left\langle H_y\left({\bf x}_1 \right) H_x\left({\bf x}_2
\right)\right\rangle^2
+ 2 \left\langle H_y\left({\bf x}_1 \right) H_y\left({\bf x}_2
\right)\right\rangle^2
\nonumber \\
\left\langle \left[H_{\perp}^2 \left({\bf x}_1 \right)- H_{\perp}^2
\left({\bf x}_2 \right)\right]^2 \right\rangle
&=& 2 \left\langle H_{\perp}^4 \right\rangle
- 2 \left\langle H_{\perp}^2 \left({\bf x}_1 \right) H_{\perp}^2 \left({\bf x}_2
\right) \right\rangle =
\nonumber \\
&=& 4 \sigma_{xx}^2 + 4 \sigma_{yy}^2 + 8 \sigma_{xy}^2 \nonumber \\
&-& 4 \left\langle H_x\left({\bf x}_1 \right) H_x\left({\bf x}_2
\right)\right\rangle^2
- 4 \left\langle H_x\left({\bf x}_1 \right) H_y\left({\bf x}_2
\right)\right\rangle^2
- 4 \left\langle H_y\left({\bf x}_1 \right) H_x\left({\bf x}_2
\right)\right\rangle^2
- 4 \left\langle H_y\left({\bf x}_1 \right) H_y\left({\bf x}_2
\right)\right\rangle^2
\nonumber \\
&=& 4 \sigma_x^2 D_{xx} - D_{xx}^2 + 4 \sigma_y^2 D_{yy} - D_{yy}^2
+ 8 \sigma_{xy} D_{xy} - 2 D_{xy}^2
\end{eqnarray}
where we have used the fact that the correlation tensor is symmetric and have
denoted the same point 2D covariance
matrix by $\sigma_{ij} \equiv \langle H_i H_j \rangle $.   The final expression
can be written
in the manifestly 2D-invariant tensor form
\begin{eqnarray}
\langle H_{\perp}^4 \rangle  - \langle H_{\perp}^2 \rangle^2
&=&  2 \sum_{i,j=1}^2 \sigma_{ij} \sigma_{ji} \\
\langle H_{\perp}^2 \left({\bf x}_1 \right) H_{\perp}^2 \left({\bf x}_2 \right)
\rangle
- \langle H_{\perp}^2 \rangle^2
&=&  2 \sum_{i,j=1}^2 \xi_{ij} \xi_{ji} \\
 \langle \left[H_{\perp}^2 \left({\bf x}_1 \right)- H_{\perp}^2 \left({\bf x}_2
\right)\right]^2 \rangle
&=& \sum_{i,j=1}^2 \left( 4 \sigma_{ij} D_{ji} - D_{ij} D_{ji} \right)
\end{eqnarray}

The correlation function of the synchrotron intensity signal
$ I \left({\bf X}\right) \propto \int dz H_\perp^2\left({\bf X},z\right) $
is then given by
\begin{eqnarray}
\left\langle I \left({\bf X_1} \right) I \left({\bf X_2} \right)
\right\rangle
 - \left\langle I\left({\bf X_1} \right) \right\rangle^2 &\propto&
\int\! dz_1 \int\! dz_2 \left[\left\langle H_\perp^2\left({\bf
X_1},z_1\right)H_\perp^2\left({\bf X_2},z_2\right)\right\rangle
- \left\langle H_\perp^2\left({\bf X_1},z_1\right)\right\rangle \left\langle
H_\perp^2\left({\bf X_1},z_2\right)\right\rangle \right]
= \nonumber \\
&=& 2 \int\! dz_+ \int\! dz  \; \xi_{ij}\left({\bf X}
,z\right)\xi_{ij}\left({\bf X} ,z\right) ~.
\end{eqnarray}
Correspondingly, the structure function is
 \begin{eqnarray}
\langle \left [ I \left({\bf X_1} \right) - I \left({\bf X_2} \right)
\right]^2 \rangle
&\propto&\left\langle \left [\int dz_1 H_\perp^2\left({\bf X_1},z_1\right)
-\int dz_2 H_\perp^2\left({\bf X_2},z_2\right)\right]^2 \right\rangle
=  \\
&=& \int \! dz_+ \! \int \! dz \left\{ 4 \sigma_{ij} \left[ D_{ji}\left({\bf X}
,z\right) - D_{ji}\left(0 ,z\right) \right]
- \left[ D_{ij}\left(X,z\right) D_{ij}\left(X,z\right) - D_{ji}\left(0,z\right)
D_{ij}\left(0,z\right)\right] \right\} ~.
\nonumber
\end{eqnarray}

\section{Geometrical functions}
\label{sec:Gfunctions}
The synchrotron correlation function in approximation of
the current paper is expressed via the trace of the 2D projection of the
spectral tensor that describes the given type of waves.
We have encountered the multipole expansions of the following angular functions
that represent the 2D angular dependence of the traces of the Fast,
Alfv\'en and Slow projected spectra
\begin{eqnarray}
G^{F}_p(\theta) &=& \frac{1}{2 \pi} \int_0^{2\pi} d \psi e^{-i p \psi}
\frac{\sin^2\!\psi}{1 - \cos^2\!\psi\sin^2\!\theta} \quad,\\
G^{A}_p(\theta) &=& \frac{1}{2 \pi} \int_0^{2\pi} d \psi e^{-i p \psi}
\frac{\cos^2\!\theta}{1 - \cos^2\!\psi\sin^2\!\theta} \quad.
\end{eqnarray}
The two angular functions are related by $G_p^A(\theta) + \sin^2\theta
G_p^F(\theta) = \delta_{p0}$.
In Figure~\ref{fig:Gfunctions} we plot the behaviour of the multipoles up to
$p=6$ as the functions of the symmetry axis orientation angle $\theta$.
\begin{figure}[ht]
\centerline{\includegraphics[width=0.32\textwidth]{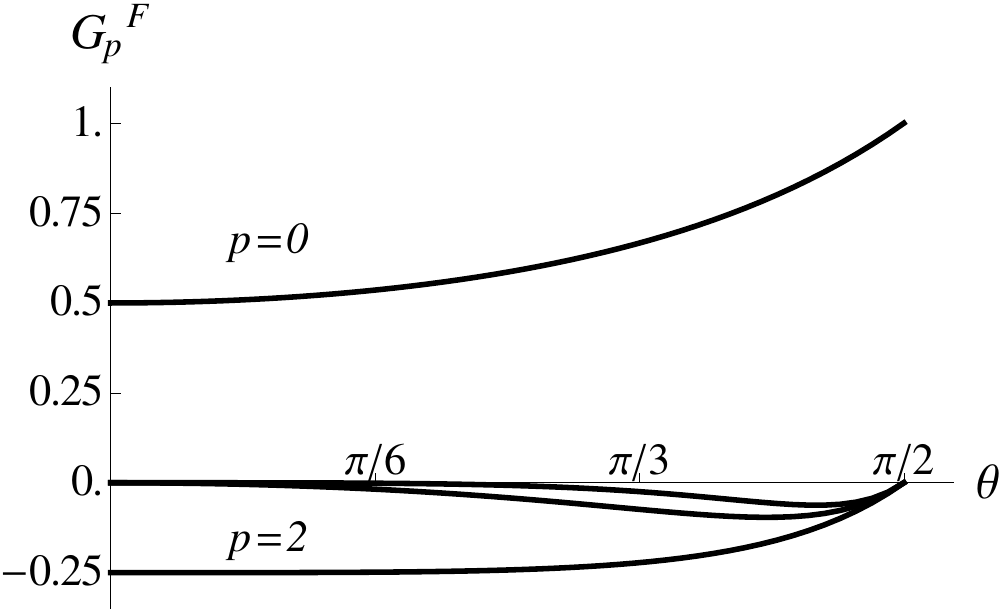}\hspace{1cm}
\includegraphics[width=0.32\textwidth]{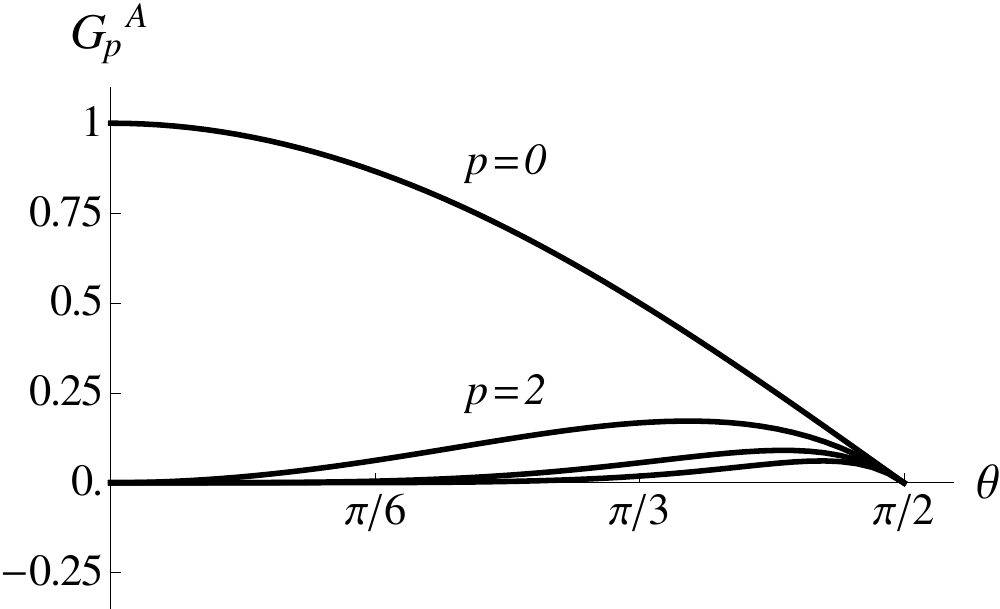}}
\caption{Low multipole behaviour for multipole decomposition of the traces
of the spectral tensors for the Fast and Slow (left, general $F$-type tensor)
and Alfv\'en (right) waves.}
\label{fig:Gfunctions}
\end{figure}

\section{Correlations of the polarization components of the synchrotron radiation}
\label{sec:zcomponents}
\subsection{Correlation of Stokes parameters}
Synchrotron intensity is just one of the characteristics of synchrotron emission. Studies of the statistics of synchrotron
polarization can be very promising (see Waelkens et al 2009). Below we discuss how the approach in the main 
part of the paper be generalized to study other Stocks parameters and their combinations. In what follows we
use the realistic model of MHD turbulence that we advocate in the paper. For the sake of simplicity we assume
that $\gamma=2$. A study of $\gamma$ dependence will be 
done in the subsequent papers.  

Complete description of the radiation incoming from the direction ${\bf X}$ on the sky
is encoded in the polarization matrix
\begin{equation}
{\bf P}({\bf X})=\left(\begin{array}{c c}
         I + Q & U + i V \\
         U - i V & I - Q
        \end{array}
\right)
= I({\bf X}) \left(\begin{array}{c c}
              1 & 0 \\
              0 & 1 \\
             \end{array}
\right)
+ \left(\begin{array}{c c}
         Q({\bf X}) & U({\bf X}) + i V({\bf X}) \\
         U({\bf X}) - i V({\bf X}) &  - Q({\bf X})
        \end{array}
\right)
\end{equation}
where $Q, U , V$ are well-known Stokes parameters.

Here we shall limit our consideration to the linear polarization, described by $Q$ and $U$.
We are interested in the correlation between two matrixes 
$\left\langle {\bf P}({\bf X}) {\bf P}({\bf X^\prime})\right\rangle$. Such correlation is convenient
to separate into parts based on transformation rules under rotation of the coordinate frame.
This gives rise to the
\begin{equation}
\mathrm{scalar:} \quad \left\langle I({\bf X}) I ({\bf X^\prime})\right\rangle \qquad
\mathrm{``vector``:} \quad \left(
\begin{array}{c}
 \left\langle I({\bf X}) Q ({\bf X^\prime})\right\rangle \\
 \left\langle I({\bf X}) U ({\bf X^\prime})\right\rangle
\end{array}
\right)
 \qquad
\mathrm{``tensor``:} \quad \left(
\begin{array}{c c}
 \left\langle Q({\bf X}) Q ({\bf X^\prime})\right\rangle &
\left\langle Q({\bf X}) U ({\bf X^\prime})\right\rangle \\
 \left\langle U({\bf X}) Q ({\bf X^\prime})\right\rangle &
 \left\langle U({\bf X}) U ({\bf X^\prime})\right\rangle
\end{array}
\right)
\end{equation}
parts. Here we put ''vector`` and ''tensor`` in the quotation marks, since they rotate with twice the 
angle $\Phi$ of the rotation of the coordinate system 
\begin{equation}
\left(
 \begin{array}{c}
 Q^\prime \\
 U^\prime
\end{array}
\right)
=
\left(
\begin{array}{c c}
\cos 2 \Phi & \sin 2 \Phi \\
-\sin 2 \Phi & \cos 2 \Phi \\
\end{array}
\right)
\left(
 \begin{array}{c}
 Q \\
 U
\end{array}
\right) \quad .
\end{equation}
The ''vector`` terms describe cross-correlation between the intensity and linear polarization, while
the ''tensor`` part describes correlation between polarizations themselves. 

For synchrotron radiation in $\gamma=2$ case, following \cite{2009MNRAS.398.1970W},
\begin{eqnarray}
I({\bf X}) &\propto& n_e \int dz \left( H_{xx}^2 + H_{yy}^2 \right) \\ 
Q({\bf X}) &\propto& p \; n_e \int dz \left( H_{xx}^2 - H_{yy}^2\right) \\ 
U({\bf X}) &\propto& p \;n_e \int dz \; 2 H_{xx} H_{yy} 
\end{eqnarray}
where $p=\sqrt{Q^2+U^2}/I$ is the degree of linear polarization and $n_e$ is the density of the electrons,
both assumed constant over the line-of-sight range where the signal forms. These expressions are in
the laboratory frame $x,y$ in which $Q$ and $U$ parameters are defined.

Evaluation of the structure functions that involve polarization 
are similar to the computations that led to Eq.~(\ref{eq:Dsync_full}) for intensity.  
Using for brevity notation 
$D_{IU}({\bf R}) = \left\langle \left[ I({\bf X_1}) - I({\bf X_2}) \right] 
\left[ U({\bf X_1}) - U({\bf X_2}) \right] \right \rangle $ 
and so on, the expressions for individual correlation components are
\begin{eqnarray}
 D_{IQ} &\propto&  4 p \int dz \times \label{eq:DIQ_gen}\\
&&  \left[ \sigma_{xx}  D_{xx}({\bf R},z) - \sigma_{yy} D_{yy}({\bf R},z)
- \frac{1}{4} \left( D_{xx}({\bf R},z)^2 + D_{yy}({\bf R},z)^2 \right)
\right] 
- \left[ \vphantom{\frac{1}{2}} {\bf R} \to 0 \right] \nonumber \\
D_{IU} &\propto& 4 p \int dz \times \label{eq:DIU_gen} \\
&&  \left[ \sigma_{xy} \left( D_{xx}({\bf R},z) + D_{yy}({\bf R},z) \right)
+ \left(\sigma_{xx} + \sigma_{yy} \right) D_{xy}({\bf R},z) 
- \frac{1}{2} D_{xy} ({\bf R},z) \left( D_{xx}({\bf R},z) + D_{yy}({\bf R},z) \right)
\right] 
- \left[ \vphantom{\frac{1}{2}} {\bf R} \to 0 \right] \nonumber \\
D_{QQ} &\propto&  4 p^2 \int dz \times \label{eq:DQQ_gen} \\
 && \left[ \sigma_{xx} D_{xx}({\bf R},z) + \sigma_{yy} D_{yy}({\bf R},z) 
- 2 \sigma_{xy}  D_{xy}({\bf R},z) 
- \frac{1}{4}  \left( D_{xx}({\bf R},z)^2 + D_{yy}({\bf R},z)^2 - 2 D_{xy} ({\bf R},z)^2 \right)
\right] 
- \left[ \vphantom{\frac{1}{2}} {\bf R} \to 0 \right] \nonumber   \\
 D_{UU} &\propto& 4 p^2 \int dz \times  \label{eq:DUU_gen} \\
&&  \left[ \sigma_{yy} D_{xx}({\bf R},z) + \sigma_{xx} D_{yy}({\bf R},z) 
+  2 \sigma_{xy}  D_{xy}({\bf R},z) 
- \frac{1}{2}  \left( D_{xx}({\bf R},z) D_{yy}({\bf R},z) + D_{xy} ({\bf R},z)^2 \right)
\right]  
- \left[ \vphantom{\frac{1}{2}} {\bf R} \to 0 \right] \nonumber \\
D_{QU} &\propto& 4 p^2 \int dz \times \label{eq:DQU_gen}\\
&& \left[ \sigma_{xy} \left( D_{xx}({\bf R},z) - D_{yy}({\bf R},z) \right) 
+ \left(\sigma_{xx}-\sigma_{yy} \right)  D_{xy}({\bf R},z)
- \frac{1}{2} D_{xy}({\bf R},z) \left( D_{xx}({\bf R},z) - D_{yy} ({\bf R},z) \right)
\right]
 - \left[ \vphantom{\frac{1}{2}} {\bf R} \to 0 \right] \nonumber 
\end{eqnarray}
where $\left[ {\bf R} \to 0 \right]$ signifies the same integrand as in the main expression but evaluated
at zero separation ${\bf R}$ between the line of sights.  We remark that $U$ or $Q$
correlations with intensity $I$ are proportional to the first power of the degree of polarization,
while polarization cross-correlations are quadratic. For weakly polarized signal it may be easier to measure
the former.

The trace of the polarization correlation ''tensor`` provides the frame invariant measure
\begin{equation}
D_{QQ}+D_{UU} \propto  4 p^2  \int dz \left[ 
(\sigma_{xx}+\sigma_{yy}) (D_{xx}({\bf R},z) + D_{yy}({\bf R},z)) - 
\frac{1}{4} \left( D_{xx}({\bf R},z) +  D_{yy}({\bf R},z) \right)^2 \right]
- \left[ \vphantom{\frac{1}{2}} {\bf R} \to 0 \right] 
\end{equation}

In Eqs~(\ref{eq:DIQ_gen}-\ref{eq:DQU_gen}) the 2D components of the magnetic field correlation (structure)
tensor $D_{ij}$ refer to the laboratory frame, in which the Stokes $Q$ and $U$ are measured. Let us
rewrite the same expressions using $D_{ij}$ in the frame aligned with the symmetry axis 
$\hat \Lambda$,  where theoretical calculations for axisymmetric turbulence are more straightforward.
From here onwards, $D^+ = D_{xx} + D_{yy}$, $D^- = D_{xx} - D_{yy} $ and $D^\times = D_{xy}$ refer to components
in such a symmetry frame, while $\Phi$ is the angle between $\hat \Lambda$ and the laboratory $x$-axis.
Limiting ourselves to the linearized approximation, applicable, as in
Eq.~(\ref{eq:Dsync_approx}), at small scales,  we find
\begin{eqnarray}
 D_{IQ} &\propto&  2 p \; (\sigma_{xx} + \sigma_{yy} ) \int dz 
\left[ \left( D^-({\bf R},z)  + \epsilon D^+({\bf R},z) \right) \cos 2\Phi
-2 D^\times ({\bf R},z) \sin 2\Phi
\right] 
- \left[ {\bf R} \to 0 \right] \label{eq:DIQ_inv} \\
 D_{IU} &\propto&  2 p \; (\sigma_{xx} + \sigma_{yy} ) \int dz 
\left[ \left( D^-({\bf R},z)  + \epsilon D^+({\bf R},z) \right) \sin 2\Phi
+2 D^\times ({\bf R},z) \cos 2\Phi
\right] 
- \left[ {\bf R} \to 0 \right]  \label{eq:DIU_inv} \\
D_{QQ} &\propto&  2 p^2 (\sigma_{xx} + \sigma_{yy} )  \int dz 
\left[  D^+({\bf R},z) + \epsilon D^-({\bf R},z) \cos 4\Phi - 2 \epsilon D^\times({\bf R},z) \sin 4 \Phi
\right] 
- \left[ {\bf R} \to 0 \right] \label{eq:DQQ_inv} \\
D_{UU} &\propto&  2 p^2 (\sigma_{xx} + \sigma_{yy} )  \int dz 
\left[  D^+({\bf R},z) - \epsilon D^-({\bf R},z) \cos 4\Phi + 2 \epsilon D^\times({\bf R},z) \sin 4 \Phi
\right] 
- \left[ {\bf R} \to 0 \right]  \label{eq:DUU_inv}\\
D_{QU} &\propto& 2 p^2 \; \epsilon \; (\sigma_{xx} + \sigma_{yy} ) \int dz 
\left[ D^-({\bf R},z) \sin 4\Phi + 2 D^\times ({\bf R},z) \cos 4\Phi \right]
 - \left[  {\bf R} \to 0 \right]
\label{eq:DQU_inv}
\end{eqnarray}
The frame invariant combinations are
\begin{equation}
D_{QQ}+D_{UU} \propto  4 p^2 \; (\sigma_{xx}+\sigma_{yy})  \int dz  
D^+({\bf R},z) 
- \left[ {\bf R} \to 0 \right] ~,
\label{eq:DQQ+UU_inv}
\end{equation}
and, for completeness, the intensity structure function from Eq~(\ref{eq:Dsync_approx})
\begin{equation}
 D_{II} \propto  2  \; (\sigma_{xx} + \sigma_{yy} ) \int dz 
\left[ \left( D^+({\bf R},z)  + \epsilon D^-({\bf R},z) \right) 
\right] 
- \left[ {\bf R} \to 0 \right] \label{eq:DII_inv} 
\end{equation}
The magnetic field variance anisotropy parameter, $\epsilon$, is defined in Eq.~(\ref{eq:epsilon}).

\subsection{Angular properties of the polarization correlations}

Let us turn to the angular dependence of the polarization functions. 
Our study of the intensity of the synchrotron in the main text can be reformulated
as the following rules of correspondence between the
different parts of the magnetic correlation tensor and the multipoles of the synchrotron correlations they
contribute to
\begin{eqnarray}
\int dz  D^+ ({\bf R},z) \quad &\to&  \quad  {\tilde  D}_{n} \propto \sum_{s=-\infty}^\infty E_s G_{n-s}(\theta)  \\
\int dz D^-({\bf R},z) \quad  &\to& \quad   {\tilde  D}_{n} \propto 
 \sum_{s=-\infty}^\infty -\frac{1}{2} \left(E_{s+2} + E_{s-2} \right) G_{n-s}(\theta)  
\end{eqnarray}
where $E_s$ are coefficients of multipole expansion of the projected power spectrum $E(K,\psi)$ wrt the sky orientation
angle of the wave vector $\psi$ and $G_p(\theta)$
are multipoles of one of the geometric functions $G(\theta,\psi)$ (see Appendix~\ref{sec:Gfunctions})
that describes the specific turbulent mode.
This rule reproduces Eqs.~(\ref{eq:DnRE}) and (\ref{eq:Dn_Fscaling}) when applied, correspondingly,
to $E$ and $F$-type modes of the general axisymmetric tensor Eq.~(\ref{eq:Hij2D}).  

To study polarization one also needs the additional component
\begin{equation}
\int dz D^\times({\bf R},z) \quad \to \quad {\tilde  D}_{n} \propto  
\sum_{s=-\infty}^\infty \frac{i}{4} \left(E_{s+2} - E_{s-2} \right) G_{n-s}(\theta)  \, .
\end{equation}
While $D^+$ and $D^-$ combinations give rise to real $\tilde D_n$ coefficients and, thus, 
symmetric $\sim \cos n \phi$ dependence of the synchrotron correlations, $D^\times$ contributes the imaginary
part to $\tilde D_n$ and antisymmetric $\sim \sin n \phi$ signal correlation
behaviour. 
Antisymmetric contribution
is not present if the mode tensor is isotropic, when $G_{n-p} = \delta_{np}$.
Presence of the
antisymmetric, $\sin n\phi$, behaviour in the correlation patterns of
polarization is a direct indication
of the $F$-term contribution to the spectral tensor structure. It is expected at
some level
for Alfv\'en and Fast modes.

 The resulting multipole decomposition of the polarization structure 
functions is
\begin{eqnarray}
 (\tilde D_{IQ})_n &\propto&  2 p \; A C_n(m) R^{1+m} \sum_{s=-\infty}^\infty
\left[\left( -\frac{1}{2} \left( E_{s+2} + E_{s-2} \right)   + \epsilon E_s
\right) \cos 2\Phi
+ \frac{i}{2}  \left( E_{s+2} - E_{s-2} \right)  \sin 2\Phi
\right] G_{n-s}^{(A,F,S)}
 \\
 (\tilde D_{IU})_n &\propto&  2 p \; A C_n(m) R^{1+m} \sum_{s=-\infty}^\infty
\left[\left( -\frac{1}{2} \left( E_{s+2} + E_{s-2} \right)   + \epsilon E_s
\right) \sin 2\Phi
- \frac{i}{2}  \left( E_{s+2} - E_{s-2} \right)  \cos 2\Phi
\right]  G_{n-s}^{(A,F,S)}
 \\
(\tilde D_{QQ})_n &\propto&  2 p^2 \; A C_n(m) R^{1+m} \sum_{s=-\infty}^\infty 
\left[  E_s - \frac{\epsilon}{2} \left( E_{s+2} + E_{s-2} \right) \cos
4\Phi + \frac{i \epsilon}{2}  \left( E_{s+2} - E_{s-2} \right) \sin 4 \Phi
\right]  G_{n-s}^{(A,F,S)}
 \label{eq:DQQ_n} \\
(\tilde D_{UU})_n &\propto&  2 p^2 \; A C_n(m) R^{1+m} \sum_{s=-\infty}^\infty 
\left[  E_s + \frac{\epsilon}{2} \left( E_{s+2} + E_{s-2} \right) \cos
4\Phi - \frac{i \epsilon}{2}  \left( E_{s+2} - E_{s-2} \right) \sin 4 \Phi
\right]  G_{n-s}^{(A,F,S)}
 \label{eq:DUU_n} \\
(\tilde D_{QU}) &\propto& 2 p^2 \; A C_n(m) R^{1+m} \sum_{s=-\infty}^\infty 
\left[  -\frac{1}{2} \left( E_{s+2} + E_{s-2} \right) \sin 4\Phi 
- \frac{i}{2}  \left( E_{s+2} - E_{s-2} \right) \cos 4\Phi \right]
 G_{n-s}^{(A,F,S)}
\label{eq:DQU_n}
\end{eqnarray}
where $G_{n-s}^{(A,F,S)}(\theta)$ stands for one of the particular geometrical mode functions, Alfv\'en, Fast or Slow,
or their combinations. It depends, parametrically on the angle $\theta$ between the mean magnetic field direction and
the line of sight.

Polarization provides new avenues to link properties of the magnetized
turbulence to observables.
The most direct measure of the turbulence spectral anisotropy is provided by the trace of the 
synchrotron polarization matrix
\begin{equation}
\left(\tilde D_{QQ+ UU}\right)_n  \propto  4 p^2 \;  A_{(A,F,S)} C_n(m) R^{1+m}
 \sum_{s=-\infty}^\infty E_s G^{(A,F,S)}_{n-s}(\theta)
\end{equation}
It may be possible to determine the direction of the axis of symmetry (mean magnetic field) by measuring separately
symmetric ($\cos$) and antisymmetric ($\sin$) patterns of the correlations. For example 
$\tan 2 \Phi = \mathrm{Re} (D_{IU})_n/ \mathrm{Re} (D_{IQ})_n $ can be compared with the direction obtained from elongation
of iso-correlation contours of intensity.  And when symmetry direction is determined, $\epsilon$ parameter
can be estimated from $\mathrm{Im} (D_{QU})_n/ \mathrm{Im} (D_{IU})_n = p \epsilon$ evaluated in the symmetry frame $\Phi=0$, if there is
a non-vanishing imaginary contribution to correlation multipoles from $D_{xy}$.

We may conclude that angular and scaling dependencies of the correlations of synchrotron polarization and intensity, 
Eqs~(\ref{eq:DIQ_inv}-\ref{eq:DII_inv}),
contain wealth of information about orientation of the symmetry axis, $\epsilon$ parameter
and full projected correlation tensor of the magnetic field.  Redundancies build into
different signals allow for multiple cross-checks. Detail study of the possibilities that synchrotron polarization
opens for studies of anisotropic magnetized turbulence is a subject of the future work.

\subsection{Correlations of the line-of-sight component of the magnetic field}

Faraday rotation measures provide a valuable source information about magnetic 
turbulence. However, the models of turbulence adopted in earlier studies were 
models of isotropic turbulence. Below we calculate the statistics of Faraday rotation
measures which is based on a realistic MHD turbulence model. The measures we
deal with do not depend on the cosmic ray spectral index.

Study of intensity and linear polarization of the synchrotron
led us to the discussion of the orthogonal to the line-of-sight
components of the magnetic field.
Let us in conclusion collect the results for the projected 
correlation of the line-of-sight component 
$D_{zz}({\bf R}) = \int dz \langle H_z({\bf x}_1) H_z({\bf x}_2 \rangle$.
This quantity describes the correlation of Faraday phase 
$\langle \phi({\bf X}_1) \phi({\bf X}_2) \rangle$  and is important in
polarization studies. We present the results
for the line-of-sight component  here to complete our statistical formalism.

From Eq.~(\ref{eq:Hij2D}) one immediately obtains the sky-projected structure
function between the line-of-sight ($z$) components of the magnetic field as 
\begin{eqnarray}
D_{zz}({\bf R}) = \int dz \langle H_z H_z \rangle \propto \frac{1}{(2\pi)^2}
\int d^2 K
e^{i {\bf K \cdot R }}
\left[\vphantom{\frac{ \cos^2\psi \hat K_i \hat K_j +
\hat \Lambda_i \hat \Lambda_j - \cos\psi
(\hat K_i \hat \Lambda_j+\hat K_j \hat \Lambda_i)}{1 - \cos^2\psi \sin^2\theta}}
E({\bf K})  +   
F({\bf K}) \; \frac{  \cos^2\!\theta}{1 - \cos^2\!\psi \sin^2\!\theta}
\right]\quad .
\label{eq:Hzz2D}
\end{eqnarray}
Spectral representation of different turbulence modes can be 
expressed again using combinations of the geometrical functions introduced in
Appendix~\ref{sec:Gfunctions}. Namely, using
Eqs.~(\ref{eq:Alfven_tensor_loc},  \ref{eq:slowtensor},
\ref{eq:Alfven+Slow_tensor}, \ref{eq:Fspectrum})
\begin{equation}
\begin{array}{llll}
\text{Alfven}  & & F({\bf K}) = -E({\bf K})
 & D_{zz} \propto  E({\bf K}) \sin^2\!\theta G^F \\
\text{Slow~(high}~\beta) & E({\bf K}) = 0, & & D_{zz} \propto F({\bf K})  G^A \\
\text{Alfven + Slow~(strong)} && F({\bf K}) = 0
 & D_{zz} \propto E({\bf K}) \\
Fast~(low~\beta) & E({\bf K}) = 0, & F({\bf K}) = F(K) & D_{zz}
\propto F(K)  G^A  \\
Fast~(high~\beta) & E({\bf K}) = 0, & F({\bf K}) = F(K) \left(1
-\cos^2\!\psi
\sin^2\!\theta \right) & D_{zz} \propto F( K)
\cos^2\!\theta
\end{array}
\label{eq:Dzz}
\end{equation}

Multiple decomposition of the angular dependence of the $D_{zz}({\bf R})$ is
described by convolution of the multiple decompositions of the spectral and
geometrical functions in Eq.~(\ref{eq:Dzz}). Variation of the direction of the
mean magnetic field along the line of sight leads to isotropization of the
$D_{zz}$, in a similar way that we have studied for the synchrotron
intensity.

\end{document}